\RequirePackage{rotating}
\documentclass[a4paper,fleqn,usenatbib]{mnras}
\usepackage{newtxtext,newtxmath}
\usepackage[T1]{fontenc}
\usepackage{ae,aecompl}
\usepackage{ulem}  

\usepackage{eso-pic}

\AddToShipoutPictureBG*{%
  \AtPageUpperLeft{%
    \hspace{0.75\paperwidth}%
    \raisebox{-3.5\baselineskip}{%
      \makebox[0pt][l]{\textnormal{DES 2018-0390}}
}}}

\AddToShipoutPictureBG*{%
  \AtPageUpperLeft{%
    \hspace{0.75\paperwidth}%
    \raisebox{-4.5\baselineskip}{%
      \makebox[0pt][l]{\textnormal{FERMILAB-PUB-18-674}}
}}}%

\usepackage{graphicx}	
\usepackage{amsmath}	
\usepackage{amssymb}	
\usepackage{lscape} 
\usepackage{rotating}
\usepackage{lineno}

\newcommand*{\Msun}{\ensuremath{\, M_{\odot}}}

\newcommand*{\LamMCMF}{$\lambda_{\mathrm{MCMF}}$}
\newcommand*{\LamRM}{$\lambda_{\mathrm{RM}}$}
\newcommand*{\ProbLam}{$P_\lambda$}
\newcommand*{\ProbS}{$P_\mathrm{s}$}

\newcommand{\XMM}{XMM{\it -Newton}}
\newcommand{\Chandra}{{\it Chandra}}
\newcommand{\fcont}{{$f_\mathrm{cont}$}}
\newcommand{\fcontr}{{$f_\mathrm{cont,r}$}}

\DeclareMathOperator\arctanh{arctanh}

\title[New RASS Galaxy Clusters to $z\sim1$]{A New RASS Galaxy Cluster Catalog with Low Contamination Extending to $z\sim1$ in the DES Overlap Region}

\author[Klein et al.]{
\parbox{\textwidth}{
\Large
M.~Klein$^{1,2}$\thanks{E-mail: mklein@usm.uni-muenchen.de},
S.~Grandis$^{3,1}$,
J.~J.~Mohr$^{3,1,2}$,
M.~Paulus$^{1}$,\\
T.~M.~C.~Abbott$^{4}$,
J.~Annis$^{5}$,
S.~Avila$^{6}$,
E.~Bertin$^{7,8}$,
D.~Brooks$^{9}$,
E.~Buckley-Geer$^{5}$,
A.~Carnero~Rosell$^{10,11}$,
M.~Carrasco~Kind$^{12,13}$,
J.~Carretero$^{14}$,
F.~J.~Castander$^{15,16}$,
C.~E.~Cunha$^{17}$,
C.~B.~D'Andrea$^{18}$,
L.~N.~da Costa$^{11,19}$,
J.~De~Vicente$^{10}$,
S.~Desai$^{20}$,
H.~T.~Diehl$^{5}$,
J.~P.~Dietrich$^{3,1}$,
P.~Doel$^{9}$,
A.~E.~Evrard$^{21,22}$,
B.~Flaugher$^{5}$,
P.~Fosalba$^{15,16}$,
J.~Frieman$^{5,23}$,
J.~Garc\'ia-Bellido$^{24}$,
E.~Gaztanaga$^{15,16}$,
P.~A.~Giles$^{25}$,
D.~Gruen$^{26,17,27}$,
R.~A.~Gruendl$^{12,13}$,
J.~Gschwend$^{11,19}$,
G.~Gutierrez$^{5}$,
W.~G.~Hartley$^{9,28}$,
D.~L.~Hollowood$^{29}$,
K.~Honscheid$^{30,31}$,
B.~Hoyle$^{2,32}$,
D.~J.~James$^{33}$,
T.~Jeltema$^{29}$,
K.~Kuehn$^{34}$,
N.~Kuropatkin$^{5}$,
M.~Lima$^{35,11}$,
M.~A.~G.~Maia$^{11,19}$,
M.~March$^{18}$,
J.~L.~Marshall$^{36}$,
F.~Menanteau$^{12,13}$,
R.~Miquel$^{37,14}$,
R.~L.~C.~Ogando$^{11,19}$,
A.~A.~Plazas$^{38}$,
A.~K.~Romer$^{25}$,
A.~Roodman$^{27}$,
E.~Sanchez$^{10}$,
V.~Scarpine$^{5}$,
R.~Schindler$^{27}$,
S.~Serrano$^{15,16}$,
I.~Sevilla-Noarbe$^{10}$,
M.~Smith$^{39}$,
R.~C.~Smith$^{4}$,
M.~Soares-Santos$^{40}$,
F.~Sobreira$^{41,11}$,
E.~Suchyta$^{42}$,
M.~E.~C.~Swanson$^{13}$,
G.~Tarle$^{22}$,
D.~Thomas$^{6}$,
V.~Vikram$^{43}$
and the DES Collaboration
\\
}}

\date{Accepted XXX. Received YYY; in original form ZZZ}
\pubyear{2016}
\begin{document}
\label{firstpage}
\pagerange{\pageref{firstpage}--\pageref{lastpage}}
\maketitle


\begin{abstract}
We present the MARD-Y3 catalog of between 1086 and 2171 galaxy clusters (52\% and 65\% new)
produced using multi-component matched filter (MCMF) followup in 5000\,deg$^2$ of DES-Y3 optical data of
the $\sim$20000 overlapping 2RXS X-ray sources. 
Optical counterparts are identified as peaks in galaxy richness
as a function of redshift along the line of sight toward each 2RXS source within a search region informed
by an X-ray prior.  All peaks are assigned a probability \fcont\ of being a random superposition.
The clusters lie at $0.02<z<1.1$ with more than 100 clusters at $z>0.5$. 
Residual contamination is 2.6\% and 9.6\% for the cuts adopted here. 
For each cluster we present the optical center, redshift, rest frame X-ray luminosity,
$M_{500}$ mass, coincidence with NWAY infrared sources and estimators of dynamical state.  
About 2\% of MARD-Y3 clusters
have multiple possible counterparts, the
photo-z's are high quality with $\sigma_{\Delta z/(1+z)}=0.0046$, and 
$\sim$1\% of clusters exhibit evidence of X-ray luminosity boosting from emission by cluster AGN.
Comparison with other catalogs (MCXC, RM, SPT-SZ, Planck) is
performed to test consistency of richness, luminosity and mass estimates.
We measure the MARD-Y3 X-ray luminosity function
and compare it to the expectation from a fiducial cosmology and externally
calibrated luminosity- and richness-mass relations.  Agreement is good, 
providing evidence that MARD-Y3 has low contamination and
can be understood as a simple two step selection-- X-ray and then optical-- of an underlying
cluster population described by the halo mass function.
\end{abstract}



\section{Introduction}

The endeavor to use galaxy clusters to investigate the cosmic acceleration, the standard cosmological parameters as 
well as extensions to the standard model using the amplitude of mass fluctuations has rapidly improved in the 
past years with an increased understanding of cluster properties and larger cluster samples \citep[][]{wang98,haiman01,vikhlinin09,mantz10a,rapetti10,bocquet15,dehaan16,bocquet18}.
Galaxy clusters also provide the tightest constraints on the dark matter self interaction cross section to date 
\citep{Sartoris14,robertson17}, and the efforts to understand clusters as cosmological probes in turn offers insights 
into plasma physics and galaxy evolution.

One obvious first step before  clusters can be used  as probes of different physical processes is their 
identification.
Cluster detection techniques based on the hot intra-cluster medium (ICM), such as the measurement of the X-ray flux or  the Sunyaev-Zel'dovich Effect (SZE) signature, do not provide all the information needed to make optimal use of those 
cluster candidate catalogs. Both techniques require, for a significant fraction of the sources, auxiliary data to obtain 
redshift estimates and to provide the opportunity to reduce any sample contamination. 

With increasing numbers of cluster candidates, a systematic and automated method needs to be applied to 
objectively confirm clusters and assign redshifts to those systems.
As an example, the eROSITA \citep{predehl10} all sky X-ray survey will likely detect $\sim$10$^5$ 
clusters \citep{merloni12,grandis18} together with more than three million X-ray AGN along with other sources. Cluster 
redshifts from X-ray data alone will be only available for a small fraction of sources and only to a precision of $
\Delta z/(1 + z)=\lesssim 0.1$ \citep{Borm14}, which we demonstrate in our work here is a factor $\sim$20 worse than what is achievable with state of the art optical imaging data.  
The Multi-Component Matched Filter Cluster Confirmation Tool \citep[MCMF;][]{Klein18}, is designed for use on large 
scale imaging surveys such as the Dark Energy Survey \citep[DES;][]{DESC16} to do automated confirmation 
and redshift estimation for large surveys like eROSITA.

In this work we use MCMF to confirm clusters detected in the ROSAT All-Sky Survey \citep[RASS,][]{truemper82} 
over 5000\,deg$^2$ using DES imaging data. More precisely, we use the proprietary DES Y3A2 GOLD 
catalog, which is a value-added version of the catalog recently published with the DES DR1 dataset 
\citep{DESDR1}, to investigate $\sim$20000 candidates from the second ROSAT All-Sky Survey source catalog 
(2RXS) presented in \cite{boller16}. 
As described in detail in our pilot study \citep{Klein18}, MCMF uses a red sequence (RS) galaxy technique together with an
X-ray prior and matched random pointings to obtain redshifts and exclude chance superpositions.

This paper is structured as follows. In Section~\ref{sec:data} we describe the dataset used in this work, and in 
Section~\ref{sec:method} we outline the cluster confirmation method. 
The application of the confirmation method and the properties of the resulting cluster catalog are described in 
Section~\ref{sec:application}. The conclusion of this paper appears in Section~\ref{sec:conclusions}. 
Throughout this paper we adopt a flat $\Lambda$CDM cosmology with $\Omega_M=0.3$ and $H_0=70$
\,km\,s$^{-1}$\,Mpc$^{-1}$.


\section{Data}
\label{sec:data}

\begin{figure*}
\includegraphics[keepaspectratio=true,width=2.0\columnwidth]{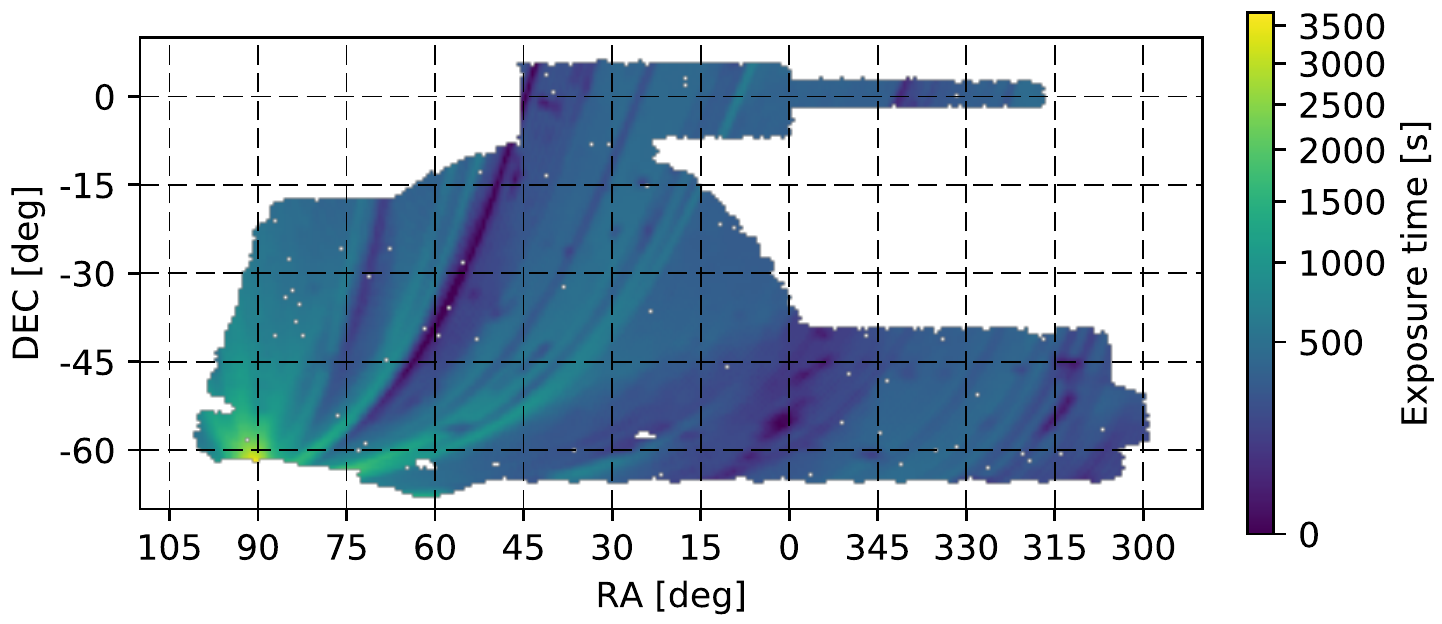}
\vskip-0.10in
\caption{RASS exposure time distribution over the DES-Y3A2 footprint in Cartesian projection. Regions with no 
DES data or which are masked in DES are shown in white. RASS exposure times are color coded from dark blue (low exposure) to bright yellow (high exposure). }
\label{fig:RASSinDES}
\end{figure*}

This paper uses data from DES and RASS. We restrict the description of the datasets here to the minimum 
needed for this paper and refer the interested reader to the papers dedicated to describing the details of the surveys.  
In Fig.~\ref{fig:RASSinDES} we show the RASS exposure time distribution over the DES footprint.

\subsection{The DES Y3A2 GOLD catalog}
This work makes use of $g$, $r$, $i$, $z$ DECam \citep{flaugher15} imaging data from DES, obtained within the first three years of the survey, between August 2013 and February 2016. The data reduction and basic data quality of the imaging data are described in detail elsewhere
\citep{DESDR1,DESdatareduction}. 
The DES Y3A2 GOLD is a value-added version of the catalog available within the public data release 1 (DR1), 
and it covers about 5000\,deg$^2$ in area with at least one exposure per filter. The typical number of overlapping exposures per 
band is 3-5. The 95\% completeness limits are 23.72, 23.34, 22.78 and 22.25 for $g$, $r$, $i$ and $z$ bands, 
respectively.
Similar to DES Y1A1 Gold  \citep{DESY1Gold} the DES Y3A2 GOLD catalog includes additional calibration steps, additional types of photometry and the flags needed for optimal usage of DES data for cosmological studies.
While the set of additional value-added products is large, we limit the discussion here to the actual quantities used 
in this work and refer the interested reader to other sources for additional information \citep[][Y3Gold, in prep]{DESY1Gold}.
 
The coadded images produced by the DESDM pipeline, in contrast to the COSMODM pipeline used in 
\cite{Klein18}, were not PSF homogenised. The argument leading to the decision to not perform PSF homogenisation 
was that this causes correlated scatter in the coadd images, which impacts the estimate of the photometric 
errors. Unfortunately, the usage of DETMODEL photometry for low noise colors, as in our previous work \citep{Klein18}, 
is untenable without homogenization due to PSF discontinuities within the coadd images \citep[for more discussion, see][]{desai12}.

As an alternative to DETMODEL, the DES Y3A2 GOLD catalog contains the multi-epoch, multi-band, 
multi-object fitting photometry "MOF". This photometry method is based on the ngmix code
\citep{DESY1Gold,2014MNRAS.444L..25S,2017ApJ...841...24S,2016MNRAS.460.2245J}, which fits a galaxy model to each 
single epoch exposure and band at the same time. The fit is performed for each source in the DES Y3A2 coadd 
catalog and includes simultaneous fitting of multiple neighboring sources for improved de-blending. The fit 
makes use of the interpolated PSF model at the location of a source for each single epoch image and therefore 
uses the full information available at a given location.
In addition to MOF, the DES Y3A2 GOLD catalog provides the single object fitting photometry "SOF". 
SOF is acquired in the same way as MOF with the only difference being that it masks neighboring sources instead of 
simultaneously fitting them. Tests have shown that SOF and MOF perform similarly well with the difference 
that the number of failures is lower in SOF.  We therefore choose SOF as our default photometry for measuring galaxy colors.

We further make use of the star-galaxy separator available in GOLD, which is 
expanded compared to Y1A1 \citep{DESY1Gold} to include MOF based extent information. In this 
work we only exclude unresolved objects to i=22.2~mag, which may result in some contamination by close binaries and single stars in the galaxy catalog, especially at 
fainter magnitudes. This could, in principal, impact cluster measurements at redshifts greater than  z=0.66, when our 
fiducial flux selection exceeds i=22.2~mag, and the color of red cluster galaxies gets closest to the stellar locus. At those redshifts 
we adopt a local background correction approach, 
which statistically accounts for any remaining stellar contamination.

The Y3A2 GOLD catalog provides bad region masking similar to that described in the Y1A1 version 
\citep{DESY1Gold}. We use that information to exclude the regions around bright stars but keep regions around 
nearby galaxies in our catalog, because we assume that some of those sources could be members of 2RXS 
detected galaxy clusters.

A last additional piece of information available in the GOLD version of Y3A2 and used in this work is the photometric calibration of the sources 
to the "top of the galaxy". This includes zero point calibrations, chromatic corrections and corrections to galactic extinction using SED based de-reddening.

\subsection{The Second ROSAT All-Sky Survey Source catalog}\label{sec:2rxs}
Similarly to our previous work \citep{Klein18}, we use the second ROSAT All-Sky Survey source catalog 
\citep[2RXS; ][]{boller16}, to produce an X-ray selected cluster catalog.
The 2RXS is based on the RASS-3 processed photon event files and uses an improved methodology compared to 
the 1RXS catalogs \citep{voges99,voges00}. The full catalog contains 135,000 sources, of which  $\sim$30\% 
are expected to be spurious sources \citep{boller16}.

Apart from count rates within a 5 arcmin radius aperture, the 2RXS catalog further includes
measurements like source extent, source variability and hardness ratio. The large RASS survey PSF with a FWHM 
of  $\sim$4~arcmin \citep{boese00} and typically low number of source counts hampers the reliable detection 
of clusters as extended sources. We therefore do not use that information for the main cluster catalog.
Similarly, source variability is only significantly detected for a small number of sources and therefore can not be used 
to remove non-cluster sources from the X-ray candidate catalog.

As in our previous analysis, we therefore examine all $\sim$20000 sources within the DES footprint, from which we 
expect $\sim$10\% to be clusters based on previous work \citep{henry06,Ebeling13,Klein18}.


\section{Cluster Confirmation Method}
\label{sec:method}

Only a small fraction ($\le10\%$) of 2RXS sources are clusters, and given the lack of extent information for all but the few lowest redshift and highest mass clusters, we require an optical confirmation to identify a 2RXS source as a cluster.  Moreover, given the large number of optical systems together with the density of 2RXS sources, the likelihood of chance superpositions is significant.  Thus, we must also characterize the probability that a 2RXS source with an optical counterpart is an actual cluster.  To this end
we use the color-magnitude-redshift dependency of passively evolving galaxies, the so called red-sequence (RS) 
\citep[][]{gladders00}, together with the spatial clustering of galaxies to identify galaxy overdensities along the line 
of sight towards each 2RXS source. We include X-ray information by estimating the number of excess galaxies 
(richness) within a redshift dependent region of interest associated with each 2RXS source.  The region of interest $r_{500}$ 
is defined by the implied X-ray luminosity and inferred mass estimate at each redshift.  To eliminate 
contamination by chance superpositions, we compare the identified overdensities of each 2RXS source with those found 
along random lines of sight with similarly sized radial apertures.  These random lines of sight exclude regions with 2RXS detections. 
The richness distribution of 2RXS sources and randoms at a given redshift allow us to estimate the probability of a chance 
superposition given the redshift, richness, implied $L_\mathrm{X}$ of each source.  We use this information to
estimate the expected fraction of random superpositions contaminating
 the 2RXS cluster catalog at a given redshift, $L_\mathrm{X}$ and above a given richness.
 
A detailed description of the optical cluster confirmation method and results of an initial application to 208\,deg$^2$ of the DES 
science verification data are presented in our previous work \citep{Klein18}.
Rather than providing a full description, we focus here on changes and improvements to MCMF with respect to our previous work.

\subsection{X-ray luminosity}\label{sec:XL}

The basis of our X-ray prior is the source count rate in the  0.1-2.4 keV band given in 2RXS, obtained within a 5$'$ 
radius around each 2RXS position.  From that we calculate a simplified estimate of the cluster X-ray luminosity 
using an APEC plasma model \citep{smith01} with fixed temperature (5\,keV) and metallicity (0.4 solar) and given 
redshift and neutral hydrogen column density. 
We further assume that this simplified luminosity $L_\mathrm{X}$ is closely related to $L_{500}$, the luminosity within a 
radius within which the mean density is 500 times the critical density of the universe at the assumed cluster 
redshift. 
The fixed size aperture used for the X-ray source counts will cause additional scatter and bias between $L_\mathrm{X}$ and  
$L_{500}$. The impact may well be small given that a change of aperture size of a factor two changes 
the luminosity by only a few $(\sim6)$ percent as well as the large intrinsic scatter in $L_\mathrm{X}$ at a given mass together with the Poisson noise in the measurement uncertainty. 
We do expect the 5 arcmin radius aperture to lead to a systematic underestimate of low redshift and massive clusters.
However, this impacts the confirmation of clusters only marginally, because we compare quantities like richness to those obtained along random lines of sight obtained with the same systematic effect.
Only when comparing to external quantities such as X-ray luminosities extracted from pointed \XMM\ or \Chandra\ observations, do we need to correct for this effect.

\subsection{Cluster mass and follow-up region of interest}

We measure the cluster matched filter richness \LamMCMF\ as a function of redshift along the line of sight towards 
each X--ray selected candidate.  \LamMCMF\ is extracted within a radius $r_{500}$ derived from an observable mass relation.  
In this work, we derive this radius using 
the estimated luminosity at that redshift and an $L_{X}$--mass scaling relation. For this analysis we 
adopt the scaling relation from the analysis of \cite{bulbul19}, which uses the SZE 
selected cluster catalog from SPT \citep{bleem15} and deep XMM observation to consistently derive multiple 
observable--mass relations.  

Within \cite{bulbul19} three different forms of the scaling relations are presented for two different
sets of priors. We choose the second form of the scaling relations presented in that paper, which has the form
\begin{eqnarray}
\label{eq:xray_type_II}
L_{500,0.5-2.0 \mathrm{~keV}} = A_\mathrm{X}\left( \frac{M_{500}}{M_\mathrm{piv}} \right)^{B_\mathrm{X}} 
\left(\frac{E(z)}{E(z_\mathrm{piv})}\right)^2
\left( \frac{1+z}{1+z_\mathrm{piv}} \right)^{\gamma_\mathrm{X}}.
\end{eqnarray}
Here, $A_\mathrm{X}$, $B_\mathrm{X}$ and $\gamma_\mathrm{X}$ are the free parameters of the scaling relation that have best fit values of $4.15\times10^{44}$~erg s$^{-1}$, 1.91 and 0.252, respectively \citep[see table 5,][]{bulbul19}. Those results use SZE based halo mass information derived from X-ray calibrated SZE cluster number counts combined with BAO data \citep[see in table 3, column 2, ][]{dehaan16}. 
The pivot mass $M_\mathrm{piv}$ and redshift $z_\mathrm{piv}$  are $6.35\times10^{14} M_\odot$ and 0.45, respectively. 

To calculate the region of interest we simply use $L_\mathrm{X}$ instead of $L_{500,0.5-2.0 \mathrm{~keV}} $ to calculate $M_{500}$ and $r_{500}$ from it.
This may seem to be a bold assumption, given that the X-ray flux is neither measured within $r_{500}$ nor in the $0.5-2.0 \mathrm{~keV}$ energy band, but a precise matching of the radius is not needed at this stage. The confirmation process relies on comparison with random lines of sight, which are obtained
in precisely the same way as for real 2RXS sources. Small differences in scaling relations largely cancel out.
In Section~\ref{sec:comparison} we show for a subset of clusters with externally published masses, that our estimated X-ray $L_\mathrm{X}$ based masses are only off by ~12\% (and, therefore, the estimated $r_{500}$ by just 4\%).

 \subsection{Radial filter}
 
 We use the clustering information in our method by applying a radial weighting $\Sigma(R)$ based on a Navarro, 
Frenk and White (NFW) profile \citep{navarro97}. 
 The projected profile that we use for the  spatial weighting is \citep{bartelmann96}
\begin{equation}
\Sigma(R) \propto \frac{1}{(R/R_s)^2-1}f(R/R_s),
\label{eqn:radfilter}
\end{equation}
where $R_s$ is the characteristic scale radius, and
\begin{equation}
f(x) = 
\begin{cases}
1-\frac{2}{\sqrt{x^2-1}}\arctan\sqrt{ \frac{x-1}{x+1} } & (x>1) \cr
1-\frac{2}{\sqrt{1-x^2}}\arctanh\sqrt{ \frac{1-x}{x+1} } & (x<1) .
\end{cases}
\end{equation}
In this work we use a scale radius $\mathrm{R}_\mathrm{s}= R_{500}/6$, which is somewhat higher than the typical 
concentration of RS galaxies found in massive clusters extending to redshift $z\sim1$ \citep{hennig17}. Tests in \citet{Klein18} indicate
that the catalog is not highly sensitive to the adopted concentration.
To avoid the central singularity of the projected NFW profile, we adopt 
a minimum radius of 0.15~Mpc, below which we set the radial weight to be constant \citep{rykoff14}.

Following previous work \citep{Klein18,rykoff14}, we define the radial weight as 
\begin{equation}
n_i (z)= C_{\mathrm{rad}}(z) 2\pi R_i \Sigma(R_i),
\end{equation}
where  $C_{\mathrm{rad}}$ has to fulfill 
\begin{equation}
1 = C_{\mathrm{rad}}\int_0^{R_{500}(z)} dR\ 2\pi R\Sigma(R).
\end{equation}

\subsection{Color-Magnitude filter}\label{sec:colmag}
\begin{figure*}
\includegraphics[keepaspectratio=true,width=0.97\columnwidth]{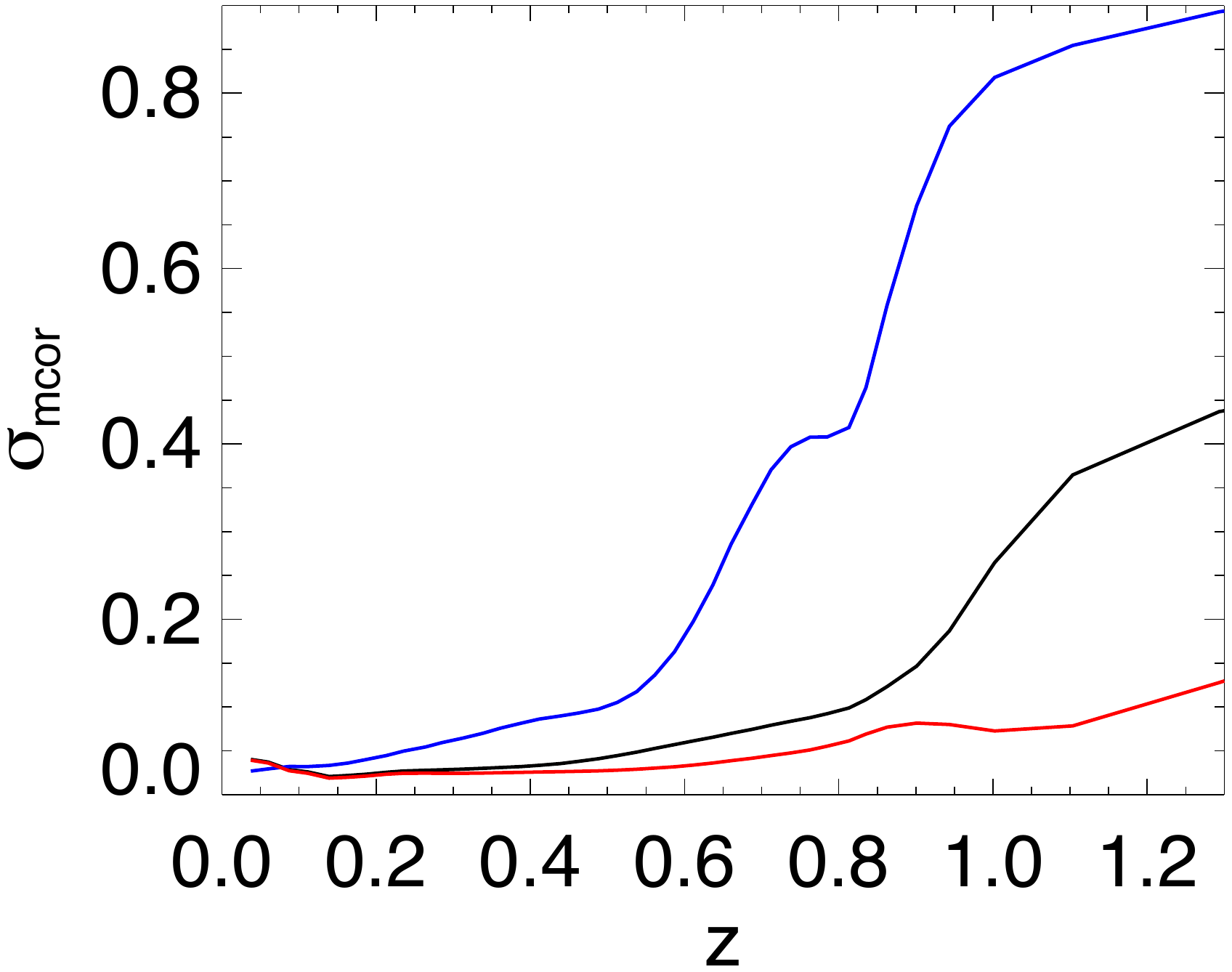}
\includegraphics[keepaspectratio=true,width=0.95\columnwidth]{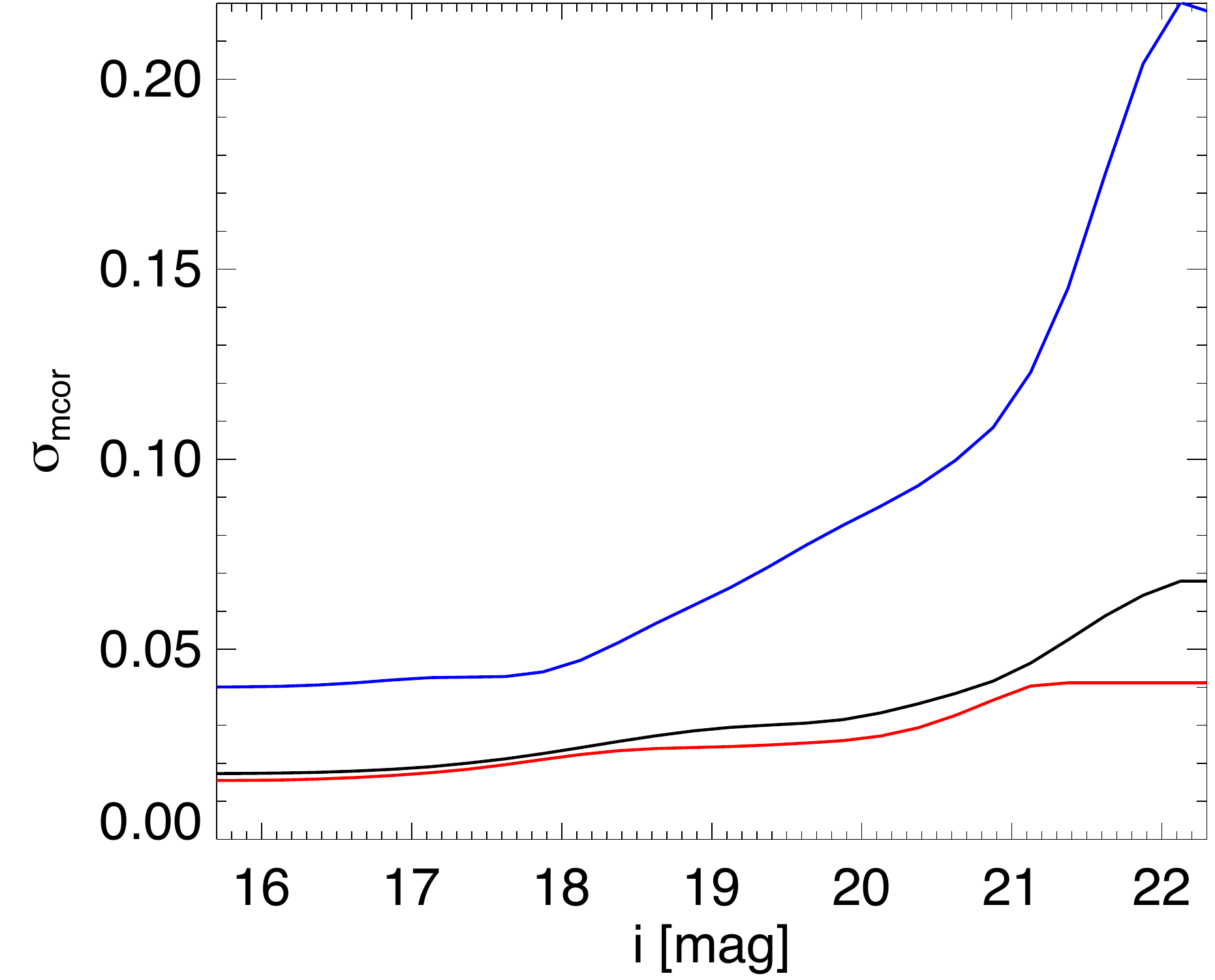}
\vskip-0.10in
\caption{Left: Measurement error corrected color width of RS galaxies at m* for $g-r$ (blue), $r-i$ (black) and $i-z$ (red). 
Right: Measurement error corrected color width versus magnitude at z=0.31. The characteristic magnitude m* is 
$i\approx 19$ mag.}
\label{fig:colorscatter}
\end{figure*}

The color-magnitude filter typically has the strongest impact on the performance of the cluster confirmation and 
redshift estimate.
We therefore recalibrate and refine our RS models by using a set of $\sim$2500 clusters and groups with 
spectroscopic redshifts (spec-z's). This catalog is a mix of three main catalogs, the redMaPPer (RM) Y1 catalog  \citep{mcclintock19},
the SPT-SZ cluster catalog \citep{bleem15} and a cross match of 2RXS sources with the MCXC  cluster catalog \citep{piffaretti11}.
We produce stacked, background subtracted $\Delta$color-magnitude histograms  within $r_{500}$ for the 
redshift range $0.01<z<1.3$.  Here $\Delta$color means that we subtract the color predicted by our initial red 
sequence model from each measured one, using the spectroscopic cluster redshift. As initial RS model we used the model adopted in our pilot study. Those RS models are assuming a simple linear relation between magnitude and color of RS galaxies and therefore consists only of a slope and a normalisation. More complex models were investigated but did not show improved performance.
We update our RS models using the 
observed offsets in normalisation and slope within a magnitude range of $-4.5<m-m^*<2$. The characteristic magnitude $m^*$ used in this work is based on a star formation model with an exponentially decaying starburst at a redshift z = 3 that has a Chabrier initial mass function and a decay time of 0.4 Gyr \citep{bruzual03}.  
After three iterations no significant offsets in the colors are found, and the process of estimating the RS models has converged.  

After calibrating the color-magnitude-redshift relation of the RS, we create a final set of stacked color-magnitude 
histograms excluding the RM clusters. Those final stacked color-magnitude histograms are then used to measure the total 
width of the RS given redshift and magnitude. The RM clusters are excluded because of the lack of a reasonably calibrated 
mass observable scaling relation when the RS models were produced.
 Based on the measurement errors for the colors of galaxies 
close to the RS, we calculate a measurement scatter corrected width $\sigma_\text{mcor}
=\sqrt{\sigma_\text{tot}^2 - \sigma_\text{meas}^2}$. 
This measurement scatter corrected RS width allows us to alter the color-magnitude filter used in our previous work to 
the following form,
\begin{equation}
 w_i(z)=\frac{\prod\limits_{j=1}^{3}{G\left(c_{i,j}-\left<c(f,z)\right>_j,\sigma_{c_{i,j}}(f,z)\right)}}{N(\sigma_{c_{i,1}}
(f,z),\sigma_{c_{i,2}}(f,z),\sigma_{c_{i,3}}(f,z))}.
 \label{eq:color_weight}
\end{equation}
 Here $G\left(c_{i,j}-\left<c(f,z)\right>_j,\sigma_{c_{i,j}}(f,z)\right)$ is the value of the normalized Gaussian Function 
at a color offset between observed color and predicted RS color given observed $i$-band magnitude $f$ of source 
$i$ and assumed redshift $z$. Similar to our pilot study, the color combinations $c_{j}$
 correspond to $(c_1,c_2,c_3)$=($g$-$r$,$r$-$i$,$i$-$z$).
 The standard deviation of the Gaussian function is $\sigma_{c_{i,j}}(f,z)=\sqrt{\sigma^2_\mathrm{mcor}(f,z) + 
\sigma^2_\mathrm{meas,i}}.$
In Fig.~\ref{fig:colorscatter} we show the measurement corrected RS width as a function of redshift at the characteristic magnitude $m^*$ and the dependency of the width on $i$-band magnitude at a fixed redshift.

\subsection{DES depths and incompleteness correction}
 
Similar to our previous work \citep{Klein18} we limit the number of galaxies investigated at a given redshift by  
selecting 
 galaxies with $m^*(z)-3 \leqq i\leqq m^*(z)+1.25$, where $m^*(z)$ is the expected characteristic 
magnitude for a cluster at redshift $z$. 
This magnitude range is modified if one of those limits encompasses the bright or faint magnitude limit of the data. 
The standard photometry within DES is not optimized to deal with bright nearby galaxies.  We therefore impose a magnitude cut of i=13.5 and ignore brighter sources.

\begin{figure}
\includegraphics[keepaspectratio=true,width=0.95\columnwidth]{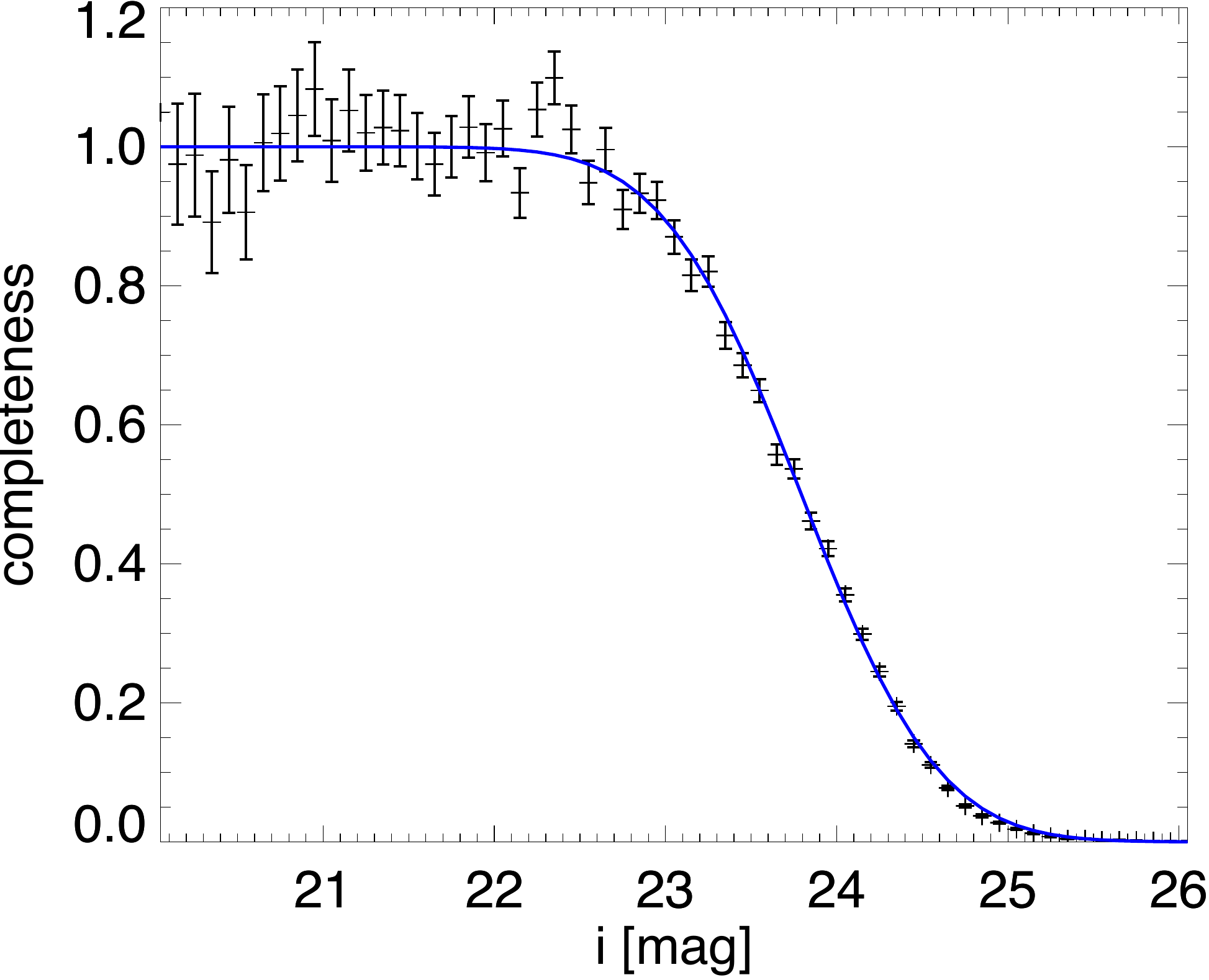}
\vskip-0.10in
\caption{Example of the completeness function around one 2RXS source: Shown is the ratio of observed over reference 
galaxy counts. The fitted completeness function yields a  50\% completeness at i=23.8 mag}
\label{fig:complete}
\end{figure}

The faint limit used at a given cluster location is the $i$-band magnitude at which the local imaging data reaches 
50\% completeness. Sources fainter than this are ignored. Similar to \cite{zenteno11},
we make the source count histogram in $i$-band magnitudes using all galaxies in a radial distance between 10 and 30 arcmin from the cluster candidate position. 
The ratio of the area normalized observed number count histogram over that of a deep reference field provides a measure of the completeness of the observed field. For the reference count histogram we match the COSMOS photo-z catalog \citep{laigle16} and the corresponding DES catalog, to create a deep (i$\approx 26$) catalog that includes DES based auxiliary information such as star/galaxy separation for the matched sources.
An example of the ratio of count rates is shown in Fig.~\ref{fig:complete}. 
Finally we fit a completeness function of the form $f_\mathrm{complet}(m)=0.5*\mathrm{erfc}(a*(m-b))$ , where $\mathrm{erfc}()$ is the complementary error function, $a$ and $b$ are fitting parameters and $m$ is the $i$-band magnitude.

At redshifts below z=0.1, the bright end of the selection range ($m^*(z)-3$) falls below the bright magnitude limit of i=13.5. For even lower redshift the magnitude range used to select galaxies would fall to zero if the faint selection limit ($m^*(z)+1.25$) were left unchanged. To avoid this, we adopt a lower limit of the faint selection to be $i$=17. This ensures that at least a magnitude range of 3.5 is used to calculated the cluster richness and redshift.

We account for differences in the used magnitude ranges and for incompleteness of the data by rescaling the 
measured richness using the correction factor
\begin{equation}
 C_\mathrm{cmp}=\frac{\int_{m^*(z)-3}^{m^*(z)+1.25} S(m^*(z),m,\alpha) dm}{\int_{lim_\mathrm{lo}}
^{lim_\mathrm{hi}} S(m^*(z),m,\alpha)*f_\mathrm{complet}(m) dm},
 \label{eq:richcor}
\end{equation}
where $S(m^*(z),m,\alpha)$ is the Schechter function \citep{schechter76}, in which $m^*(z)$ is the characteristic 
magnitude expected at redshift $z$.  The faint end slope $\alpha$ is set to  $\alpha=-1$ in our analysis.
$lim_\mathrm{lo}$ is the lower magnitude limit of i=13.5 or $m^*(z)-3$ if larger. The upper limit  $lim_\mathrm{hi}$ is 
$m^*(z)+1.25$, if greater than $i$=17 and the 50\% completeness limit range, or else the corresponding boundary values are used.
$f_\mathrm{complet}(m)$ is the completeness function and accounts for missing sources brighter than the 50\% completeness limit.

\subsection{Masking and background estimation}

To measure the cluster richness one has to account for the area within $r_{500}$ that is not covered by useful
imaging data and for the number of galaxies not related to the cluster.
As mentioned in Section~\ref{sec:data}, we use the bad region and foreground flags provided by the GOLD catalog. 
This and the imaging coverage of the DES survey can cause holes in our dataset, which need to be accounted for.
Similar to our pilot study, we use the local source density to identify regions with no data. For that we first obtain 2D 
histograms of the sources within a radius of
0.5$^\circ$, our default local cutout region from the source catalog. The bin size is chosen such that it 
contains 16 sources on average. We obtain 2D histograms with various rectangular shaped bins keeping the bin 
area constant. Empty bins are registered as masked regions and all 2D histograms are combined to one high 
resolution mask image. This method allows us to estimate the available area in a fast way without the need of 
additional input like footprint or mask maps.
The mask image created in this manner is then used to evaluate the available area inside and outside any given radius.

To estimate the number of fore- and background galaxies not associated with the cluster, we use two different background estimates. The 
local background uses all galaxies with radial distance $r_{500}<r<0.5^\circ$. The global background uses the median 
background taken from multiple randoms of 12~$\mathrm{deg}^2$ tiles covering $\sim$15\% of the DES footprint.
In this work we use the global background for redshifts of z<0.5 and otherwise the local background. We do so because DES data are typically
complete and star/galaxy separation is clean for the magnitude range used up to this redshift. Positional 
dependencies that impact the background counts are therefore not expected. 
At magnitudes higher than $i$=22.2 we do not perform point source exclusion, and the completeness starts increasingly to 
differ from one.  Both of these effects make field to field variations more relevant for our richness estimate. The $i$=22.2 magnitude limit is 
reached at z>0.6, and, therefore, starting the usage of the local background at z>0.5 is a conservative approach.

\subsection{Identifying cluster candidates and estimating redshifts}
We define our filtered richness \LamMCMF\ as in \cite{Klein18} as
\begin{align}
\label{eq:richn}
 \lambda_{\mathrm{MCMF}}(z)=&\frac{C_{\mathrm{cmp}}(z)A_{\mathrm{tcl}}(z)}{A_{\mathrm{cl}}(z)}\left( \sum_{i} 
w_i(z) n_i(z)  \right. \nonumber \\
 &\left.  - \frac{A_{\mathrm{cl}}(z)}{A_{\mathrm{BG}}(z)} \sum_{j} w_j(z)\right) ,
\end{align}
the sum of the color and the radial weight over all cluster member galaxy candidates minus the scaled background, where $j$ 
runs over all background galaxies that fulfil the same color and magnitude cuts as for the cluster candidate galaxies. Here 
the elements $A_{\mathrm{cl}}$ and $A_{\mathrm{BG}}$ correspond to the unmasked cluster and background 
areas and $A_{\mathrm{tcl}}$ to the total area within $r_{500}(z)$.  \LamMCMF\ is calculated for the redshift range 
0.01<z<1.31 in steps of $\delta z=0.005$.   For each \LamMCMF\ estimate we calculate the uncertainty $\Delta$\LamMCMF\ assuming Poisson statistics as
\begin{align}
\label{eq:erichn}
\Delta \lambda_{\mathrm{MCMF}}(z)=&\frac{C_{\mathrm{cmp}}(z)A_{\mathrm{tcl}}(z)}{A_{\mathrm{cl}}(z)}
\left( \sum_{i} w_i(z) n_i(z) \right. \nonumber \\
 &\left. + \left(\frac{A_{\mathrm{cl}}(z)}{A_{\mathrm{BG}}(z)}\right)^2 \sum_{j} w_j(z)\right)^{0.5} .
\end{align}

\begin{figure}
\includegraphics[keepaspectratio=true,width=0.95\columnwidth]{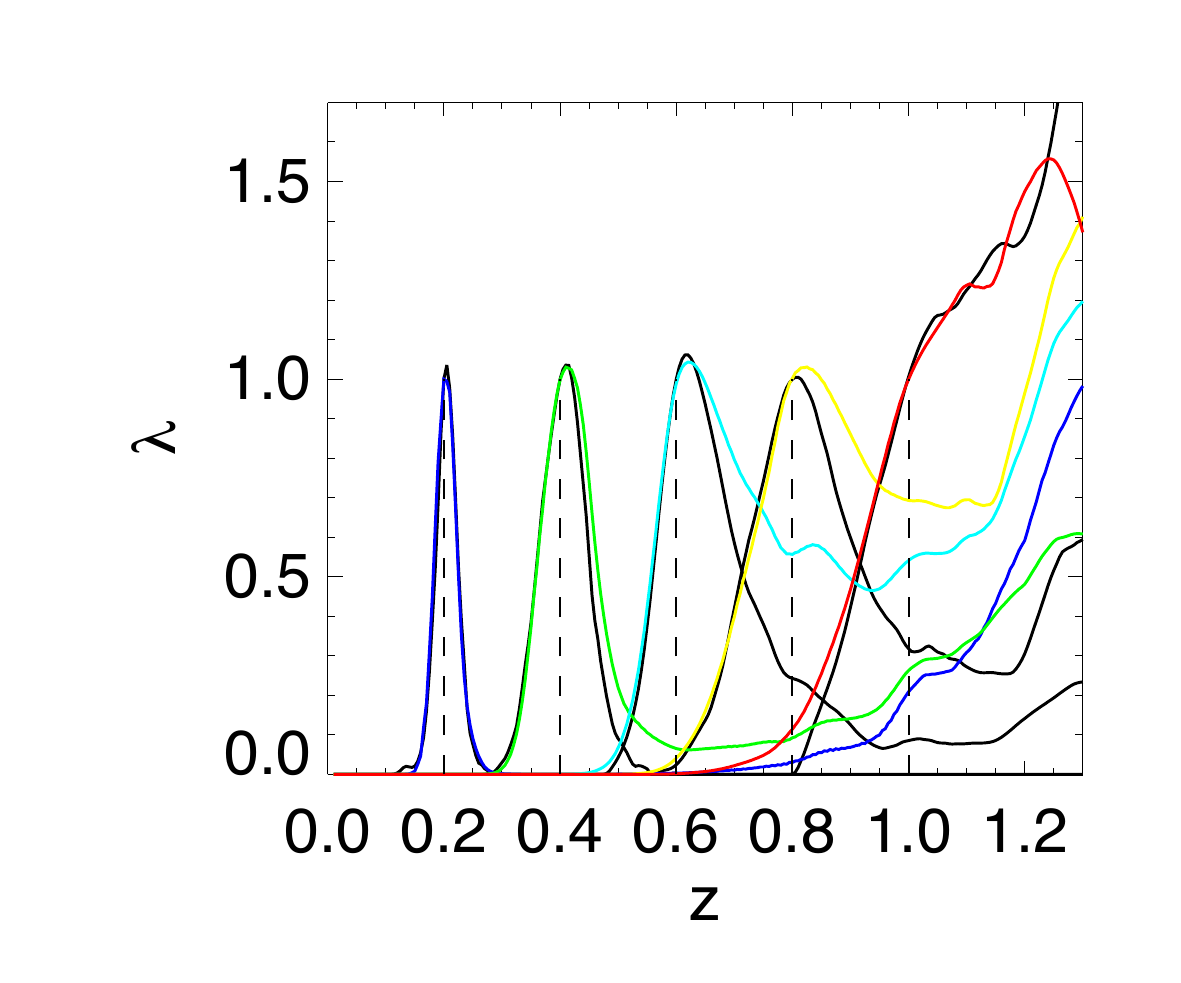}
\vskip-0.10in
\caption{Profiles of richness $\lambda$ as a function of redshift  $z$ for five clusters arranged at different redshifts. 
Colored lines are based on 
mocks solely using the color weight. Black continuous lines are based on stacked clusters with spectroscopic 
redshifts. Dashed lines mark the true cluster redshifts, and the profiles are normalized to one at the true redshift.  The distributions constructed from stacks of spec-z clusters are used as templates to fit \LamMCMF(z) peaks identified along any lines of sight.}
\label{fig:lzprofiles}
\end{figure}	
In \cite{Klein18} we searched the distribution of \LamMCMF\ versus redshift
for peaks and subsequently fit those with Gaussian functions. However, the \LamMCMF(z) peak for a cluster at redshift $z_\text{cl}$ is not well described by a Gaussian centered at the cluster redshift. Assuming so can cause biases in the cluster redshift estimates.
In \cite{Klein18} we accounted for this 
effect by a linear correction of the estimated photo-z based on a cross match with clusters with spectroscopic 
redshifts.
In this work, we estimate the true shape of the \LamMCMF(z) peak using an RS model informed by the data
that includes magnitude and redshift dependent RS widths as well as variable magnitude ranges within galaxies that are considered as cluster members.
Studies of these simulated cluster galaxy populations show that there is significant skewness in the \LamMCMF\ distribution that-- if not treated properly-- would lead to systematic errors in the estimated cluster redshifts.
To illustrate these redshift dependent effects, we 
plot in Fig.~\ref{fig:lzprofiles} using color coded lines the  \LamMCMF\ distributions for mock observations of five clusters at different redshifts.  

To avoid redshift bias and to improve the identification of peaks, we therefore make use of the large number of clusters ($\sim$1000) from our combined spec-z catalog to create stacked  \LamMCMF(z)  profiles over the full range of redshifts we explore. These profiles are used to create redshift dependent templates that are fitted to the observed 
\LamMCMF(z) peaks along any line of sight. 
Fig.~\ref{fig:lzprofiles} contains these templates (black lines), which have similar-- but not identical-- character to the color-coded curves that mark the \LamMCMF(z) from mocks. The advantage of using the stacked profiles over mock based 
profiles is that stacked profiles include all effects that impact the average profile shape, such as the change in aperture size, 
the change with redshift, the radial weighting, the impact of blue cluster members and the masking of background sources by cluster members.

Similar to \cite{Klein18}, we search each line of sight for multiple peaks, 
and fit them iteratively by subtracting neighboring peaks. 
This allows us to deblend neighboring peaks where their profiles overlap.

\subsection{Quantifying the probability of random superposition (i.e., contamination)}
\label{sec:random_superposition}

In \citet{Klein18} we introduced the estimators \ProbLam~ and \ProbS~ for each candidate source. \ProbLam~ was 
defined as the ratio of the number of random sources with  richness lower than the richness of the observed candidate 
over the full number of randoms, evaluated within $|z_\mathrm{MCMF}-z_\mathrm{rand}|<0.075$ around the 
redshift of the cluster candidate. \ProbS~ is derived in a similar manner using the signal to noise ratio (S) instead of 
the richness.

These estimators allow one to identify (and remove) likely superpositions of 2RXS sources with unassociated optical systems that lie along the line of sight toward the X-ray source by chance.  This method of decontaminating a 
cluster catalog is efficient, and can be used to create low contamination subsamples of clusters from highly contaminated cluster candidate lists such as the 2RXS list we adopt here.
Because this estimator mainly depends on the distribution of richness and redshifts of the random 
catalogs, one can create different sets of randoms to trace dependencies such as count rate or RASS exposure time in a straightforward manner.

\begin{figure}
\includegraphics[keepaspectratio=true,width=0.95\columnwidth]{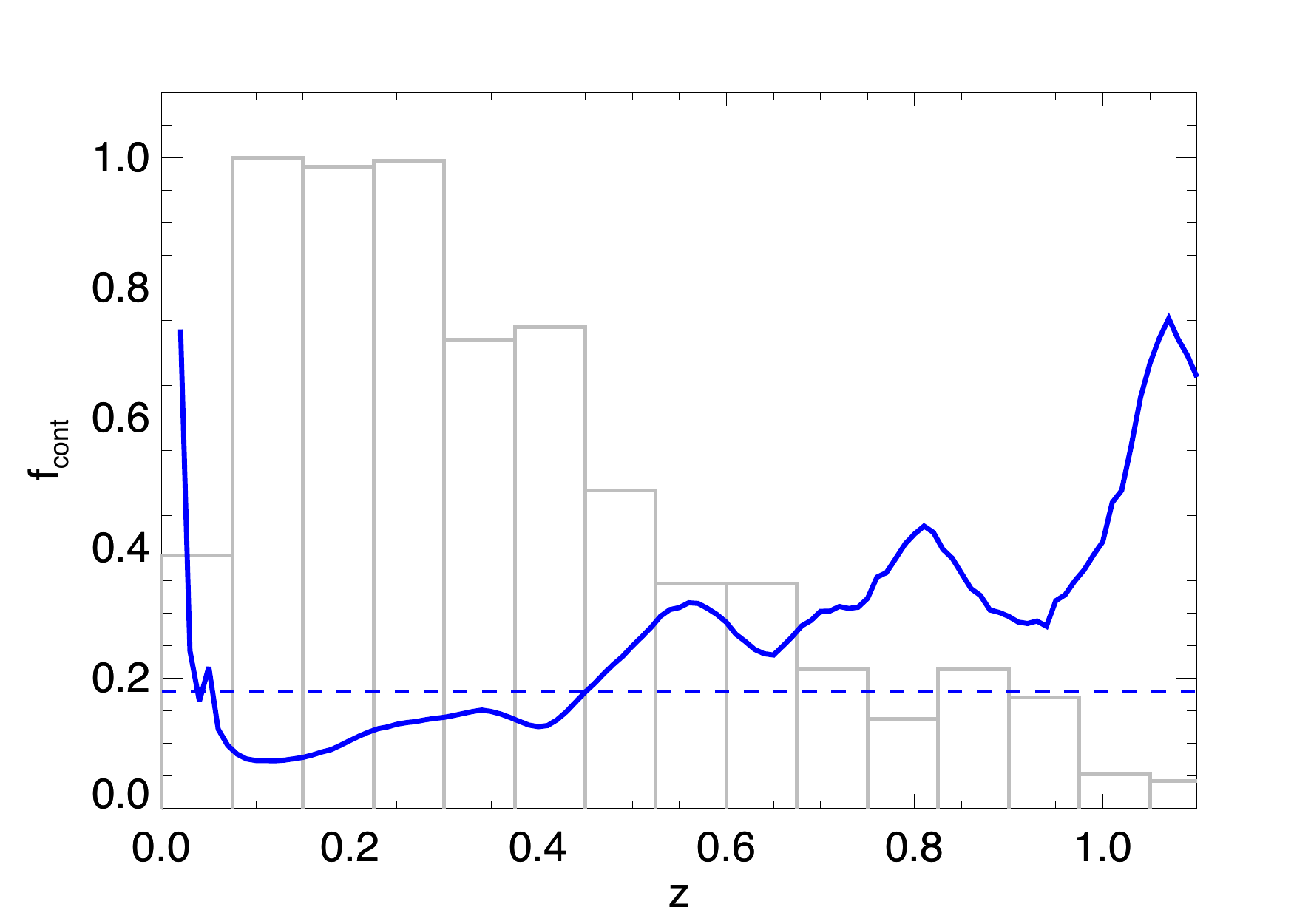}
\vskip-0.10in
\caption{Contamination versus redshift for a cut of \ProbLam >0.985: The fractional contamination for a given 
redshift calculated via  equation~(12) in \citet{Klein18} is shown as a blue continuous line, the mean contamination over all 
redshifts is shown as a blue dashed line. The normalized redshift distribution of the  \ProbLam >0.985 catalog is 
shown as a grey histogram.  Selection in \ProbLam\ ensures a particular mean contamination, but redshift trends in contamination remain.}
\label{fig:contatfixedplamdcut}
\end{figure}

The disadvantage of \ProbLam~(and \ProbS) is that it ignores the ratio of true clusters over non-clusters and its 
potential change with redshift.
The contamination fraction calculated via equation~(12) in \citet{Klein18} is therefore only providing the mean contamination 
of a \ProbLam~ cleaned sample and ignores the possibly significant variation with redshift.
Fig.~\ref{fig:contatfixedplamdcut} shows the contamination fraction as a function of redshift for a cut of \ProbLam 
>0.985 using equation~(12) in \citet{Klein18} within multiple redshift bins. This illustrates the need for an alternative estimator that allows us to construct a sample with a redshift independent contamination.

\begin{figure}
\includegraphics[keepaspectratio=true,width=0.95\columnwidth]{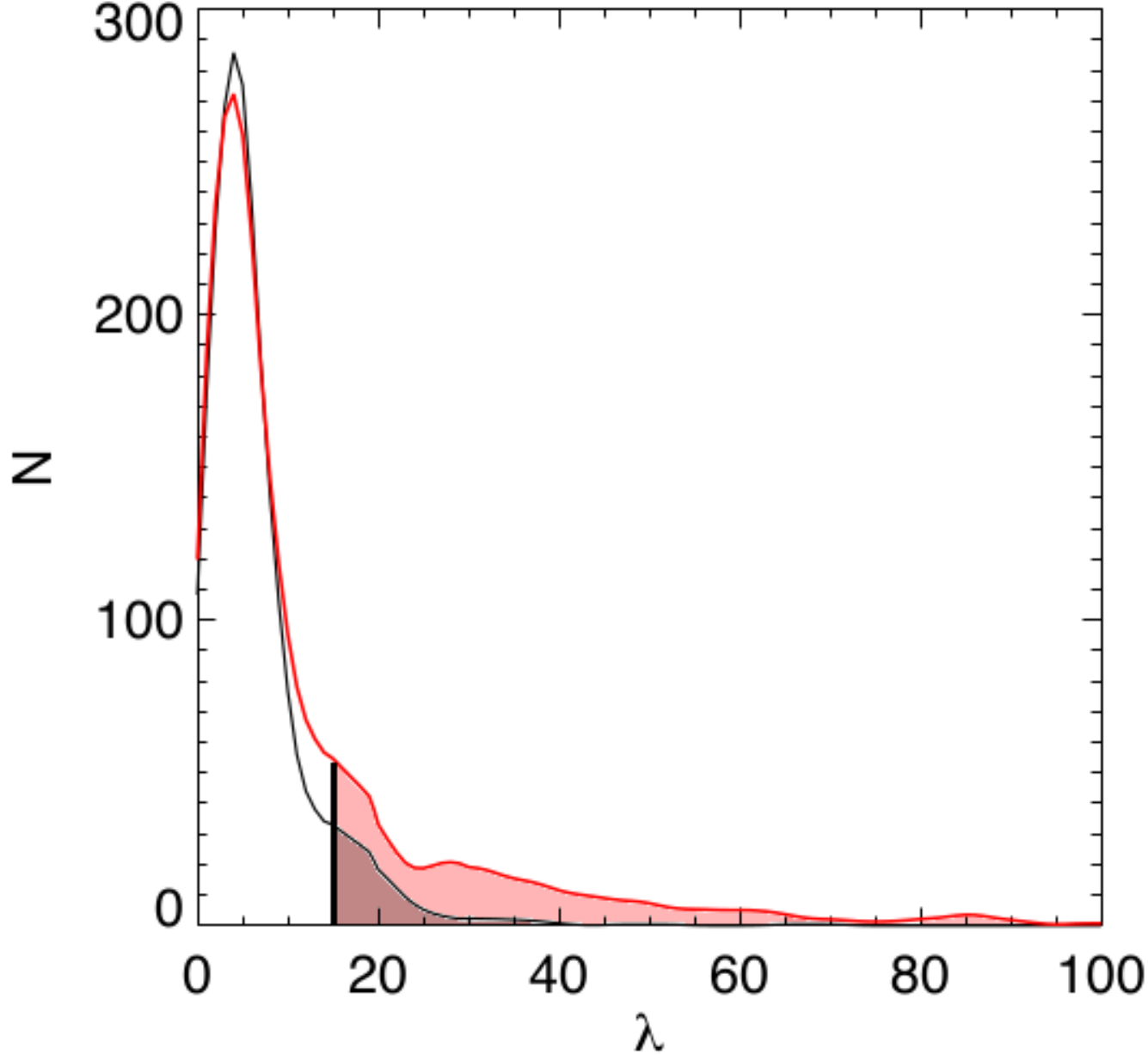}
\vskip-0.10in
\caption{The probability of a source being a random superposition \fcont\ is estimated using the smoothed richness distribution at a particular redshift of 2RXS sources (red) and of random lines of sight (black). The \fcont\ for a given cluster candidate corresponds to the ratio of the integral over 
the black curve divided by the integral over the red curve above the observed $\lambda$ of the candidate.
In this example, the observed cluster candidate at $z=0.2$ has a richness of $\lambda=15$. The corresponding integrals are indicated as shaded regions.
}
\label{fig:richdist}
\end{figure}

We therefore introduce the new estimator \fcont\ as our main selection criterion.  Cutting a candidate list at, for example, a particular value \fcont$<0.05$ then produces a cluster catalog with a fixed 5~percent
contamination fraction, independent of redshift.
We calculate \fcont\ for each source based on the richness distributions of randoms and candidates 
within $\Delta z\approx 0.025$.
Fig.~\ref{fig:richdist} shows an example of the richness distributions of 2RXS and random sources that illustrates how \fcont\ is calculated.

For a cluster candidate $i$ with richness $\lambda_i$ we calculate $f_{\mathrm{cont},i}$ as
\begin{equation}
   f_{\mathrm{cont},i}=\frac{\int_{\lambda_i}^{\infty} f_\mathrm{rand}(\lambda) d\lambda}{\int_{\lambda_i}^{\infty}
f_\mathrm{obs}(\lambda) d\lambda},
\end{equation}
where  $f_\mathrm{rand}$ and $f_\mathrm{obs}$ are the smoothed distributions of richness for the observed 
2RXS sources (obs) and random lines of sight (rand) within $\Delta z\approx 0.025$.
We provide two variants of \fcont, based on different methods of constructing the lambda distributions used to 
calculate it. The first, $f_\mathrm{cont,m}$,  uses the distribution of observed lambdas together with the 
weighted mean of multiple lambda distributions of randoms that were based on different count rates. This ensures 
that the aperture size distribution of random lines of sight are similar to those in the observed sample. By construction 
$f_\mathrm{cont,m}$ is marginalizing over the count rate in the particular redshift bin.
The second variant, \fcontr\, rescales the richnesses of each observed source and associated randoms, 
according to the expected count rate dependency of lambda. The richness distributions derived from the rescaled 
richnesses therefore account for the observed count rate of a given source that defines the size of the region of 
interest.  For the analysis that follows we adopt \fcontr\ and often refer to it simply as \fcont.

One drawback of the new estimator is the statistical limitations that come with a limited number of source candidates. 
This causes noise in richness distributions that can lead to an increase of \fcont\ for higher richnesses. 
To avoid this we make use of smoothing in lambda, redshift and for \fcont\ also in count rate space. 
Further we impose \fcont($\lambda_a$)<\fcont($\lambda_b$) for $\lambda_a>\lambda_b$.

\subsection{Determining the cluster position}
\label{sec:centering}

While the X-ray surface brightness peak is known to provide a good proxy for the 
center of a galaxy cluster, the large PSF of RASS 
and the low signal to noise of the 2RXS sources cause a large uncertainty on the X-ray position. Studies that benefit from 
good knowledge of the cluster position might therefore be negatively impacted if they adopt the 2RXS positions.
The identification of cluster centers using optical data is therefore of special importance for the 2RXS based 
cluster catalog. The performance of optically defined cluster positions in comparison to those derived from other wavelengths has been previously studied \citep{lin04b,rozo14,saro15,oguri18,hikage18}.

MCMF provides three different cluster positions or center estimates based on the optical data.  The first estimate is similar 
to that used in \citet{Klein18} and uses the peak of the density map of RS galaxies as identified using 
{\tt SExtractor} \citep{bertin96}. In contrast to our previous analysis, we choose the highest peak within 
$r_{500}$ in cases where multiple peaks are identified. This avoids 
biasing the X-ray to optical center offset distribution through the assignment of low mass optical substructures 
as the optical counterpart of an X-ray source.  This approach breaks down in some rare, low redshift cases where substructures 
are detected by X-rays and the main optical peak is assigned as the counterpart to substructure. 

The second estimate of the cluster center is a by-product of our estimator of the cluster dynamical state, described in 
detail in the following Section~\ref{sec:dynstate}.
It is based on the fit of a two dimensional King profile \citep{king62} to the RS galaxy density map. The fit is performed within a radius of $r_{500}$ extending from the X-ray center.

The third approach adopts the rBCG, where the rBCG is identified as the brightest galaxy within 1.5 Mpc 
that has all colors within 3$\sigma$ of the RS at the cluster redshift.
While the rBCG potentially provides one the most accurate optical positions for the cluster center, its automated 
identification is not always successful. Further, the identification of the rBCG requires that it be present in the catalog 
with accurate photometry.  As MCMF is pushing to low redshifts, we expect that at z<0.1 the rBCG could to be too bright and 
extended in DES to be properly measured with the standard DESDM photometry techniques. In those cases the other two estimators are still capable of correctly identifying the cluster position. 
The 2D profile fit allows one to identify the center even if parts of the cluster are masked out. 
The center derived directly from the 
galaxy density map offers the simplest and most robust estimate of the center in the absence of masking effects.
The comparison of these different center estimates for each cluster allows one to test the reliability of the center 
estimate and to identify the correctly selected rBCGs as well as cases where there are failures in one of the estimates.
For our method, we adopt the rBCG position as the cluster position in cases where it is within 60\arcsec\ of the galaxy density peak. Otherwise, we adopt the galaxy density peak as the cluster position.  All centers are separately listed in the online version of the catalog.

\subsection{Estimators of cluster dynamical state}\label{sec:dynstate}

Information on the dynamical state of a system can enable additional scientific analyses of the cluster sample. It 
allows one to examine, for example, the dependency of the cluster properties such as
the galaxy population, the dark matter distribution or the ICM properties on cluster dynamical state.
So far the majority of studies rely on dynamical states estimated from either X-ray observations \citep{mohr93,mohr95,jeltema05,nurgaliev13} or spectroscopic data \citep{dressler88,martinez12,ribeiro13,rios16}, but the earliest work on galaxy cluster dynamical state focused also on the galaxy distribution \citep{geller82}.

As demonistrated in \cite{Wen13}, the use of large optical imaging surveys with broadband photometry allows one to provide galaxy distribution based dynamical estimates for thousands of clusters.
The caveat of using imaging data compared to the other probes is the noise of the estimators (based on a few 
dozen galaxies compared, for example, to thousands of X-ray photons) and its susceptibility to line of sight projections.
Moreover, in comparison to X-ray imaging, optical imaging estimators are not able to 
distinguish between two clusters in a pre-collision or post-collision state.
However, the combination of optical and X-ray data will allow us to study merging clusters in all phases of merging.
Despite the fact that estimators based on broadband photometry should be more prone to projections and more noisy than X-ray based estimators, \citet{Wen13} reported that their estimator reaches a success rate of 94\% on a test sample of 98 clusters with known dynamical state.

For our work we adopt the set of dynamical state estimators based on \citet{Wen13}, adapting them somewhat to the available dataset. 
In contrast to \citet{Wen13}, we do not produce a final combined estimate of the dynamical state of each 
cluster based on the individual estimators. This is partially caused by the lack of a test sample and the sensitivity of the 
different estimators on different types of merger states. We believe that the measurements of the individual 
estimators are stable enough to provide them in our catalog.

In the near future, MCMF runs on other surveys will 
include substantial sub-samples with X-ray based estimates of dynamical state.
A detailed study of the performance of the individual estimators and their optimal combination will 
therefore be performed in near future. Given that the estimators are independent of the survey that is followed-up by MCMF, 
those results will be applicable to all MCMF based catalogs, including the catalog presented here. 
We therefore describe those estimators and provide the corresponding measurements already in this work, 
enabling early access to estimators that might be already useful to a some users. Certainly the estimators based on 
those in  \citet{Wen13} can be expected to behave similarly, although not identically, to the original estimators.

\begin{figure*}
\includegraphics[keepaspectratio=true,width=0.67\columnwidth]{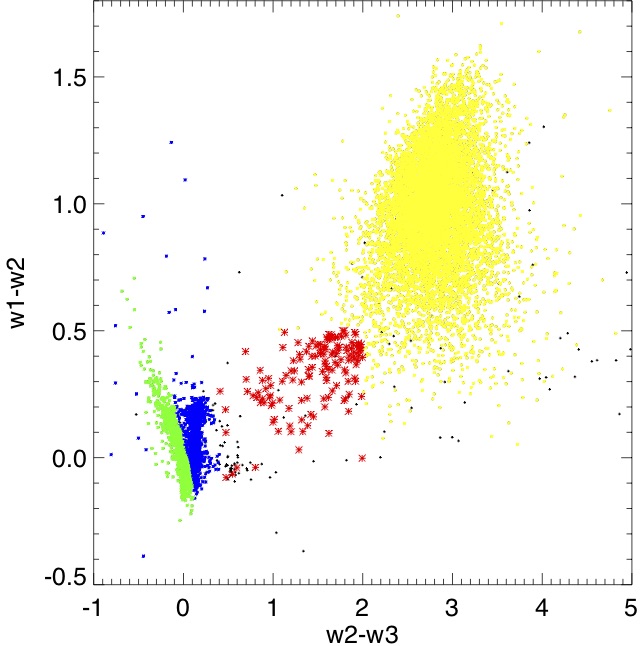}
\includegraphics[keepaspectratio=true,width=0.64\columnwidth]{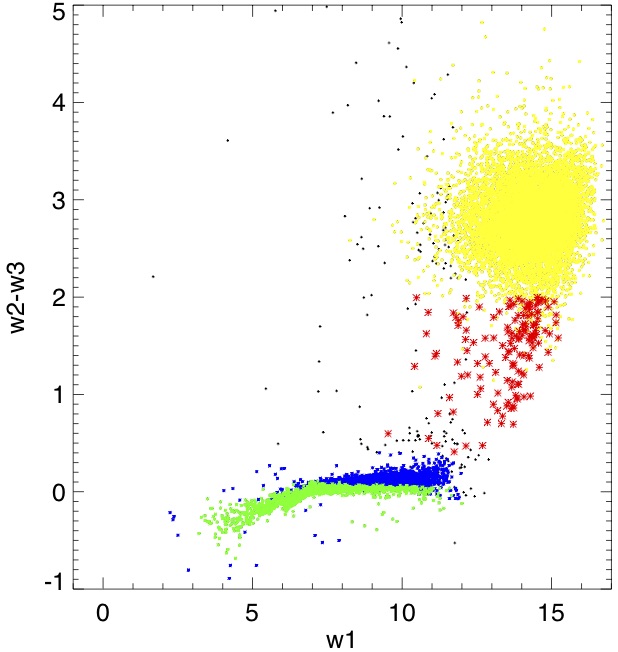}
\includegraphics[keepaspectratio=true,width=0.65\columnwidth]{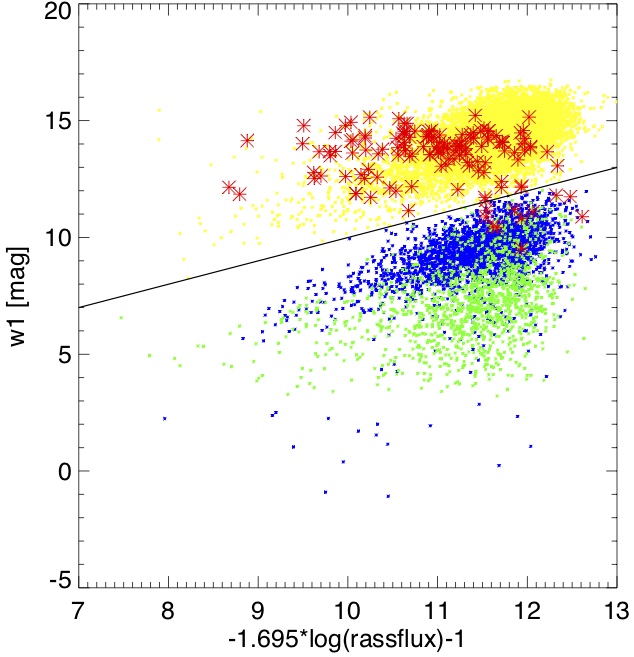}
\caption{X-ray source classification of NWAY matches to ALLWISE:  The mid-infrared color-color distribution is 
shown on the left, the color-magnitude distribution in the middle and the mid-infrared versus X-ray pseudo magnitude is 
shown on the right. Two types of stellar sources are highlighted in green and blue, while galaxies appear in red and AGN in 
yellow.  AGN and the stellar population marked in blue show a scaling between mid-infrared and X-ray flux, while 
the green population shows a weak scaling and the red galaxy population none. The one to one line, which marks the division between AGN and stars, is shown on the right plot.}
\label{fig:agnclassi}
\end{figure*} 

\subsubsection{Estimators based on \citet{Wen13}}

All estimators described in \citet{Wen13} are based on a smoothed map of optical positions and $r$-band luminosities 
of sources with photo-z's within 4\% of the cluster redshift. In this work we use the standard output of 
the MCMF pipeline, which includes density maps of RS galaxies at the cluster redshift smoothed with 
a 125~kpc Gaussian kernel. Our experience is that the dynamical indicators are quite stable to small variations of the galaxy selection and smoothing kernel scale, and therefore we adopt this single approach \citep[but see][]{Wen13}.

There are three individual estimators described in \citet{Wen13}: (1) the asymmetry factor $\alpha$, (2) the normalized 
deviation $\delta$ and (3) the ridge flatness $\beta$. 
The asymmetry factor $\alpha$ is defined as the ratio of the 'difference power' over the 'total fluctuation power' 
within $r_{500}$
\begin{equation}
  \alpha=\frac{\sum_{i,j}[I(x_i,y_j)-I(-x_i,-y_j)]^2/2 }{\sum_{i,j}I^2(x_i,y_j)},
\end{equation}
where $I(x_i,y_j)$ is the value of the density map at cluster centric position $x_i$, $y_j$.
The normalized deviation $\delta$ uses the fit of a 2D King model \citep{king62}
\begin{equation}
  I_\mathrm{2Dmodel}(x,y)= \frac{I_0}{1 + (r_\mathrm{iso}/r_0)^2},
\end{equation}
where $I_0$ is the intensity at the cluster center, $r_0$ the characteristic radius and $r_\mathrm{iso}$ is the cluster 
centric distance of an isophote with
$r_\mathrm{iso}^2= (x \cos \theta + y \sin \theta)^2 + \epsilon(-x\sin \theta + y \cos \theta)^2$.
The normalized deviation $\delta$ is then the normalized deviation of the residual map within $r_{500}$ after 
subtraction of the model
\begin{equation}
\delta=\frac{\sum_{i,j}[I(x_i,y_j)-I_\mathrm{2Dmodel}(x_i,y_j)]^2}{\sum_{i,j}I^2(x_i,y_j)}.
\end{equation}
The third estimator,  the ridge flatness $\beta$, is derived by fitting a 1D king profile $I_\mathrm{1D}=I_0/(1+(r/
r_0)^2)$ to different sectors of the galaxy density map. We define the concentration $c_\mathrm{King}$ as 
$c_\mathrm{King}=r_{500}/r_0$. We find the lowest concentration out of thirty-six $10^\circ$ wide angular 
wedges centered on the cluster and call this the 
concentration of the ridge $c_\mathrm{King,R}$. The ridge flatness is then defined with respect to the median of 
the derived concentrations as
\begin{equation}
  \beta=\frac{c_\mathrm{King,R}}{\tilde{c}_\mathrm{King}}.
\end{equation}

\subsubsection{Additional estimators}

The estimators introduced by \citet{Wen13} investigate the asymmetry and smoothness of the cluster galaxy distribution. The asymmetry of the cluster is a good tracer of dynamical youth if  
the merging structures are significantly offset or have significantly different richnesses.
If the projected distance between the merging systems is too small, a single 2D King profile might be a sufficiently good approximation of the galaxy density distribution, causing only a weak  signal of merger based on the previously described estimators.
Those systems, unless merging almost along the line of sight, 
might be found by an unusually high ellipticity of the derived king model. 
We therefore list the ellipticity found by the fit of the 2D King model as an additional indicator of the dynamical 
state.

Finally it might be of interest to identify the nearest galaxy overdensity that exceeds a certain fraction of the mass 
of the main cluster investigated.
This can be used to identify massive mergers in various stages of the merger process. 
To identify the nearest galaxy overdensity not associated with the main cluster we use of the {\tt SExtractor} based catalogs of the galaxy density map previously used to obtain the cluster position. We select the nearest RS 
overdensity that has a "$FLUX\_ISO$" measurement of at least $25\%$ of that source that is taken to be the 
main cluster. The $FLUX\_ISO$ measurement of {\tt SExtractor} can be interpreted in this context as a richness estimate 
that should scale with the mass of the structure. For all substructures reaching this threshold, we 
list in the catalog the $FLUX\_ISO$ ratios and the offset distances in units of $r_{500}$ of the main cluster.

\subsection{X-ray emitting point sources} \label{sec:xrstars}

The majority of X-ray sources listed in the 2RXS catalog are not galaxy clusters.  Rather, they are either AGN, stars or noise fluctuations. 
Reliable identification of the non-cluster sources and their multi-wavelength counterparts are challenging tasks.
Compared to cluster confirmation, the point source nature of AGN and stars allows a clear knowledge of the offset distribution between the X-ray and multi-wavelength counterpart that can be used for identification. However, the number of potential counterparts given the X-ray positional uncertainty can be large.

One way to reduce the number of chance superpositions and to find the right counterpart, is to use priors 
on the colors and magnitudes of the sources that match the source populations of true counterparts. 
Observation in the mid-infrared regime has been shown to be a valuable source to reliably identify AGN \citep{Stern12,Assef18}, 
making use of the radiation from the accretion disk as well from the dust torus around the AGN. 
Cross identification between X-ray sources and mid-infrared sources therefore seems to be promising to identify AGN.

 Recently \cite{Salvato18} used a Bayesian statistics based 
 algorithm called NWAY to associate 2RXS sources with sources from 
the ALLWISE catalog \citep{wright10}. This method makes use of priors in the mid-infrared bands to find the best counterpart for a 
given 2RXS source.
 We use this catalog to investigate the color distribution of 2RXS counterparts in ALLWISE color space and the 
correlation between ALLWISE flux and X-ray flux.
 The NWAY code calculates the probability that a 2RXS source will have any ALLWISE counterpart ($p_\mathrm{any}
$) and the probability that the given ALLWISE counterpart is the correct counterpart ($p_\mathrm{i}$).
Throughout this work, we restrict ourselves to the NWAY catalog with $p_\mathrm{any}>0.5$ and $p_\mathrm{i}
>0.8$, and we select the most probable counterpart in the case where there is more than one 
identified above these cuts. According to \citet{Salvato18}, 
these cuts should result in a catalog with only 2 to 5\% contamination by chance superposition. With these cuts, we find that $\sim$55\% of the 2RXS sources have an NWAY match.  Assuming a 30\% spurious fraction in the 2RXS catalog, this suggests that we find matches for more than 75\% of the true 2RXS sources. 
 
 In Fig.~\ref{fig:agnclassi} we show the color-color and color magnitude distribution of NWAY matches. 
We split the sources into different types based on the mid-infrared properties, following Fig. 9 in \citet{Salvato18}.
 AGN are highlighted in yellow and represent the main type of source in the 
 NWAY catalog.   X-ray emitting stars are shown in blue and green.
 A fourth population of sources that lies between stars and AGN is marked in red and is related primarily to galaxies--  
including cluster galaxies.
 The right-most panel of Fig.~\ref{fig:agnclassi} shows ALLWISE $w1$ band magnitude versus X-ray pseudo-magnitudes, 
chosen such that the one to one line splits AGN and stars.
As can be seen, AGN and one of the stellar populations follow a linear 
relation between ALLWISE magnitude and RASS pseudo-magnitude.
The stellar sources marked in green show a much higher scatter than the stars marked in blue. The red sources do not 
show a correlation between X-ray flux and ALLWISE flux.  Rather, they seem to simply scatter in ALLWISE $w1$ in a 
similar manner at all X-ray pseudo-magnitudes
In our final MARD-Y3 catalog, we list all NWAY matches that fulfil the aforementioned NWAY cuts and ALLWISE cuts 
together with their classification into the different source populations, the positional distance to the 2RXS source and 
the source distance from the corresponding mean X-ray to ALLWISE relation for that source classification.
The AGN and stellar contamination in the final cluster catalog is evaluated in Section~\ref{sec:contamination}.
 
\subsection{Flagging multiple detections of the same source} \label{sec:multiple}

The 2RXS catalog is designed as a point source catalog with respect to the RASS PSF. While the majority of 
clusters are not or are only barely resolved and therefore well captured in 2RXS, well resolved and bright sources can 
cause trouble for the algorithm. One of these failure modes is that bright and extended clusters are detected 
multiple times.
MCMF and 2RXS do not individually attempt any deblending of neighboring sources. Multiple 2RXS entries are 
therefore independently treated by MCMF and will result in multiple confirmed 2RXS clusters corresponding to the 
same real cluster.
We therefore group and flag multiple detections based on their projected separations and redshift differences. This step 
must be done prior to estimating \fcont\ to avoid a bias in the richness distribution due to multiple versions of the same 
cluster appearing in the catalog.


\begin{figure}
\includegraphics[keepaspectratio=true,width=0.95\columnwidth]{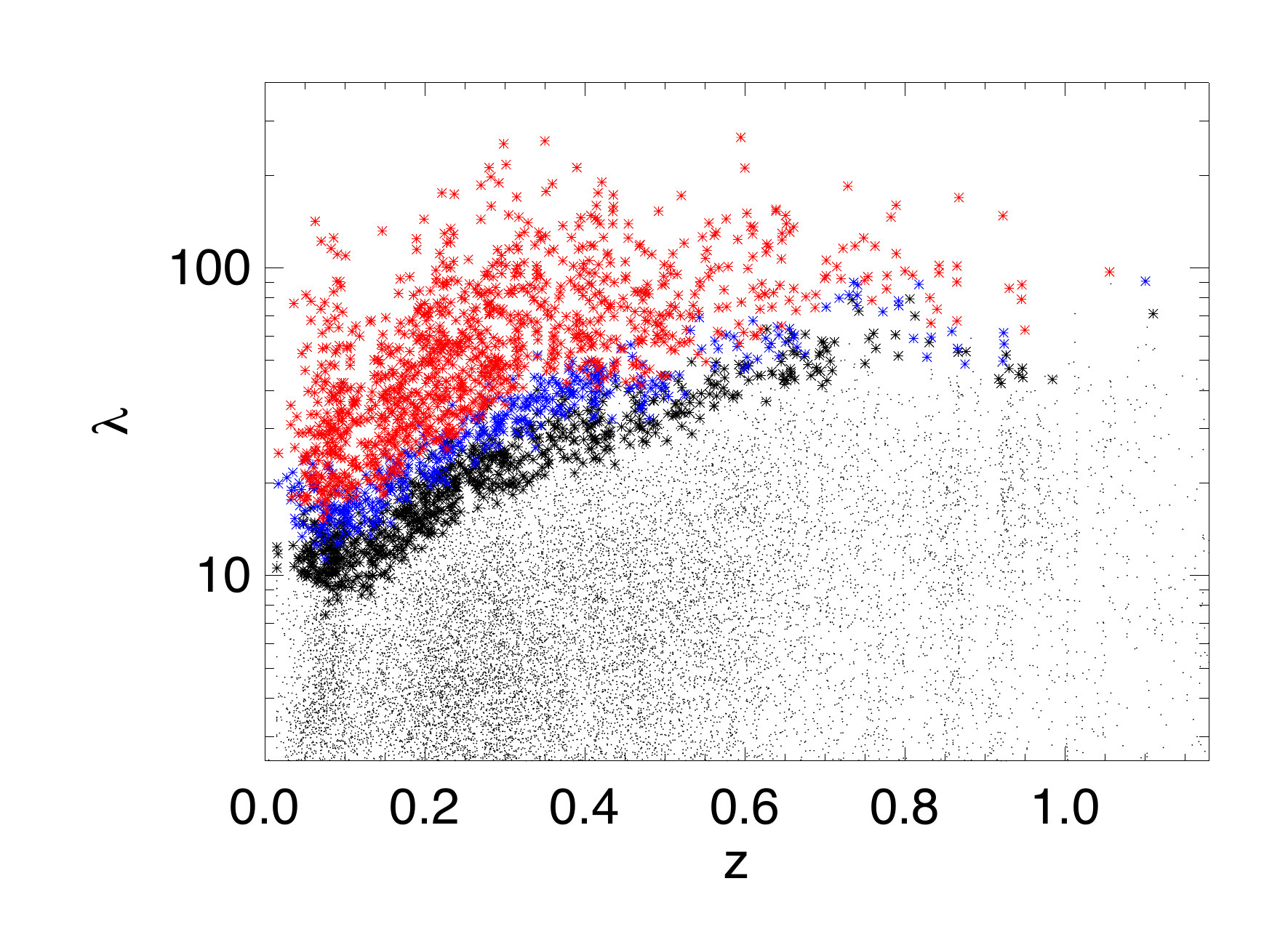}
\vskip-0.10in
\caption{Richness versus redshift of the MARD-Y3 cluster catalog for selections  \fcont$<0.2$ (black),  
\fcont$<0.1$ (blue) and  \fcont$<0.05$ (red). Multiple detections of clusters are excluded, no 
X-ray, AGN or star exclusion is applied.}
\label{fig:lambdaz}
\end{figure}

\begin{figure*}
\includegraphics[keepaspectratio=true,width=0.95\linewidth]{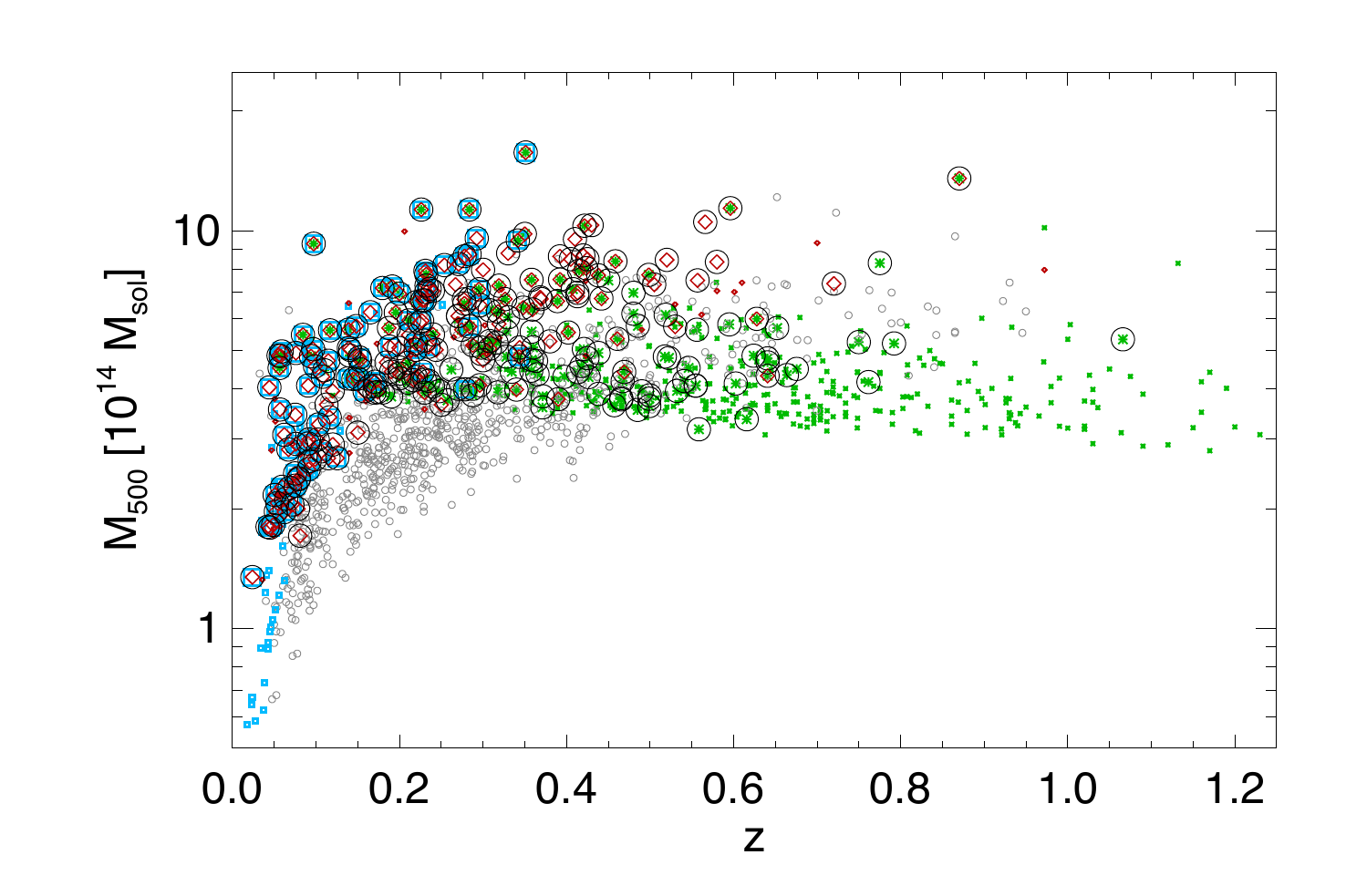}
\vskip-0.05in
\caption{Distribution of clusters in mass and redshift for four major galaxy cluster surveys overlapping the DES-Y3 footprint. Gray and black circles show the MARD-Y3 \fcont$< 0.05$  sample. Clusters from the RASS based REFLEX survey are shown as cyan squares, clusters from the Planck PSZ2 catalog are shown as red diamonds and clusters from the SPT-SZ survey are shown as green asterisk. Large symbols show matched sources. In case of matched sources we adopt redshifts and masses from the matched survey to maintain the shapes of the individual samples as much as possible. In case of matches in multiple catalogs we prefer SPT-SZ over Planck over REFLEX. We note that that the SPT-SZ survey does only cover $\sim 2400 \mathrm{deg}^2$ of the DES footprint.}
\label{fig:MassVsRedsComp}
\end{figure*}

\begin{figure}
\includegraphics[keepaspectratio=true,width=0.95\columnwidth]{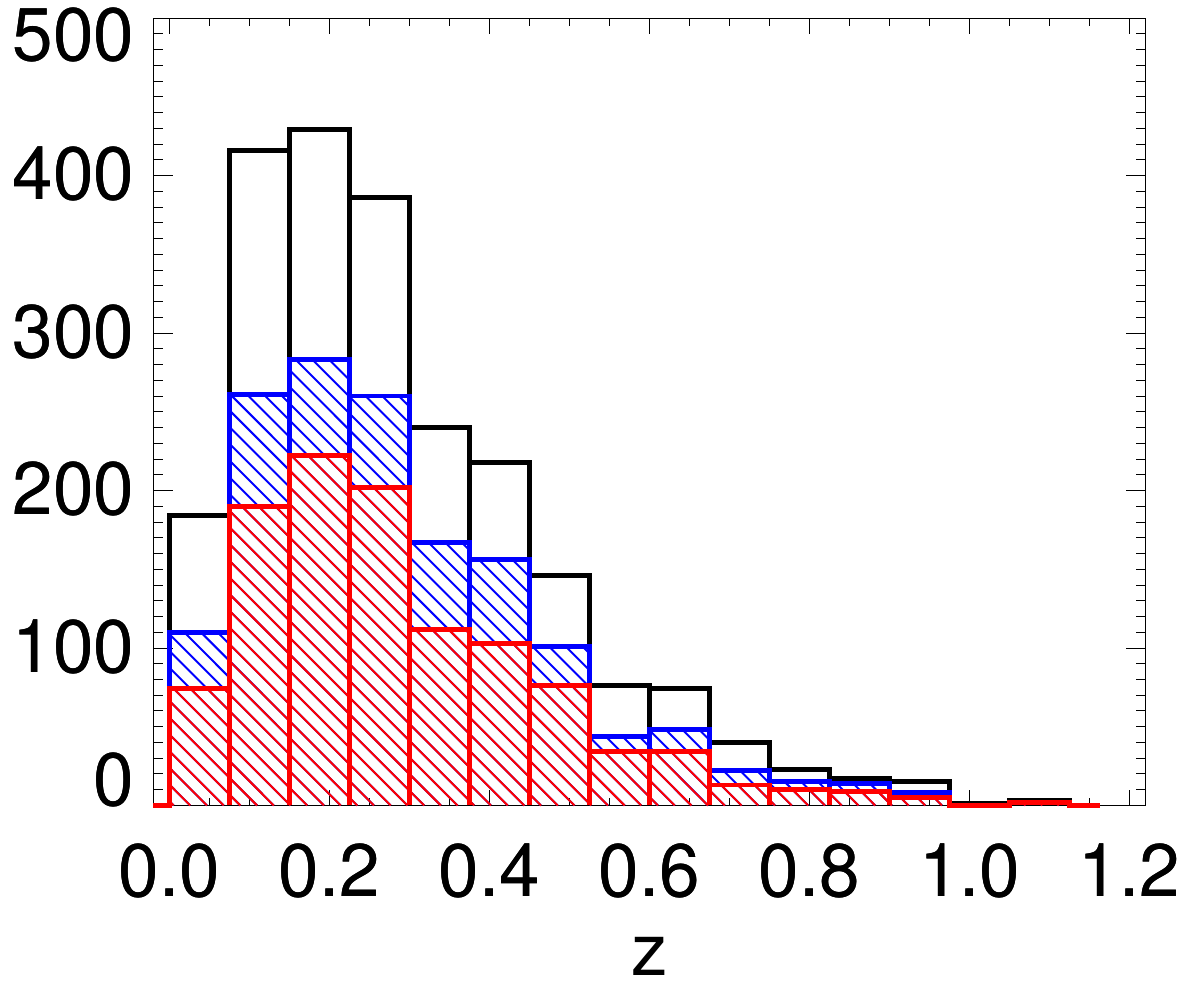}
\vskip-0.10in
\caption{Redshift distribution of the MARD-Y3 cluster catalog for selections  \fcont$<0.2$ (black),  
\fcont$<0.1$ (blue) and  \fcont$<0.05$ (red). Multiple detections of clusters are excluded, no 
X-ray, AGN or star exclusion is applied.}
\label{fig:redshiftdistri}
\end{figure}

\begin{figure*}
\includegraphics[keepaspectratio=true,width=1.15\columnwidth]{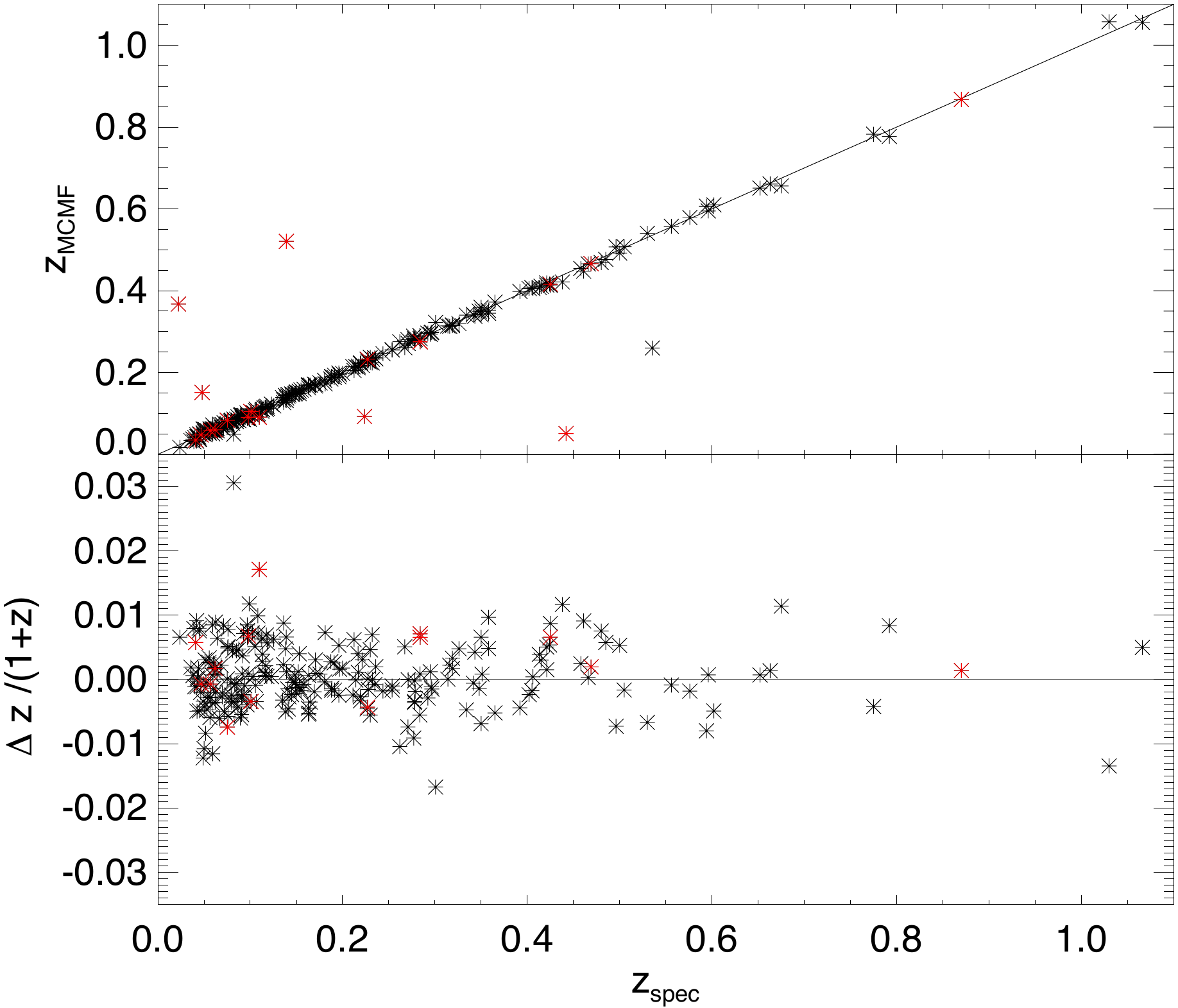}
\includegraphics[keepaspectratio=true,width=0.82\columnwidth]{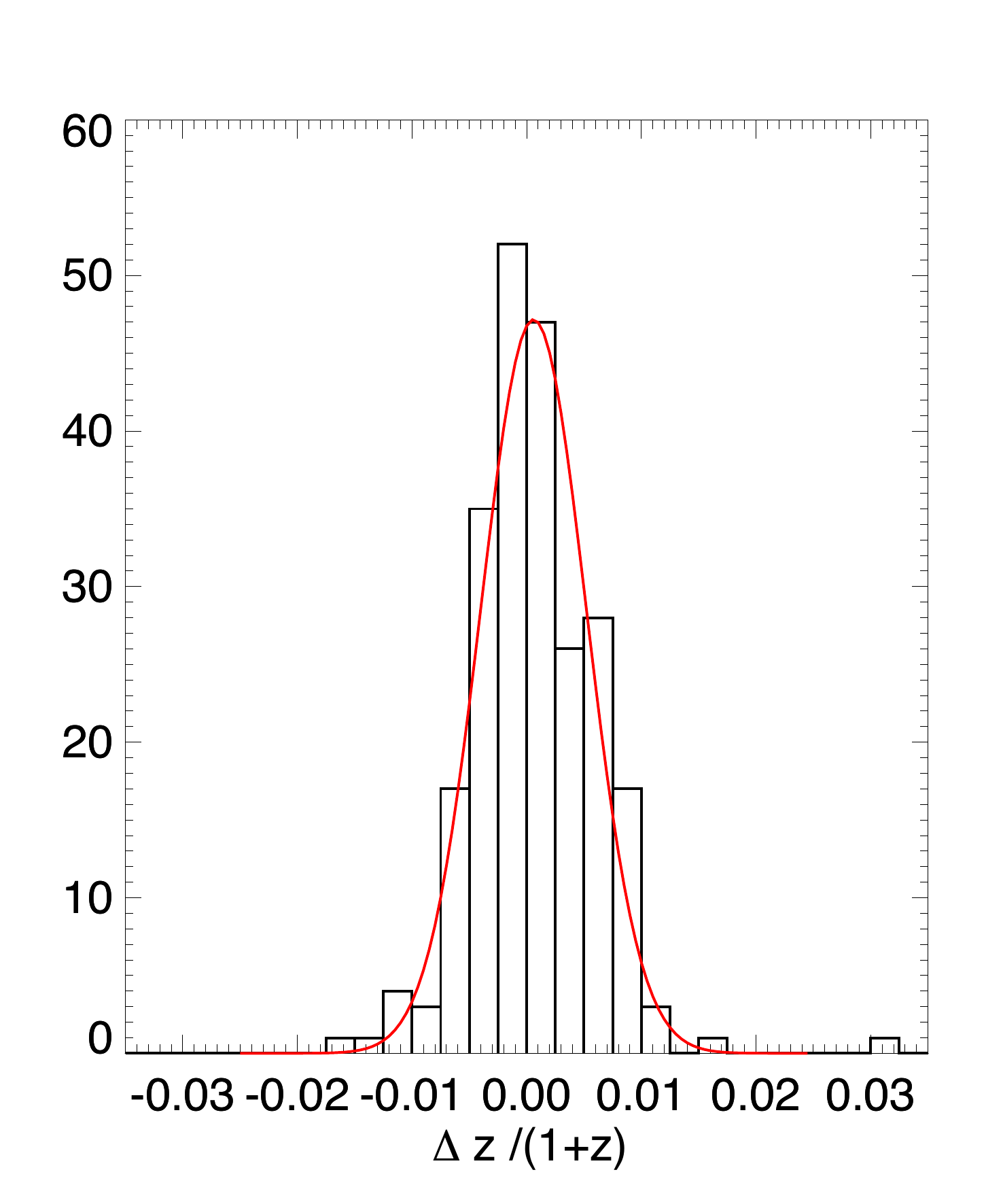}
\vskip-0.10in
\caption{Left: The MCMF photometric redshifts versus spectroscopic redshifts (above) and the redshift offset $
(z_\mathrm{spec}-z_\mathrm{MCMF})/(1 + z_\mathrm{spec})$ against spectroscopic redshifts (below). Clusters with 
multiple significant peaks in redshift are marked in red. Right: Histogram of redshift offsets $(z_\mathrm{spec}-
z_\mathrm{MCMF})/(1 + z_\mathrm{spec})$. The distribution is fit by a Gaussian (red) with RMS variation $\sigma=0.0046$ and mean $\mu=0.0006$.}
\label{fig:redshiftdist}
\end{figure*}

\section{The MCMF confirmed RASS cluster catalog using DES-Y3 data}
\label{sec:application}

The multi-component matched filter RASS cluster catalog confirmed with DES-Y3 data (MARD-Y3) is the main 
product of this paper. The DES-Y3 galaxy catalog covers the majority of the final DES footprint to a depth 
that is sufficient for confirming all RASS detected galaxy clusters. Future DES data will increase the 
imaging depth and reduce calibration systematics. Both depth and calibration are already at a level 
completely sufficient for the RASS 
confirmation, which means potentially future MCMF runs using new DES data should not significantly 
alter the results presented here.

In the following subsections we present the new cluster sample (Section~\ref{sec:sample}), examine 
the impact of AGN and stars on the cluster sample (Section~\ref{sec:contamination}), compare our 
sample to other previously published X-ray, optical and SZE selected cluster catalogs 
(Section~\ref{sec:comparison}), examine the dynamical state and its redshift evolution of the cluster 
sample (Section~\ref{sec:dynamical_state}) and then finally use the sample together with a simple 
selection function to measure the luminosity function out to redshift $z\sim0.9$ and compare it with 
the theoretical expectation for a fiducial cosmological model (Section~\ref{sec:luminosity_function}).

\subsection{Galaxy cluster sample} 
\label{sec:sample}

MCMF allows one to clean the input cluster candidate list to the desired level of contamination by chance 
superpositions using \fcont\ cuts. The main results presented in this paper are based on catalogs with 
contamination cuts  \fcont$<0.2$, $<0.1$ and $<0.05$. The catalog 
is created with a limit of  \fcont$<0.2$, but \fcont\ is listed for each cluster so that the users
can select the cluster sample with the combination of size and contamination that best suits their work.

We find 2312 clusters with \fcont$<0.2$, 1517 with \fcont$<0.1$ and 1101 \fcont$
<0.05$, where multiple detections of the same source have been excluded. These numbers do not include
additional selections such as redshift cuts or exclusion of likely AGN sources that meet the NWAY thresholds 
and are outliers in the richness-mass plane. 
Table~\ref{tab:catalogprop} contains the catalog sizes for four different \fcont\ cuts, where $N_\mathrm{src}$ is 
the number of 2RXS sources whose counterparts meet the \fcont\ cut, $N_\mathrm{cl}$ is the number after 
rejection of multiple 2RXS detections of the same source, $N_\mathrm{cl-AGN}$ is the number of clusters 
after AGN rejection on NWAY sources, and the final estimated contamination and incompleteness introduced 
by the AGN rejection are listed as $f_\mathrm{cont,fin}$ and $f_\mathrm{ex}$ (see discussion of this 
contamination rejection in Section~\ref{sec:contamination}).
The selection by \fcont\ is illustrated in the richness-redshift plane using color coded points in Fig.~\ref{fig:lambdaz}. 

The distribution
of clusters in redshift and mass for the \fcont$<0.05$ rejecting AGN and multiple detections is shown on Fig.~\ref{fig:MassVsRedsComp}. For comparison we further show Planck PSZ2 \citep{planck25-27}, SPT-SZ \citep{bleem15} and REFLEX \citep{boehringer04} clusters overlapping the DES-Y3 footprint. For visualisation purposes we account for mean mass offsets between surveys and use the corrected masses of these surveys in case of matched sources. We use a generous 300\arcsec\ matching radius for Planck and REFLEX and 200\arcsec\ for SPT clusters and require a maximum redshift difference of $\delta z=0.2$. A detailed comparison between surveys is performed in Section~\ref{sec:comparison} and Appenix A.

The redshift distribution of the cluster catalog for different cuts in \fcont\ is shown in Fig.~\ref{fig:redshiftdistri}. The full cluster catalog up to
 \fcont$=0.2$ will be made available online at the VIZIER archive\footnote{http://vizier.u-strasbg.fr}. An example table showing the most important MCMF derived quantities is shown in Table~\ref{tab:clust_table} in the Appendix.
\begin{table}
\centering
\caption{Catalog properties after applying \fcont\ selections. From left to right we list the \fcont\ upper limit, 
the number of 2RXS sources below the cut $N_\mathrm{src}$,  the number of clusters after excluding 
multiple detections of the same source $N_\mathrm{cl}$, the number of clusters after exclusion of NWAY 
matches that differ at $>2\sigma$ from the lambda-mass relation $N_\mathrm{cl-AGN}$, the percentage 
of expected final contamination $f_\mathrm{cont,fin}$ and the percentage of expected true sources 
excluded by the cut in lambda-mass $f_\mathrm{ex}$ .}.
\label{tab:catalogprop}
\begin{tabular}{|c|c|c|c|c|c|}
\hline 
\fcont\ cut  & $N_\mathrm{src}$ &  $N_\mathrm{cl}$ &  $N_\mathrm{cl-AGN}$ &  $f_\mathrm{cont,fin}$ &  $f_\mathrm{ex}$ \\ 
\hline 
0.20 & 2950 & 2312 & 2171 & 9.6\% & 0.6\%\\ 
0.15 & 2485 & 1896 & 1812 & 6.7\% & 0.4\% \\ 
0.10 & 2017 & 1517 & 1466 & 5.6\% & 0.4\%\\ 
0.05 & 1507 & 1101 & 1086 & 2.6\% & 0.2\% \\ 
\end{tabular} 
\end{table}

\subsubsection{Photo-z performance}
\label{sec:PhotozPerformance}

For the MARD-Y3 catalog with \fcont$<0.2$ we find 242 clusters with known spec-z's
within a matching radius of 150\arcsec. Fig.~\ref{fig:redshiftdist} compares the spec-z
with the MCMF photo-z that shows the lowest contamination \fcont\ for a given cluster.  Highlighted in red are 
clusters with at least one additional peak in redshift that has an \fcont\ or contamination fraction less than 0.2 higher than the 
peak corresponding to the lowest \fcont. Out of 19 sources with at least one significant 
additional peak in redshift, we find 14 consistent with the spec-z. In three other sources, 
the peak with the second lowest contamination fraction is consistent with the spec-z. 
One source has a fourth peak consistent with the spec-z, and so we exclude it from our main catalog. 
The remaining and only outlier with multiple significant peaks which does not 
show a significant peak at the spec-z is at $z=0.022$.  This cluster is likely at the lower redshift limit 
for DES, and the majority of cluster members are considered as too bright and extended to be well measured with 
the standard photometry.
The only outlier not showing a significant peak was checked by visual inspection of the DES images. 
We find that the cluster with spec-z does not correspond to the cluster found by MCMF.  It is about 
150\arcsec\ away from the center of the MCMF cluster. We do find a second peak in redshift at the 
spec-z with \fcont$=0.45$, but compared to the main peak contamination fraction \fcont$=0.005$, 
this is not a significant peak. We therefore consider this as a cluster mismatch rather than a failure of MCMF.

We measure the scatter between photo-z and spec-z by fitting a Gaussian function to the histogram of $\Delta z / 
(1 +z_{spec})$ measurements, where $\Delta z=z_\mathrm{spec}-z_\mathrm{MCMF}$.
We find a standard deviation $\sigma$=0.0046 (or 0.46\%) and a mean of $\mu$=0.0006. The histogram and 
the fit are shown in Fig.~\ref{fig:redshiftdist}.
We also split the sample into three redshift bins (0<z<0.2, 0.2<z<0.43 and z>0.43), selected to contain $\approx 
1/3$ of the MARD-Y3 sources for \fcont$<0.2$. We find standard deviations of 0.42\%, 0.45\% and 0.71\% 
for the different bins, based on 150, 68 and 24 clusters, respectively.
We do see an increase of the scatter in the highest redshift bin $z>0.43$, which we can explore better once a larger
spec-z sample is available in this redshift range.

To investigate the scatter as a function of richness we limit the sample to $z<0.43$ to avoid redshift dependent 
effects. We measure the scatter between photometric and spec-z's within 
three richness bins ($0<\lambda<50$, $50<\lambda<100$, $100>\lambda$). We do not find any significant trend 
with richness, indicating that the remaining photo-z scatter is not driven by the number of cluster members.

\begin{figure}
\includegraphics[keepaspectratio=true,width=\columnwidth]
{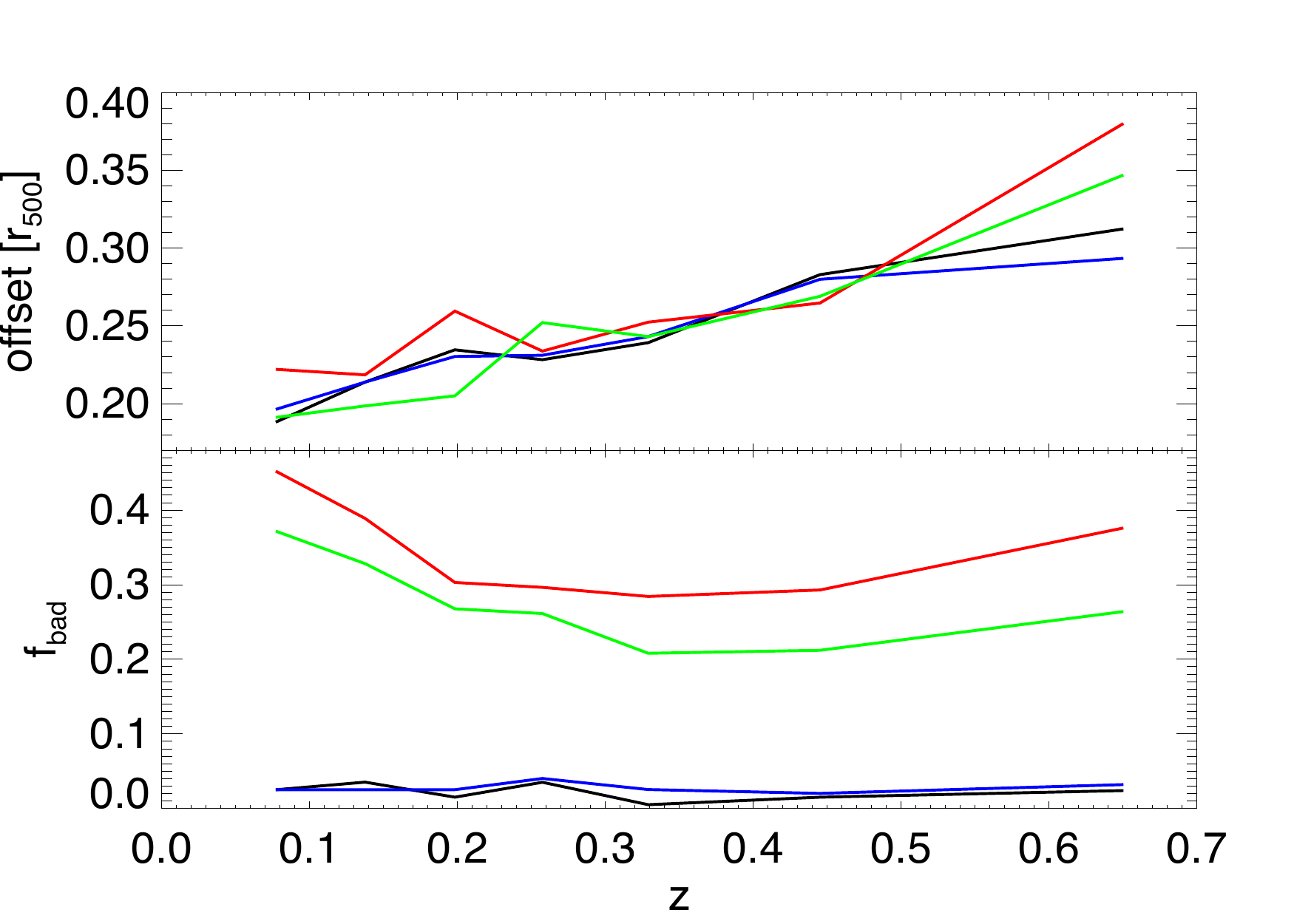}
\vskip-0.10in
\caption{Redshift dependence of the median offset between the X-ray and the optical center (top) for different center estimators: 
the RS galaxy density peak (blue), the rBCG position (red) and the 2D fit position to the RS galaxy density (green). 
The default center is shown as a black line. The fraction of sources (bottom) with offsets larger than $r_{500}$ or unsuccessful center estimate.}
\label{fig:offsetsvsz1}
\end{figure}

\begin{figure}
\includegraphics[keepaspectratio=true,width=0.91\columnwidth]{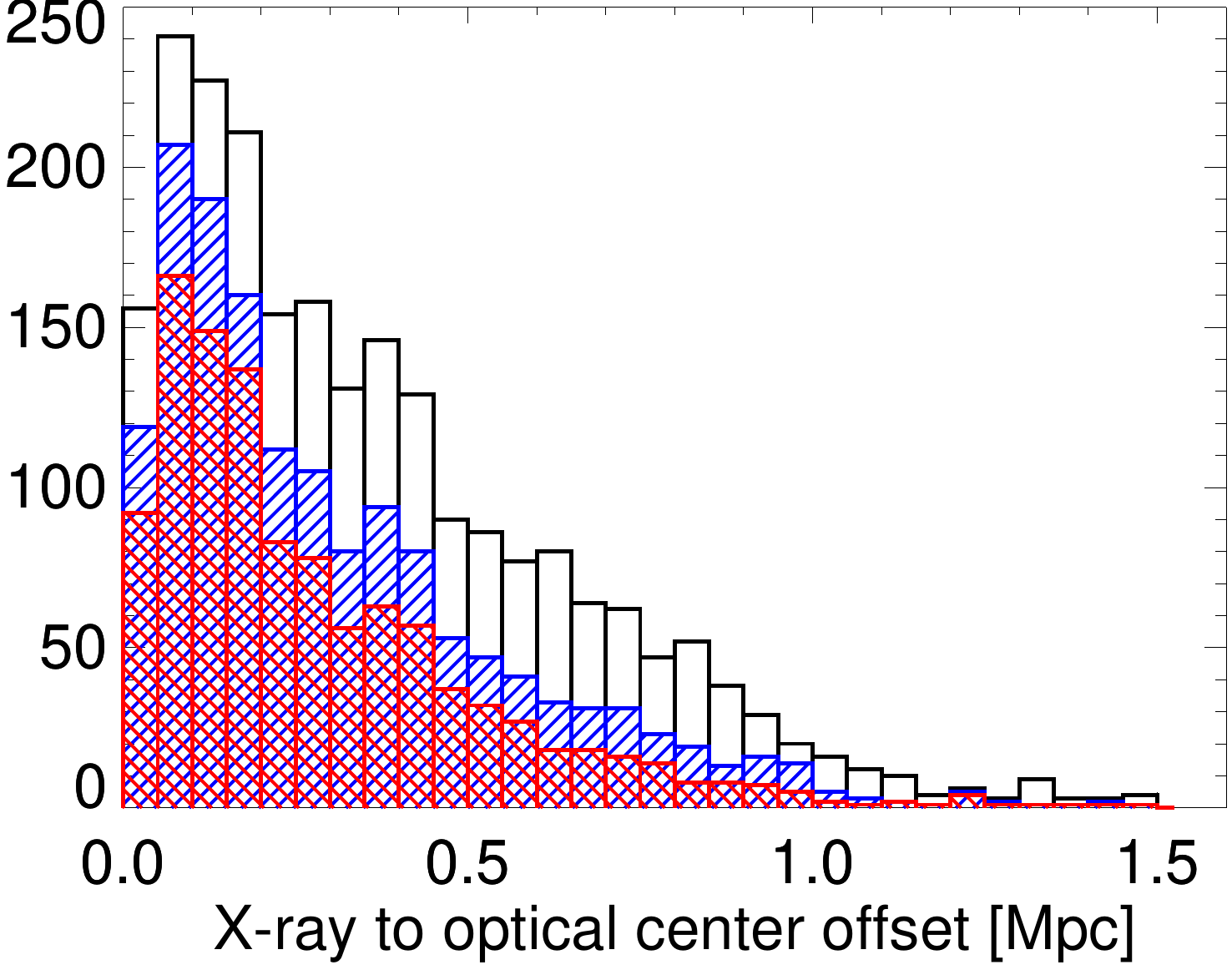}
\vskip-0.10in
\caption{X-ray to optical center offset distribution for the \fcont$<0.2$ (black), \fcont$<0.1$ (blue) and 
\fcont$<0.05$ (red) samples.}
\label{fig:offsets}
\end{figure}

\begin{figure}
\includegraphics[keepaspectratio=true,width=\columnwidth]{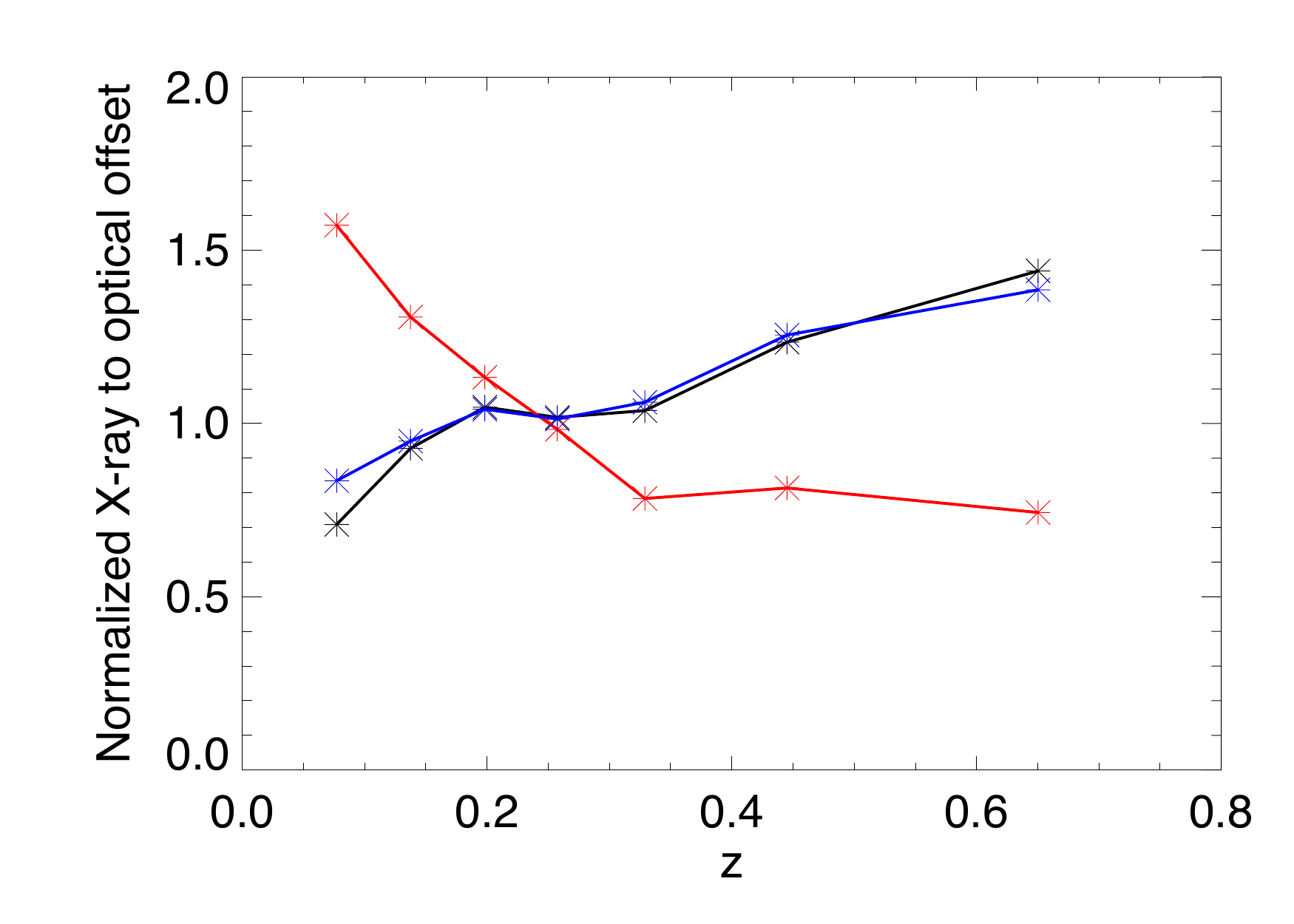}
\vskip-0.10in
\caption{ Redshift dependence of the median offset between the X-ray and the optical centers measured in Mpc (black), 
$r_{500}$ (blue) and angular distance (red) for sources with \fcont$<0.1$. Each line is separately 
normalized by the corresponding median offsets of the full sample, which are 0.21~Mpc, 0.23~$r_{500}$ and 56\arcsec.}
\label{fig:offsetsvsz2}
\end{figure}

\subsubsection{Cluster position measurements}
\label{sec:positions}

To explore the performance of the different center estimators we investigate the median offset of sources 
in units of $r_{500}$ as a function of redshift. Further, we measure the fraction of badly or unsuccessfully measured 
sources by listing the fraction of sources with offsets larger than $r_{500}$. 
The results are shown in Fig.~\ref{fig:offsetsvsz1}.
While the 2D fit method tends to give the smallest offsets, it tends to fail in 20-35\% of the cases. The rBCG 
identification seems to be unsuccessful in at least 30-40\%.
The galaxy density peak and the default centering show the lowest fraction of badly centered sources and a 
reasonable performance in positional accuracy.  As a reminder, the default center is the rBCG 
position unless it is offset more than 60\arcsec\ from galaxy density peak, in which case the 
galaxy density peak position is adopted (see Section~\ref{sec:centering}). 

The offset distribution between the 2RXS position and the default MCMF position is shown in Fig.~\ref{fig:offsets} for three 
different cuts in \fcont. We see a distribution peaked at $\sim$0.15~Mpc with a 
tail extending to $\sim$1.0~Mpc. 
The scatter between the optical and the 2RXS positions is mainly driven by the X-ray source positional uncertainties.  
At high redshift the cluster 
appears as an unresolved source in RASS, and thus the offset distribution is similar to that of a point source, 
reaching a constant angular value. At low 
redshift the cluster is resolved in the X-ray, and the offset distribution broadens.
This effect is shown in Fig.~\ref{fig:offsetsvsz2}, where we plot the median offset between the 2RXS and the default 
MCMF positions in different units as a function of redshift.
While the median offset in Mpc or $r_{500}$ is rising with redshift, the offset measured in angular units remains 
constant for $z>0.3$ at a level corresponding to $\approx 45$\arcsec.

\begin{figure}
\includegraphics[keepaspectratio=true,width=\columnwidth]{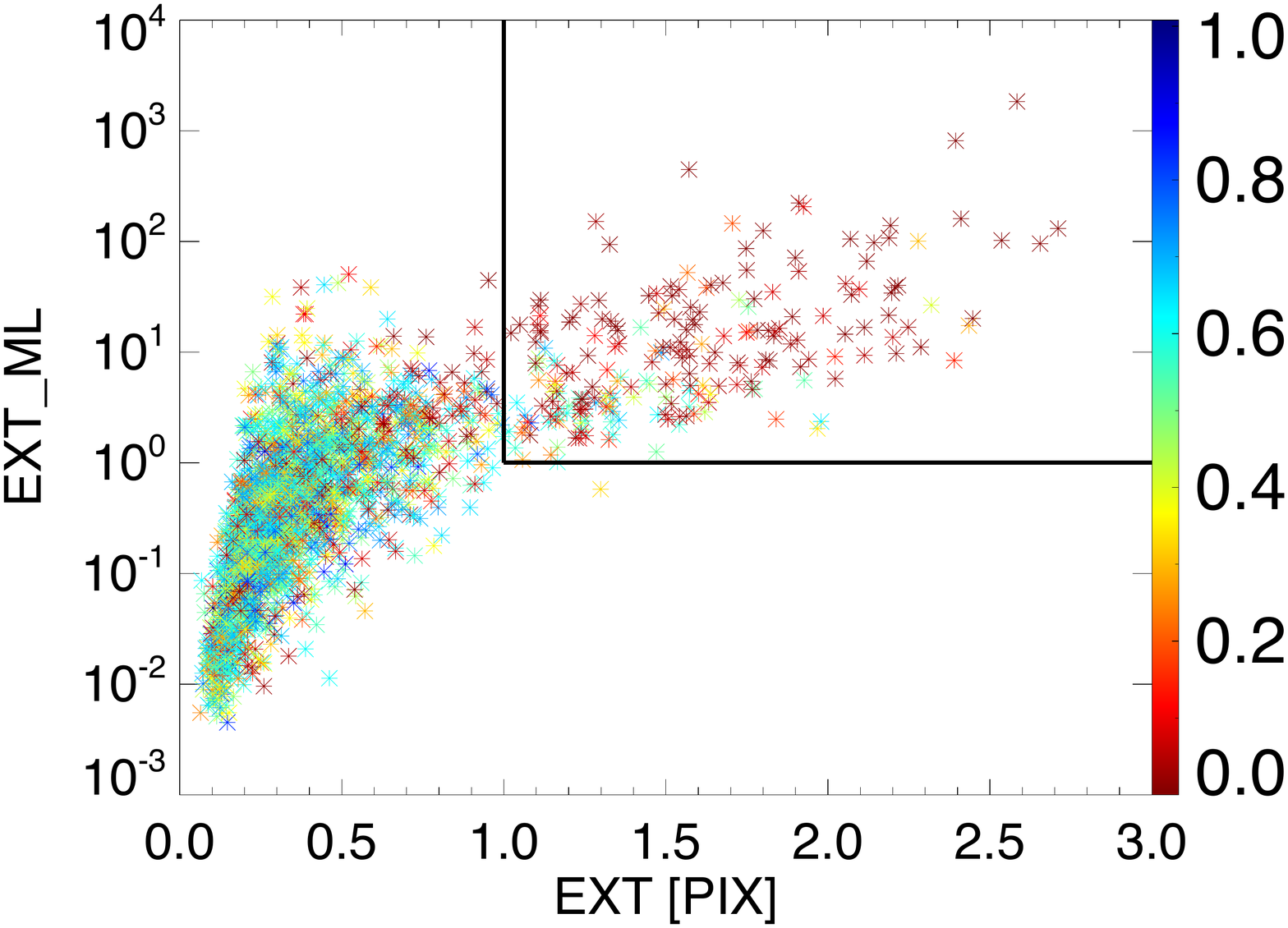}
\vskip-0.10in
\caption{Extent likelihood (EXT\_ML) versus extent (EXT) for sources with $z<0.3$ and EXT\_ML>0. The 
contamination estimator \fcont\  of each source is used as color coding. Black 
lines indicate the EXT>1 pixels ($45\arcsec$)
and EXT\_ML>1 region used to identify extended sources.}
\label{fig:extended1}
\end{figure}

\subsubsection{Extended sources in the 2RXS catalog}\

Typically, non-cluster sources in X-ray cluster surveys are excluded by requiring the
sources to show angular extent.  Working well above the noise threshold, this leaves 
typically $\sim$10\% residual contamination in X-ray selected cluster
catalogs \citep{vikhlinin98}, which can then be reduced through optical followup.
As mentioned in Section~\ref{sec:2rxs} the RASS survey PSF is large and therefore 
a pre-selection of cluster 
candidates based on extent is not possible for all but the brightest sources at low redshift.  
Because of this, the 2RXS catalog was created with a focus on point source detection, but source
extent estimates are still included.

Here we wish to explore the extended source subsample from 2RXS using the MCMF observable \fcont.
Fig.~\ref{fig:extended1} shows the 2RXS extent likelihood EXT\_ML versus 
extent EXT distribution of MCMF sources with $z<0.3$ and EXT\_ML >0. The \fcont\ estimator is used to color 
code the points. The majority of sources with low \fcont\  and measured extent are occupying a region of EXT>1 and 
EXT\_ML>1.

Within this region we find 220 sources with $z<0.3$, and 154 (70\%) of those have \fcont$<0.2$ and
 therefore would be classified as clusters both from an X-ray extent and an optical counterpart perspective.
We visually inspected 46 sources with \fcont$>0.4$, finding only one case to be an obvious cluster and a 
second case to be a less compelling cluster. In case of the less obvious counterpart we find the central 
region lacks DES color information in at least one band, causing the relevant region to be masked and thereby artificially
reducing the richness and increasing the \fcont. The obvious missing 
cluster is MACSJ0257.6-2209, which has been missed due to a DES photometric calibration flaw (see more detailed discussion
in the following section).
Between $0.2<$\fcont$<0.4$ we find 20 sources. In 11 cases we do find an optical counterpart.  All but 
one of those clusters are at $z<0.07$ and most of those systems are $z<0.05$. The missing one at higher redshift lacks 
data at the cluster center, which likely causes an under estimate of the richness and overestimate of 
\fcont.

We conclude that 70\% of the 220 2RXS extended sources with EXT>1 and 
EXT\_ML>1 are included in the MARD-Y3 cluster sample.  Of the remaining 66 extended source systems with \fcont$>0.2$, we find 
13 clusters. Eleven of those clusters have $0.2<$\fcont$<0.4$, and ten have redshifts $z<0.07$. Thus, those systems could be 
recovered or added to the MARD-Y3 catalog by requiring \fcont$<0.4$ and $z<0.07$ in addition to the selection in X-ray extent and extent likelihood.  Moreover, no additional non-cluster sources (i.e., contamination) would be added.  Finally, our analysis 
indicates that 53 of the 220 sources (24\%) with EXT>1 and EXT\_ML>1 are not clusters of galaxies.


\begin{figure}
\includegraphics[keepaspectratio=true,width=0.95\columnwidth]{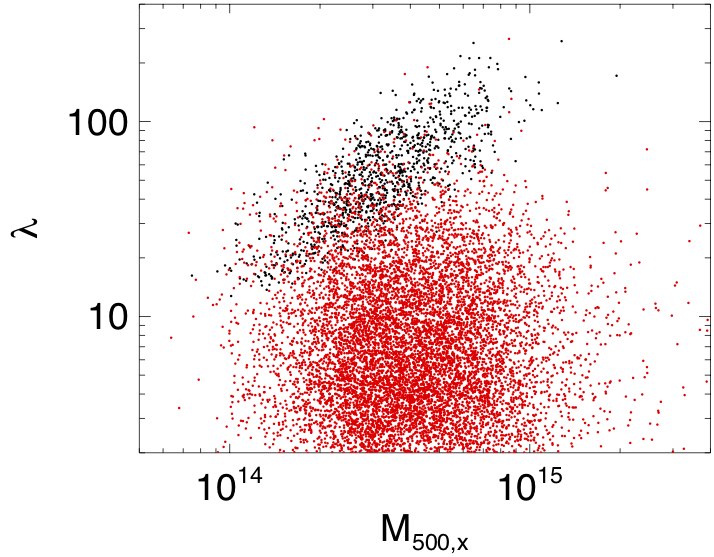}
\vskip-0.10in
\caption{Distribution in $\lambda$ and X-ray interfered mass (assuming MCMF redshift) for NWAY matches 
classified as AGN and stars (red) 
and \fcont$<0.1$ clusters without NWAY match (black).}
\label{fig:NWmatchesInMassRichness}
\end{figure}

\begin{figure}
\includegraphics[keepaspectratio=true,width=0.95\columnwidth]{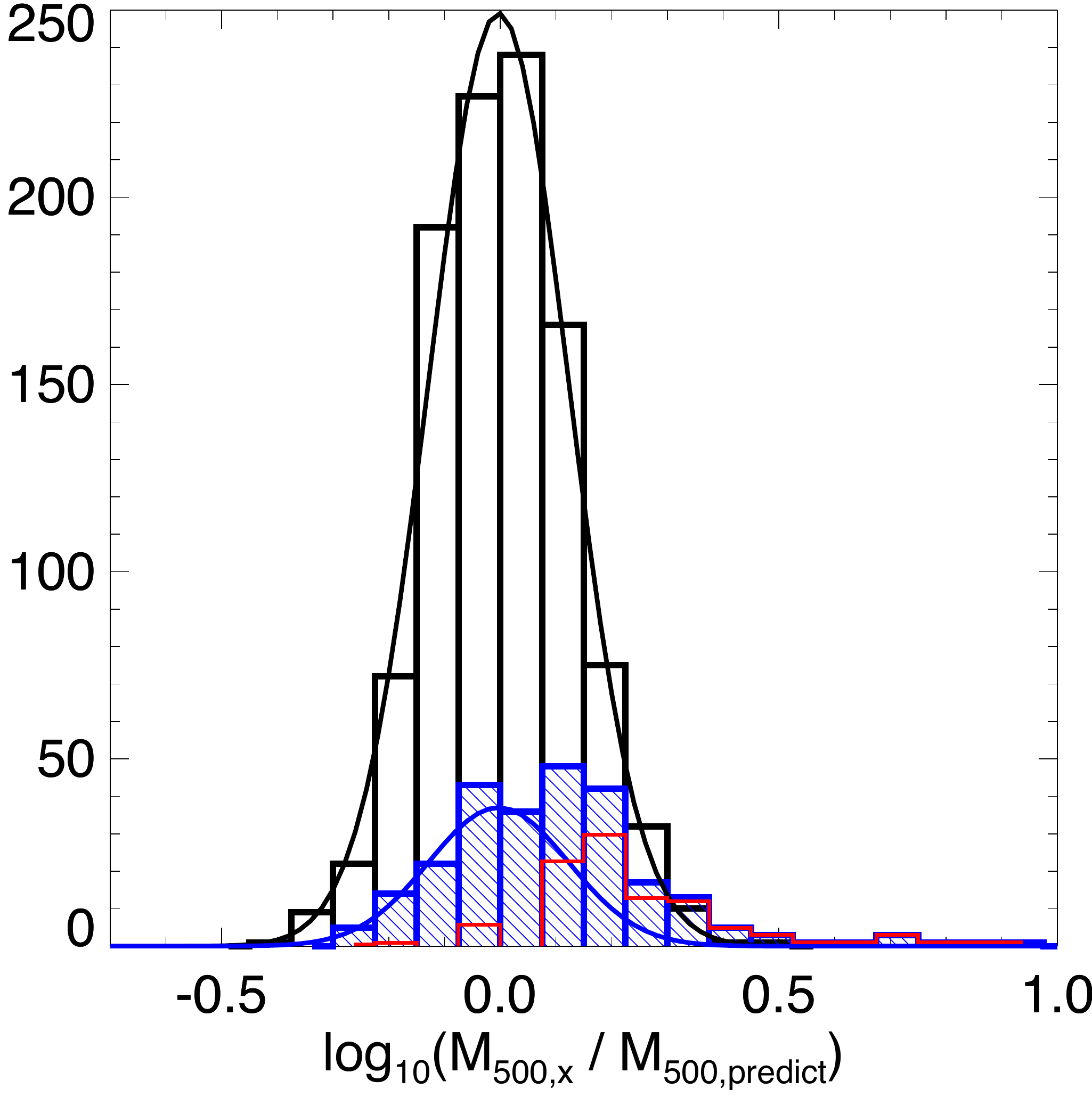}
\vskip-0.10in
\caption{Distribution of offsets from the mean scaling relation as derived from the \fcont$<0.1$ sample 
without NWAY match (black histogram) and with NWAY match (blue histogram). The black line shows the 
best fit Gaussian to the offset distribution, and the blue line show the Gaussian fit to the blue distribution 
for offsets below zero, keeping the position and width of the Gaussian fixed. 
The fitted Gaussian accounts for 59-63\% of the 
sources of the blue histogram. The difference between the blue Gaussian fit and the blue histogram is 
shown in red, indicating a tail of objects with NWAY matches that are outliers to higher mass.}
\label{fig:fc01lamMassoffset}
\end{figure}

\subsection{Catalog contamination by AGN and Stars}
\label{sec:contamination}

The Bayesian matching code NWAY allows one to reliably identify the most probable ALLWISE counterpart to the given 
2RXS source. However, it does not provide information on the nature of the source. The MARD-Y3 
catalog is created by adopting an \fcont\ threshold, which effectively excludes random superpositions and 
leads to a cluster catalog with contamination by random superpositions at the selected level. 
If we instead use the entire 2RXS catalog and apply no \fcont\ selection, then  
55\% of the 2RXS sources in our footprint have an NWAY counterpart matching our NWAY selection criteria. That sample  
is composed of 17\% class 1 stars (marked in blue in Fig.~\ref{fig:agnclassi}), 8\% class 2 stars (green), 
74\% AGN (yellow) and 1\% galaxies (red).

For an \fcont\ selection threshold \fcont$<$0.2 ($<$0.1, $<$0.05), the fraction of clusters with NWAY 
counterparts is 28\%, 20\% and 15\%, respectively.  This highlights both that the \fcont\ selection is effective 
at removing chance superpositions from our sample and that there is a residual population of NWAY sources associated  (either randomly or physically) 
with real galaxy clusters.

In general, we find a larger fraction of NWAY matches in our cluster sample  
than the expected fraction of random superpositions.
One reason for this is that an \fcont\ cut is a redshift dependent cut in richness, so 
by construction the source density is higher at the location of \fcont\ selected clusters compared 
to the typical source density, and this enhances the probability of finding an ALLWISE counterpart near this position. 
It is also important to note that the classification of NWAY sources we have adopted is far from 
perfect.  For example, a significant 
fraction of the NWAY matched sources could simply be associated with cluster member galaxies.
Finally, there are AGN associated with cluster positions, because of AGN in the clusters 
themselves and because the positions of AGN and clusters are correlated due to their connections to the 
distribution of large scale structure in the Universe \citep{Miyaji11,Koutoulidis13,Krumpe18}.

Of particular interest are cluster AGN with X-ray luminosities that are comparable to the cluster X-ray luminosity, including
cases where the AGN is the dominant source of X-rays \citep{Biffi18}. The probability of a 
cluster hosting an AGN increases with decreasing cluster mass and with redshift \citep{Allevato12,Oh14,Koulouridis18}. 
Therefore, one worries that AGN could enhance the detection of low mass clusters in a redshift dependent 
manner in an X-ray selected sample, thereby complicating the selection function.

We show in Fig.~\ref{fig:NWmatchesInMassRichness} the distribution of NWAY matched 2RXS sources in $\lambda$ and 
mass, where the mass estimate is derived from the inferred 2RXS luminosity using an X-ray 
luminosity-mass-redshift scaling relation appropriate for galaxy clusters \citep{bulbul19}.  For comparison, 
we show the $\lambda$-mass distribution of cluster candidates with 
\fcont$<0.1$ that do not have an NWAY match (black points) .
The majority of NWAY matches are well separated from the lambda-mass relation of clusters, but there 
are clearly some that overlap with the region occupied by clusters.
Introducing \fcont\ selection to clean the cluster catalog of random superpositions removes the NWAY matched sources 
selected at low $\lambda$.  This excludes most NWAY matches that are classified as AGN and stars, but a 
remaining fraction between 15\% and 28\% of the cluster sample still has an NWAY match. 
Restricting the cluster catalog to just those sources that are richer than the richest 
sources in the random catalogs (i.e. an \fcont$<$0.01 selection) results in about 10\% NWAY matches in 
the resulting cluster sample.  This indicates that a fraction $1\over3$, $1\over2$ and $2\over3$ of NWAY 
matches in the \fcont\ selected cluster samples with \fcont$<$0.2, 0.1 and 0.05, respectively, are associated 
(either through superposition of NWAY source with actual cluster or through actual physical association) with the clusters in the sample.  

To investigate the contamination of the cluster sample by NWAY X-ray sources in greater detail, 
we investigate the distribution of NWAY matches and non-matches in the \fcont$<$0.1 cluster sample in the
$\lambda$-mass plane. The NWAY matched sources are offset somewhat with respect to those sources
without an NWAY counterpart. Looking at the scatter around the best fit scaling relation, 
as shown in Fig.~\ref{fig:fc01lamMassoffset}, we can see 
that the offset distribution of the sources without an NWAY match is reasonably well 
described by a Gaussian distribution (black histogram). The 
distribution of the sample with NWAY matches (blue histogram) is smaller and includes a 
tail of sources whose estimated masses (from X-ray luminosities) are systematically higher than expected if they
were drawn from the distribution of the non-matched sources.  This is one way of visualizing the fact that a fraction of the NWAY matched 
sources have X-ray luminosities that are biased high with respect to the expected luminosity given their richness.

To estimate the number of such biased sources, we adopt the location and width of the 
Gaussian fit to the distribution from the clusters without NWAY matches (black line) and fit 
it to the blue distribution within the log-mass ratio 
range -0.4 to 0.1, while allowing only the normalization to change.  The 
result represents those clusters with NWAY matches that have no apparent bias in their 
X-ray fluxes.  This population is shown as a blue 
Gaussian curve in Fig.~\ref{fig:fc01lamMassoffset}.  Differencing the blue histogram of 
all sources with NWAY matches from those with 
NWAY matches that have no flux bias, we can isolate the subsample of systems that 
are biased.  This subsample is shown with the red histogram.
This analysis indicates that 59-63\% of the NWAY matched sources show no difference from the 
clean, non-NWAY-matched sample, while the remainder (37-41\%) exhibit different 
properties in the richness-mass (i.e., X-ray luminosity) plane.
This corresponds to $\sim$7\% of the full \fcont$<0.1$ sample.
Notably, this is comparable to the expected contamination by random superposition (10\%) for this sample, suggesting that the biased 
sample could largely be explained by the expected random superposition between NWAY identified stars and AGN and physically 
unassociated optical systems with sufficient richness to make it into our sample.

\subsubsection{Additional AGN exclusion filter}
With the goal of developing an additional cleaning step that would remove likely random superpositions 
from our cluster sample using the NWAY matching information, we explore the usage of 
various estimators such as positional offsets between ALLWISE and 2RXS source locations,
MCMF and 2RXS source locations and ALLWISE and MCMF source locations.  None of these
provided a clean selection of sources with obviously aberrant behavior in the richness-mass plane.
The estimator that worked best is the MCMF to ALLWISE position offset, because it indicates that the ALLWISE match 
is likely associated with the cluster (and therefore also consistent with the 2RXS position).
We estimate that with this estimator we could achieve contamination below 20\% for cluster or non-cluster samples when
using this estimator.

Our conclusion is that the simplest way to reduce contamination of the sample by NWAY sources 
is to use the offset between the observed richness and inferred mass (from the X-ray luminosity) and
the best fit scaling relation extracted from non-contaminated sources.  
The shape of the scatter distribution for the non-contaminated distribution
(e.g., the black Gaussian in Fig.~\ref{fig:fc01lamMassoffset}), allows one to estimate 
the incompleteness in the parent population that is introduced by  any cut that is applied to exclude outliers.   As an example, 
a cut of $\log_{10}M_\mathrm{500,x}/M_\mathrm{500,predict}>0.2$ excludes $57\%$ of non-clusters but only excluded 
$5\%$ of the true underlying cluster population.  This cut would also lead to a $\sim$12\% contaminated non-cluster sample.
We note that a $5\%$ exclusion of true sources with an NWAY match corresponds to a $0.6\%$ exclusion of true clusters in the total 
cluster sample, while the contamination by non-clusters is significantly reduced.
We adopt this method to reject non-cluster sources, and we include a qualifier in our master catalog that provides the 
offsets in sigma from the scaling relation for sources with NWAY matches. The standard cut used in this work excludes 
sources that show masses more than two sigma higher than the scaling relation prediction. The expected impact and 
remaining contamination for this cut is shown in Table~\ref{tab:catalogprop}.

\subsubsection{Cluster X-ray flux boosting by AGN}
We can use this test to investigate the fraction of clusters impacted by AGN within the cluster. 
By repeating the test with sufficiently low \fcont\ almost all NWAY matches need to be associated with the cluster, either 
being a normal cluster galaxy or a X-ray AGN.
 For that, we select clusters with \fcont$<0.01$ and $0.075<z<1$ , finding 361 sources, 33 with NWAY matches. Using the same approach
 as described above, we identify an excess of 7 sources, corresponding to 21\% of the 
 NWAY matches that do not described the same distribution as the 
clusters without an NWAY match. The strict cut on \fcont\ allows for $\sim$3 chance super positions, given that 2RXS 
contains $\sim$30\% spurious sources and assuming those will not have a NWAY match, we expect 2 mass or luminosity biased sources
in the NWAY matched sample.
Looking at sources with $\log_{10}M_\mathrm{500,x}/M_\mathrm{500,predict}>0.2$, we find seven sources, while we 
expect 1.5 from the distribution of non-matches and up to two from the cut in \fcont.
We visually inspect all seven sources, finding three cases where the NWAY match is consistent with sources classified as QSOs. 
All three are rBCGs of the clusters identified by MCMF. 
Further, we find one cluster that suffers from severe masking, another that has an X-ray 
emitting star projected near the rBCG and two unclear cases.


\begin{figure}
\includegraphics[keepaspectratio=true,width=0.95\columnwidth]{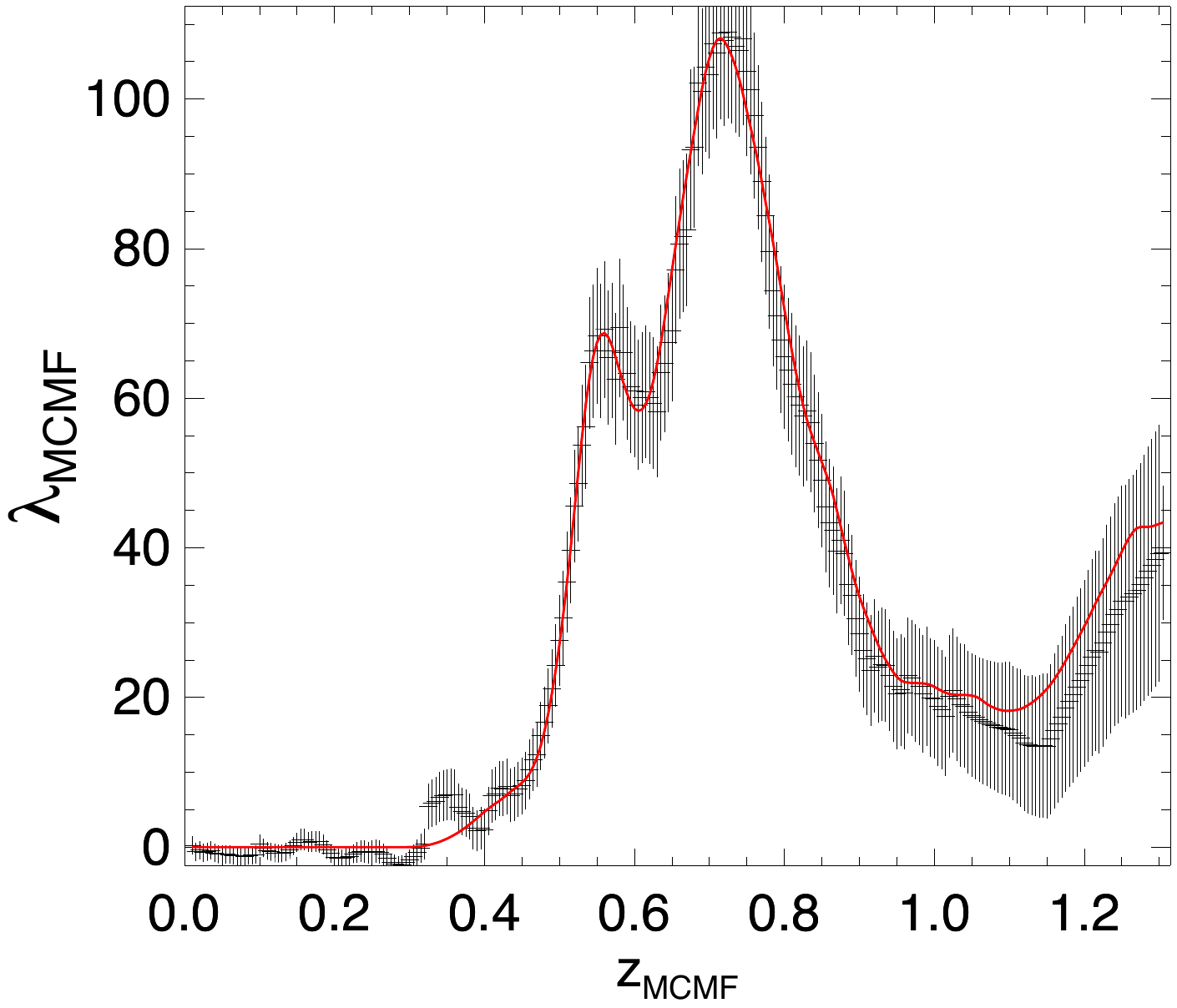}
\includegraphics[keepaspectratio=true,width=0.95\columnwidth]{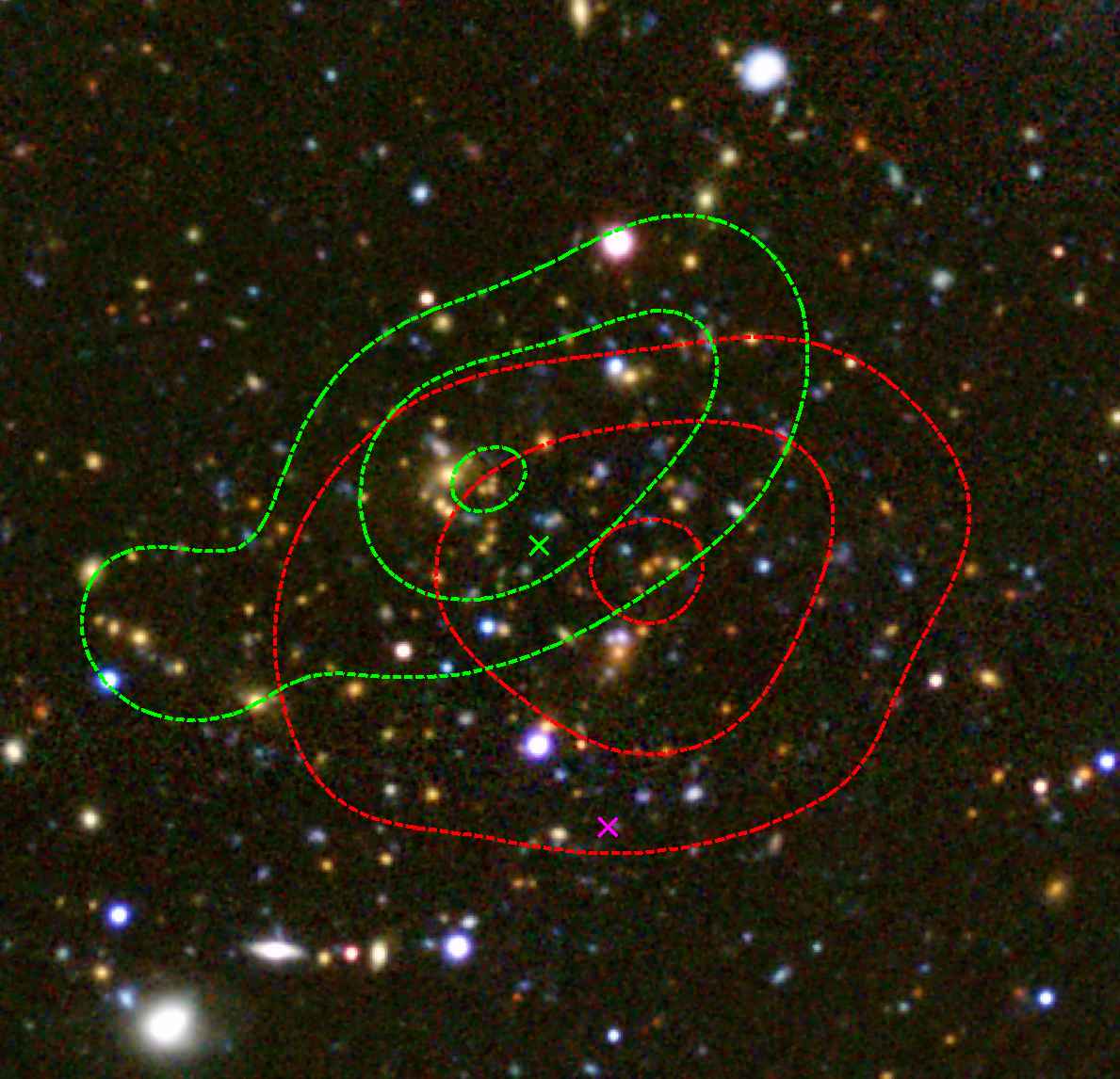}
\vskip-0.05in
\caption{Top panel: Lambda versus z as produced by MCMF for 
MARD J020216.7-540216 (SPT-CL J0202-5401). The red line represents the best model fit for two clusters at z=0.54 and z=0.70. Bottom: 
DES $grz$ pseudo-color image of the central 3x3 arcmin region, centered on the SPT position (green cross). The 
2RXS position is marked with a magenta cross. Green contours show the galaxy density for RS galaxies at z=0.54 and
red contours are used for z=0.7.}
\label{fig:SPTdoublepeak}
\end{figure}

\begin{figure}
\includegraphics[keepaspectratio=true,width=\columnwidth]{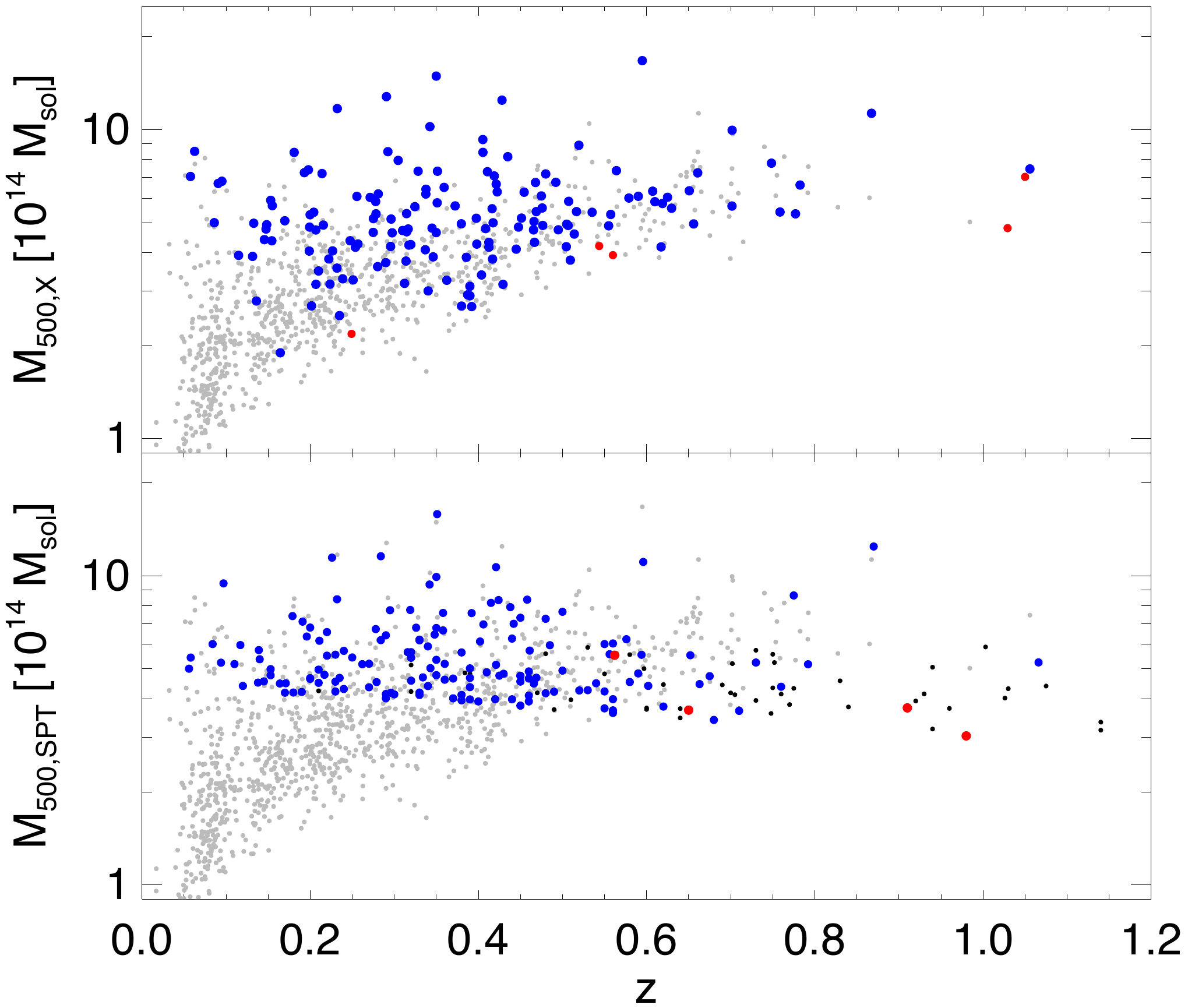}
\vskip-0.10in
\caption{Plot of SPT-SZ and overlapping MARD-Y3 (shown in gray in both panels) clusters in the mass-redshift plane.  Top panel: 
SPT-SZ matches to MARD-Y3 are shown in blue, and matches to sources in the 2RXS catalog with \fcont$>0.2$ are in red.  The red points lie 
near the lower mass limit, as expected given that richness and X-ray luminosity are physically correlated cluster properties.
 Bottom panel: 
 Similar to the top panel, but SPT-SZ sources using SPT based masses and redshifts.  SPT-SZ clusters without 2RXS counterparts are shown in black.
 MARD-Y3 fails to find high redshift, low mass SPT-SZ systems, as expected, but there are also unmatched SPT-SZ clusters in regions where
 MARD-Y3 clusters have been found.}
\label{fig:SPTmatchesMassRedshift}
\end{figure}

\subsection{Comparison to other cluster catalogs}
\label{sec:comparison}

Comparing the MARD-Y3 cluster catalog to other cluster catalogs enables us to assess the 
performance characteristics of MCMF and that of the other methods of cluster finding.
For the comparison we restrict ourselves to four large cluster catalogs: MCXC \citep{piffaretti11}, redMaPPer \citep{rykoff14}, SPT-SZ \citep{bleem15} and PLANCK PSZ2 \citep{planck25-27}. Further we limit this comparison to very simple tests, related to 
redshifts, consistency of mass proxies and missing sources in either of the catalogs.
A more stringent test on the MARD-Y3 catalog will be performed in the near future (Grandis et. al., in prep).

A detailed comparison of the characteristics can be found in the Appendix.  In the following we focus on the main lessons learned from these comparisons.
We find good agreement between MCMF derived photo-z's with a typically scatter of $\sigma_{\Delta z/ (1+z)}\approx0.005$, when we 
use the spec-z samples from the other catalogs.
Further, we find excellent agreement between our redshifts and the photo-z's given in the RedMaPPer catalog, although we see a clear
 bias in RedMaPPer photo-z's near the catalog redshift limit at $z\sim0.1$. 
We find a small number of outliers in these comparisons and determine that the main reason is that these sky positions 
have multiple clusters along the line of sight. 
One example MARD J020216.7-540216 (SPT-CL J0202-5401) is shown in Fig.~\ref{fig:SPTdoublepeak}, where MCMF 
finds two peaks in redshift at z=0.54 and z=0.7. 
The optical image reveals two distinct clusters separated by only 25\arcsec. 
We find that about 2 to 3\% of MARD-Y3 clusters show a second peak in redshift with \fcont\ less than 0.1 higher than the 
main counterpart (see also additional examples in Fig.~\ref{fig:comMCMFZmultipeak}).

Using the good agreement between redshifts to produce cleaner matched catalogs, 
we investigate our observed quantities such as luminosity, mass 
and richness with those listed in the external catalogs.
By comparing luminosities given in the ROSAT based MCXC catalog to those calculated 
by us using the 2RXS count rate, 
we identify a clear bias at low redshift due to the fixed aperture (5$'$ radius) used for the flux extraction in 2RXS.
We use the comparison of the 2RXS and MCXC fluxes to apply a redshift dependent aperture correction to our fluxes.  
As discussed further in the Appendix~\ref{app:MCXCcomp} (see also Fig.~\ref{fig:compMCXClum}), with this correction our X-ray luminosities
show good agreement with those from MCXC.

Comparison to the SPT-SZ catalog allows us to compare mass estimates based on 2RXS count rates and the \citet{bulbul19} 
luminosity-mass scaling relation with those from SPT. We find a median mass ratio $M_{500,\mathrm{L,x}}/M_{500,\mathrm{SPT}}$ 
of 1.07 for the uncorrected luminositities $L_\mathrm{X}$ and 1.02 for the aperture corrected luminosities described above (see Section~\ref{app:SPTcomp} and Fig.~\ref{fig:correctedMass}). For $M_{500,\mathrm{SPT}}$ we make use of the scaling relation given in \cite{dehaan16} that makes use of SPT clusters, BAO and BBN. 
Given that the \citet{bulbul19} luminosity-mass scaling relation is based also on SPT-SZ derives masses but using  \XMM\ 
observed luminosities in the 0.5-2.0 keV band, we use the offset in masses to estimate an additional correction factor$f_{0520}=0.96\pm0.02$ 
to convert from our RASS based, aperture corrected luminosities to the higher quality \XMM\ luminosities. 

Updating our mass estimates using this correction, we compare our MARD-Y3 masses to
those in the Planck catalog. We find a median mass ratio of $M_\mathrm{500,X}/M_\mathrm{500,Planck}= 1.19$, indicating a 
19\% offset with the Planck hydrostatic equilibrium based masses (see Section~\ref{app:Planckcomp} and Fig.~\ref{fig:PlanckMasscomp}).
Given the discussion above, this also indicates a 19\% offset
between the SZE derived masses in SPT-SZ that were employed in
\citet{bulbul19}, which are
consistent with those from weak lensing and dynamically derived SPT-SZ cluster 
masses \citep{dietrich19,stern19,Capasso19}.
A range of other galaxy and cosmic microwave background weak lensing
studies have also demonstrated that the hydrostatic 
equilibrium based Planck masses systematically underestimate the true cluster masses, driving the apparent
tension between the Planck SZE cluster and CMB anisotropy constraints 
\citep[e.g.,][see their Fig.~32]{vonderlinden14,hoekstra15,planck18-cosmolegacy}.

A comparison of the MCMF richness to the richness given in the RM catalog indicates a median ratio of 
 $\lambda_\mathrm{MCMF}/\lambda_\mathrm{RM}=1.087$ for \fcont$<0.1$ and a scatter of 24\%.
 It is interesting to note that we see a reasonable scaling between richnesses even for \fcont$>>0.1$, 
 where a large fraction of our sources are random superpositions.  There is further discussion in 
 Section~\ref{app:RMcomp} (see Fig.~\ref{fig:compRMlambda}).

We probe for clusters that are not matched in the MARD-Y3 or in the reference catalog. In the MCXC catalog, we
 find only one cluster that is clearly missed by MCMF:  MACSJ0257.6-2209 
 (see Fig.~\ref{fig:comMISPLK} and associated discussion in Section~\ref{app:MCXCcomp}). 
 The reason is a local failure of the 
 calibration of the MOF based $r$-band 
photometry that likely affects less than 0.25 \% of the DES area. MCMF in its current implementation is sensitive to large 
offsets in relative photometric zero points between bands.
However, besides two missing clusters due to missing data in the cluster core, we do not find any hint of unexpected 
incompleteness of our MARD-Y3 catalog, but we find some evidence for contamination in the MCXC catalog. 
We find that for $z>0.15$ about 10-15\% of our clusters do not show a RM counterpart 
due to more restrictive masking used in RM. We do not find missing  clusters if we consider the difference in masking. 
The fraction of clusters missing in RM below $z=0.15$ increases significantly 
due to their redshift cut of $z=0.1$.

Finally, we match to the  SPT-SZ cluster catalog and examine the matched and missing clusters
in the mass-redshift plane (see Fig.~\ref{fig:SPTmatchesMassRedshift}).
In the top panel the \fcont$<0.2$ MARD-Y3 sources are plotted in gray, and those with 
SPT-SZ matches are shown in blue. SPT-SZ sources that are matched with 2RXS sources 
having \fcont$>0.2$ (i.e., sources that did not make the MARD-Y3 cut) are in red.
The masses and redshifts in the top plot are from MARD-Y3.
The low \fcont\ SPT-SZ matched sources (in red) all lie close to the effective mass 
limit of the MARD-Y3 sample, as would be expected given the luminosity-richness 
(or equivalently the mass-richness) relation for our cluster sample 
(Fig.~\ref{fig:compMCXClamlum} shows the luminosity-richness
relation for a subsample of the MARD-Y3 catalog).  There are clearly MARD-Y3 clusters
above the mass threshold that did not make it into the SPT-SZ catalog.

The bottom panel is similiar but also shows all SPT-SZ sources without a 2RXS 
match (black points).
Redshifts for the SPT-SZ sources in this panel come from \cite{bleem15}, and masses are based on the scaling 
relations given in \cite{dehaan16} using SPT+BAO+BBN (their Table 3, results column 2).

In the bottom panel, one can see that all SPT-SZ systems near the SZE selection threshold at a 
redshift $z>0.6$ fail to make it into the MARD-Y3 catalog.  This is expected, because the X-ray
fluxes of these sources lie below the 2RXS detection threshold.  However, there are cases of SPT-SZ clusters
without matches that lie in regions of mass-redshift space where MARD-Y3 clusters exist and there are MARD-Y3 clusters above the mass limit of SPT that do not have a match.
This is expected given the scatter in observable-mass relation. The SPT sample is $50\%$ complete at $4-5\times10^{14} ~\mathrm{M}_\odot$ and, therefore, finding unmatched MARD-Y3 clusters in this regime is expected.
The luminosity based masses provided in MARD-Y3 can be expected to be more noisy than those from other works, because of the scatter introduced by the flux measurement within a fixed aperture and because of the low significance of the detection, causing the measurement error to contribute significantly to the scatter in mass. That scatter may indeed play a role is indicated when looking at the richness and X-ray based masses for matches and non-matches. We find good agreement for between both mass estimates for sources matched with SPT-SZ. For sources without SPT-SZ match, we find an offset that corresponds to $\sim1\sigma$ of the scatter between both mass estimates. We find similar scatter between both mass estimates for matched and non-matched clusters. An offset between both sub-samples is expected in both cases, contamination by non clusters and by the impact of the SPT-SZ selection function on the matching fraction, but the size of the scatter between mass proxies should be enhanced for a sub-sample that is significantly more contaminated. The similarity in scatter therefore indicates similar size of contamination in both samples. As a last check, we visually inspected all SPT-SZ non-matches with $z>0.6$, finding no obvious case of contamination of that sample.

A more quantitative interpretation of Fig.\ref{fig:SPTmatchesMassRedshift} within the context of both the SZE and X-ray observable mass relations and their scatter \citep[as carried out for SPT-SZ and RM catalogs; see][]{saro15} is challenging, and will be carried out in a future paper (Grandis et al., in prep.).  
The topic of completeness , contamination and consistency with SPT clusters will be further addressed in Sec.\ref{sec:luminosity_function}, where we present the galaxy cluster X-ray luminosity function and its consistency with cosmological predictions informed by SPT-SZ clusters.

As a last catalog based test, we assess how many systems are indeed newly discovered systems. As there is no complete meta catalog of all known clusters, we restrict to all clusters and groups listed in NED that do have a redshift estimate. We match the MARD-Y3 to all those system, requiring a maximum offset of 1~Mpc from the optical position and a maximum redshift difference between MCMF and NED of 10\%.
For the \fcont$<0.2$ sample, excluding multiples and potentially AGN contaminated clusters, we find 762 matches. For the  \fcont$<0.1$ and 
 \fcont$<0.05$ samples we find 617 and 523 matches respectively.
 Given the number of clusters listed in Table\ref{tab:catalogprop} this indicates
that 65\%  (58\%,52\%) of the \fcont$<0.2 (0.1,0.05)$ sample are new galaxy clusters.


 \begin{figure}
\includegraphics[keepaspectratio=true,width=0.47\textwidth]{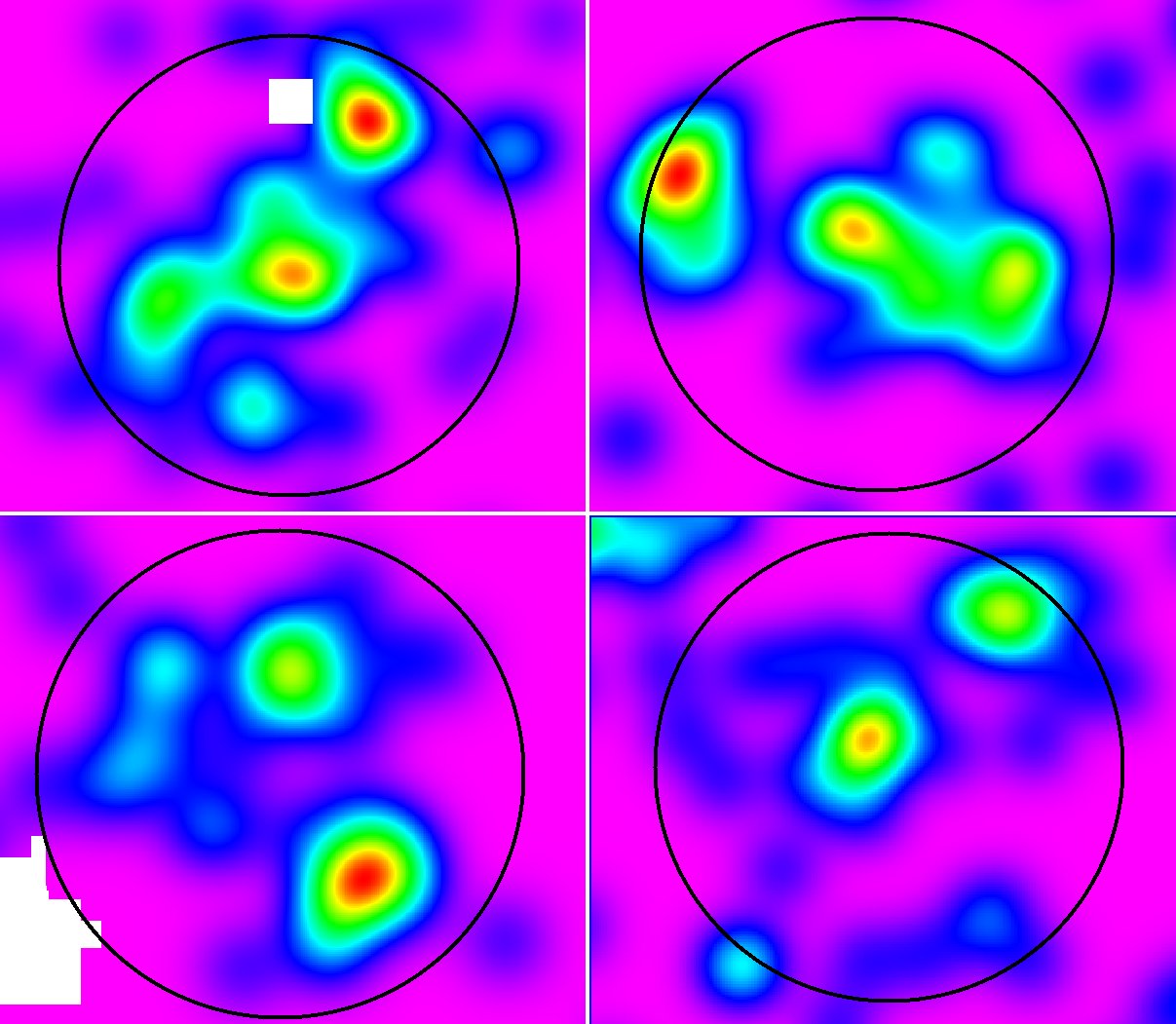} 
\vskip-0.05in
\caption{RS galaxy density maps for four MARD-Y3 clusters selected to be
unrelaxed systems according to our dynamical state estimators.  The black circles mark a region of 1~Mpc 
radius around the 2RXS position of each cluster.  Density maps like these are available for all MARD-Y3 clusters.} 
\label{fig:fourmostunrelaxed}
\end{figure}

\begin{figure}
\includegraphics[keepaspectratio=true,width=0.47\textwidth]{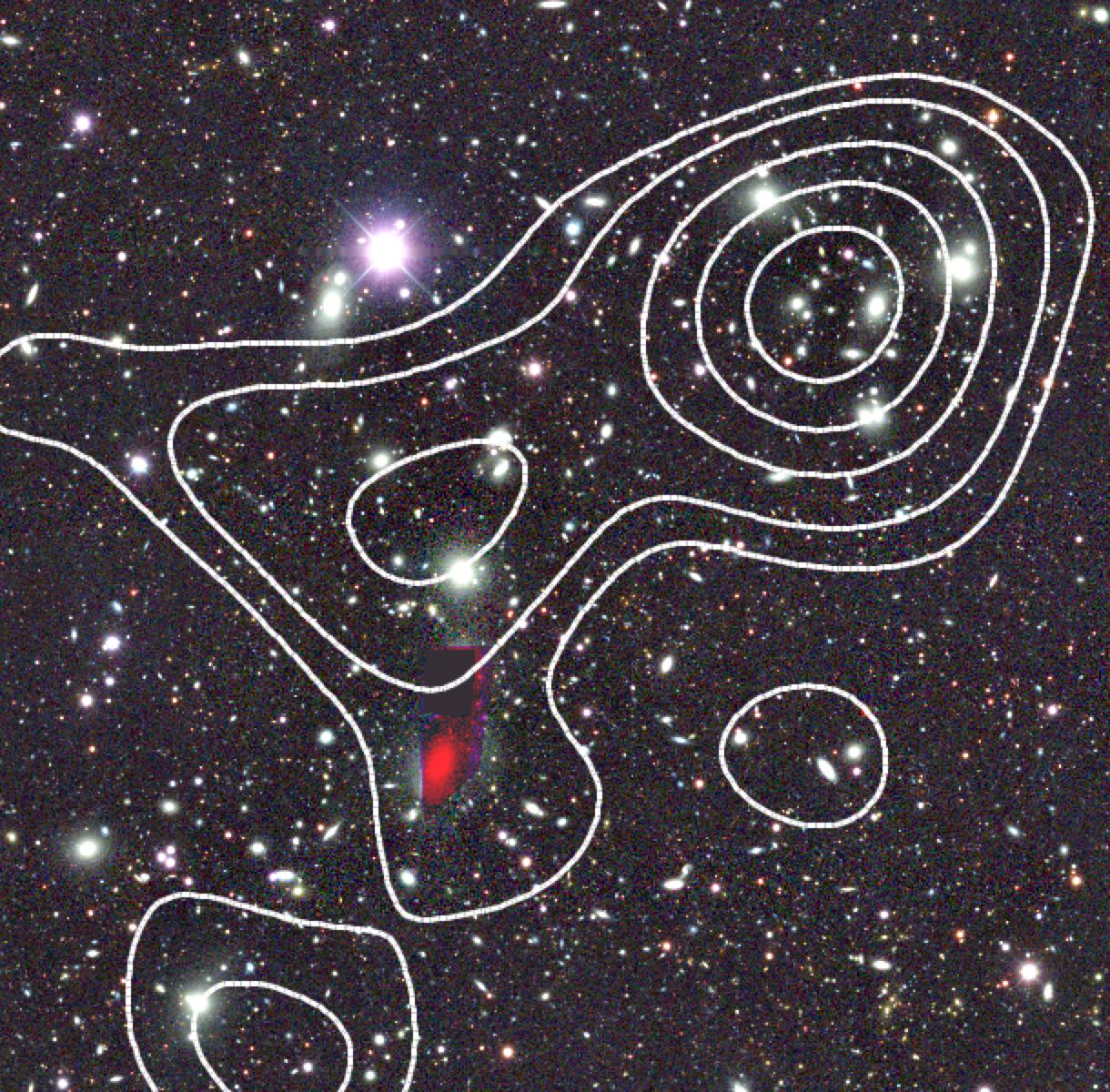}
\includegraphics[keepaspectratio=true,width=0.47\textwidth]{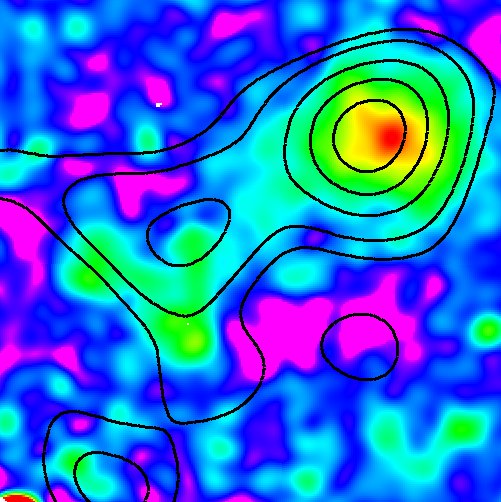}
\vskip-0.05in
\caption{A514 (z=0.073), the most disturbed system according to $\gamma$ with \XMM\ observations. Top 
panel: DES g,r, z color composite image of a 17x15 region around the cluster. Bottom panel: Smoothed \XMM\
surface brightness map of the same region. Contours show the MCMF RS galaxy density contours. } 
\label{fig:A514}
\end{figure}

 \begin{figure}
\includegraphics[keepaspectratio=true,width=0.47\textwidth]{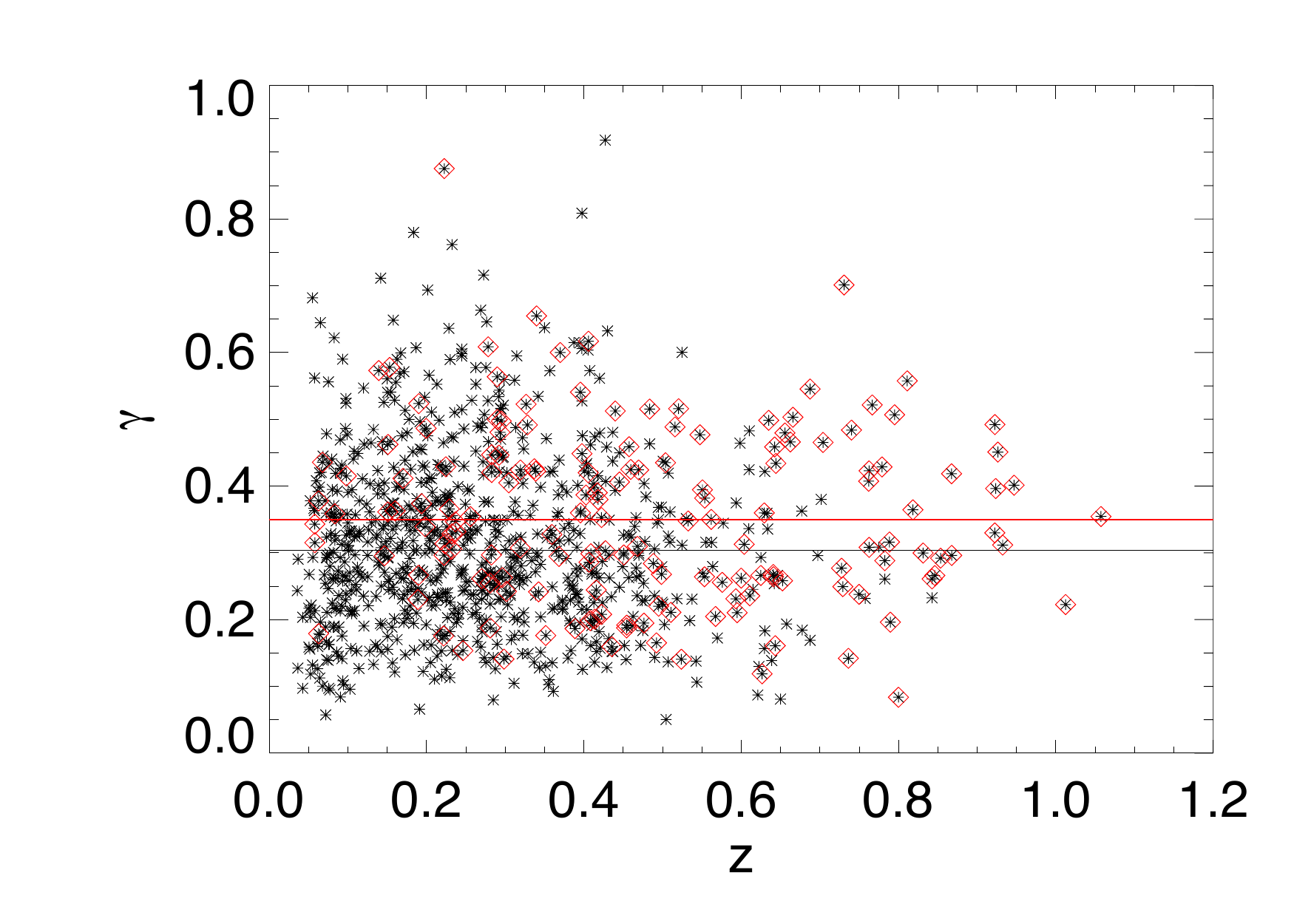}
\vskip-0.10in
\caption{Combined dynamical state estimator $\gamma$ versus redshift for 890 clusters with \fcont$
<0.1$. Clusters with $M_{500,X}>5x10^{14} \mathrm{M}_{\odot}$ are marked as red diamonds. The continuous 
lines show the median value of the corresponding clusters.} 
\label{fig:dynestvsz4}
\end{figure}

\subsection{Galaxy cluster dynamical state estimators}
\label{sec:dynamical_state}

The estimators described in Section~\ref{sec:dynstate} are probing the dynamical state 
in different ways and are therefore sensitive to different merger time scales and configurations. 
Of course, these estimators based on galaxy distributions are noisy due to, among other things, Poisson statistics, 
making it challenging to use them to
select and order systems by merger state.  With this in mind, we examine the distribution of dynamical state estimator for
the MARD-Y3 sample. 
We build a simple combination of the \cite{Wen13} estimators, $\gamma = (\alpha + \delta + 1 -  \beta) / 3$ for 
this initial investigation.  Relaxed clusters will show a small value of $\gamma$, whereas 
merging system will show a high $\gamma$.
For our initial tests we use the \fcont$<0.1$ sample and exclude those systems 
where the 2D fit failed. This yields a sample of $\sim$890 clusters.

We investigate the visual appearance of the galaxy density maps and $g$, $r$, $z$ pseudo-color images for 
the most extreme clusters selected with the $\gamma$ estimator. We find that the systems that show a high $\gamma$
are indeed undergoing merger activity.  Fig.~\ref{fig:fourmostunrelaxed} contains the galaxy density maps of 
the four most unrelaxed systems selected with $\gamma$.
Furthermore, we examine the most disturbed cluster that has existing high resolution X-ray imaging data: Abell 514.
This cluster has previously been identified as  a merging 
cluster \citep{2008A&A...490..537W}. We find good agreement between our RS galaxy density map and the 
\XMM\ X-ray surface brightness map (see Fig.
\ref{fig:A514}), indicating that our galaxy density map indeed follows the morphology of the merging cluster.

Finally, we look at the redshift and mass dependence of the $\gamma$ measurement for the MARD-Y3 sample.  
Fig.~\ref{fig:dynestvsz4} contains the distribution of $\gamma$ as a function of photo-z's. We find a median $\gamma$ 
of 0.3, and no significant evidence of variation with redshift. We repeated the same task by applying a mass cut of $M_{500,X}
>5\times10^{14} \mathrm{M}_{\odot}$. We find a shift in the median value from 0.3 to 0.35, but no redshift trend is visible.
Whether the offset between the full sample and the mass limited sample is of physical nature or a side effect of the 
dynamical state estimator (e.g. due to an increased number of cluster galaxies) is a question that awaits further investigation, 
for instance by comparing with alternative estimators.


 \begin{figure*}
\includegraphics[keepaspectratio=true,width=0.95\textwidth]{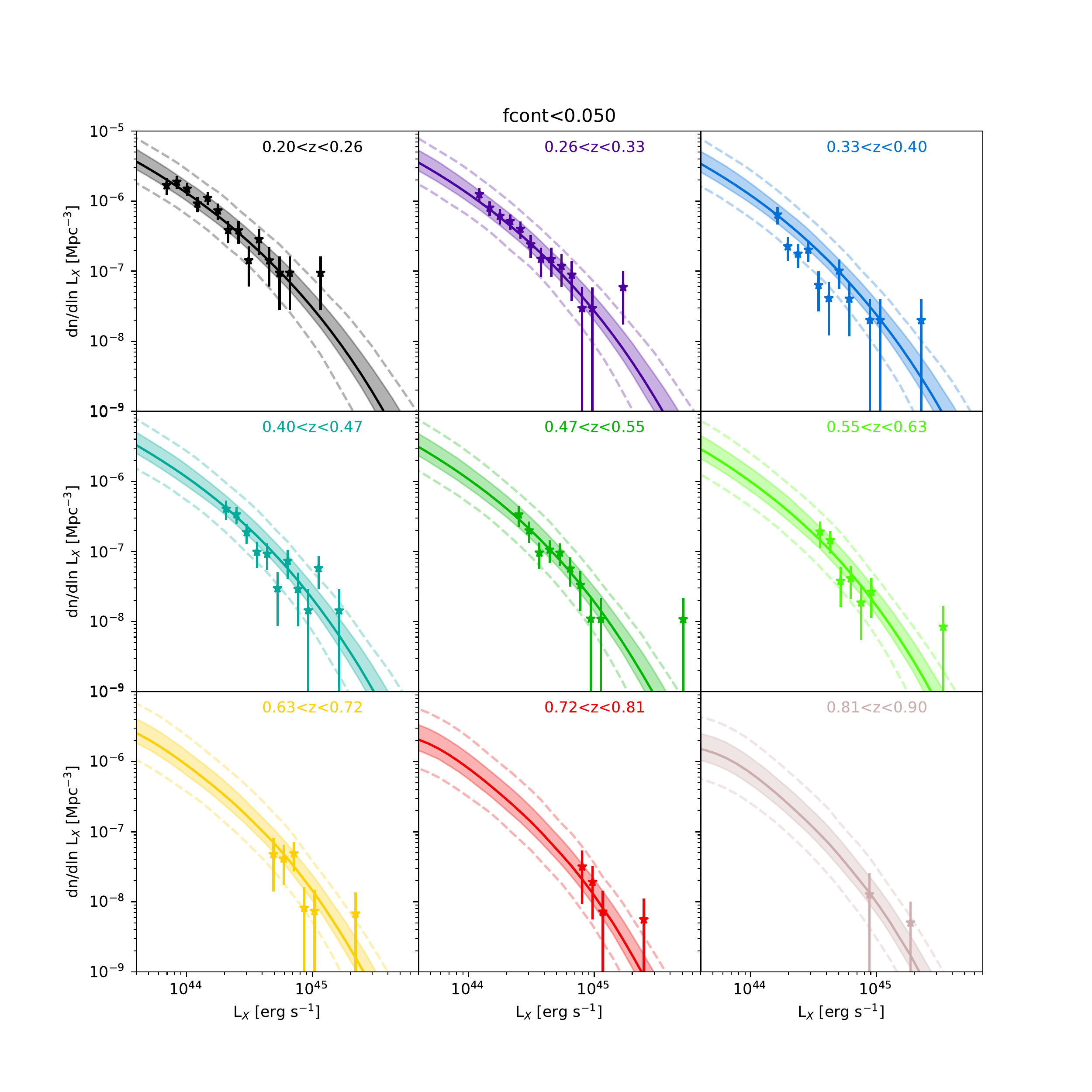}
\vskip-0.40in
\caption{In each of nine redshift bins, we plot the completeness corrected luminosity function measurements (points with error bars)
from our cluster catalog (using $f_\text{cont}<0.05$), where  
the error bars are Poisson only.  For comparison, we plot the theoretical luminosity function (solid line) expected in a fiducial 
cosmology \citep[e.g.,][]{dehaan16} when adopting the X-ray luminosity--mass--redshift scaling relation from \citet{bulbul19}.  Note that $L_\mathrm{X}$=$L_\mathrm{500,0.5-2.0}$ in this figure.  The model uncertainties are derived using the luminosity-mass-redshift scaling relation uncertainties (statistical plus systematic) in each redshift bin, where the 1$\sigma$ region is shaded and the 2$\sigma$ region is marked with dashed lines. The shape and redshift evolution of the measured cluster luminosity function agrees with the model predictions.}
\label{fig:costestLF}
\end{figure*}

\subsection{Galaxy cluster X-ray luminosity function}
\label{sec:luminosity_function}

The MARD-Y3 catalog is the product of following up about 20000 X-ray sources to produce a clean 
cluster catalog of 1000-2000 sources. As described above, we apply a search for optical counterparts 
along the line of sight toward each source and then apply an \fcont\ cut to exclude $\geq90\%$ of the 
sources, because they do not have sufficiently high probabilities of being real clusters.
As a test that the resulting cluster catalog can be described by a simple two-step selection function that is the 
combination of X-ray selection to enter the 2RXS catalog followed by optical confirmation to enter the final cluster catalog, 
we model this selection process and use it to investigate the X-ray luminosity function and to compare it with the prediction from a
fiducial cosmology.  
As fiducial cosmology, we adopt the cosmology derived from the combination of SZE selected clusters 
together with BAO and Planck CMB anisotropy \citep[][second results column of Table 3]{dehaan16};

To avoid having to model the contamination from random superpositions and to still have a large sample, we explore
the luminosity function using the \fcont$<0.05$ MARD-Y3 catalog.  This catalog, after filtering out the 
AGN contamination associated with NWAY matches (see Section~\ref{sec:comparison} and ~\ref{sec:contamination}) still contains 
$\sim$2.6\% contamination.  For convenience, 
we distribute this sample of 1086 clusters (see last row of Table~\ref{tab:catalogprop})  into bins
 of redshift and luminosity. For the X-ray luminosity we adopt $L_\mathrm{500,0.5-2.0}$, 
 the luminosity in the rest frame 0.5-2.0 keV band with an aperture correction derived by cross comparison 
 with the MCXC sample and additional small shift from comparison with the overlapping 
 SPT-SZ sample observed with \XMM\ (see Section~\ref{sec:comparison} and Appendix.\ref{app:comparison} for discussion).  
 
We restrict our analysis to those clusters with redshift $z\ge0.2$, because the aperture corrections to the 
luminosity are less important here.
The number of clusters in each redshift bin is divided by the cosmological 
volume in that bin, given the survey footprint and our fiducial cosmological model.  This produces a function of the 
space density of sources (and associated Poisson uncertainties) as a function of luminosity in each redshift bin 
(i.e., the observed and not yet selection-corrected luminosity function).

We then estimate the impact of the two stages of selection on this observed luminosity function so that we 
can correct the observed luminosity function to the selection-corrected observed luminosity function.  Note that this is a departure from the
standard approach of forward modeling to the data that we have employed in our past cosmological analyses and
forecasts \citep{bocquet15,dehaan16,grandis18}, but our goal here is to produce a simple figure to support a visual 
consistency test rather than carrying out a robust cosmological analysis.

Considering that the 2RXS X-ray catalog has been selected in detection likelihood rather than in flux or luminosity, 
we use the catalog to determine a scaling between the rest frame 0.5-2.0~keV flux $f=L/(4 \pi 
D_L^2)$  and the 2RXS detection likelihood as a function of redshift and exposure time. 
With that information, we estimate the probability that a cluster of a given flux has a detection likelihood larger than 6.5
and average the resulting X-ray completeness over the solid angle weighted RASS exposure time distribution appropriate for our catalog.
Furthermore, we take into account that the 
\fcont\ selection is equivalent to a \LamMCMF(z)$>\lambda_\text{min}(z)$ cut, as described in 
Section~\ref{sec:random_superposition} above. 
We estimate the impact of the optical selection by computing the probability that a cluster of luminosity $L_\mathrm{500,0.5-2.0}$ 
and redshift $z$ has a richness larger than $\lambda_\text{min}(z)$. To compute these probabilities, we adopt a core included
0.5 to 2.0~keV luminosity-mass-redshift scaling relation \citep{bulbul19}, 
a richness-mass-redshift scaling relation \citep{saro15}, taking into account the ratio \LamMCMF/\LamRM = 1.087 derived in Section~\ref{sec:comparison}, and the underlying  
halo mass function \citep{tinker08} within our fiducial cosmology. The completeness corrected luminosity function is shown in 
Fig.~\ref{fig:costestLF} as stars with corresponding Poisson error bars.

The predicted luminosity function is computed by combining the halo mass function with the luminosity-mass-redshift 
scaling relation to determine the number density of clusters in logarithmic luminosity bins for any given redshift. 
The model uncertainty on the luminosity function is derived by marginalizing over the published uncertainties of the 
parameters of the luminosity-mass-redshift scaling relation and the richness mass relation \citep{bulbul19,saro15}.
The predicted luminosity function with its model uncertainties is shown as the color-coded solid line, shaded region 
(1$\sigma$) and dashed lines (2$\sigma$ region) in Fig.~\ref{fig:costestLF}. We find that the prediction is 
statistically consistent with the measured number densities. 
This consistency test implies that the MARD-Y3 catalog is a cluster catalog that can be described 
by a simple and separable two-step X-ray and optical selection process, and that the catalog could potentially 
be useful for cosmological studies. We carry out a more extensive study of the cluster sample and selection model in a 
companion publication (Grandis et al., in prep).


\section{Conclusions}
\label{sec:conclusions}

In this work we present the complete optical follow-up of RASS-2RXS X-ray sources within the DES footprint. We 
apply the the multi-component matched filter cluster confirmation tool (MCMF) to $\sim$20000 2RXS sources to 
create the MCMF confirmed cluster catalog, MARD-Y3, consisting of between 1086 and 2171 clusters, depending on the 
desired level of residual contamination from random superpositions of unassociated optical systems with 
2RXS sources (\fcont\ cuts ranging 0.05 to 0.2 with AGN rejection leading to final contamination 
ranging from 2.6\% to 9.6\%; see Table~\ref{tab:catalogprop}).  
This large sample of new X-ray selected clusters extends to redshift $z\sim1$ with a median redshift of $z\sim0.25$ 
(see Fig.~\ref{fig:redshiftdistri}).  
Thus, in comparison to the previously published REFLEX \citep{boehringer04} 
catalog or even the unpublished REFLEX II selection over the same region, our sample has 8$\times$ and 4$\times$ 
more clusters, respectively, and extends to much higher redshift, containing $\sim$100 clusters at $z>0.5$. Matching to all clusters listed in NED we find that $52$ to $65\%$ do not have a counterpart within 1Mpc and $|\Delta z|/(1+z)<0.1$ indicating that the majority of our clusters are indeed new systems.

The photo-z performance of MCMF using DES Y3A2 Gold $g$, $r$, $i$, and $z$-band SOF photometry has 
further improved compared to our pilot study \citep{Klein18} and shows a scatter of 
$\sigma_{\Delta z / (1 +z_{spec})}=0.0046$  around the spec-z's for the subsample of 242 
clusters with available spec-z's.  This performance extends over the full redshift range of the catalog 
$0\le z\le1.1$ (see Fig.~\ref{fig:redshiftdist}), with an indication that at $z>0.43$ the typical scatter is 
$\sim$1.5$\times$ larger (see Section~\ref{sec:PhotozPerformance}).

Using the catalog of matched ALLWISE \citep{wright10} and 2RXS \citep{boller16} sources produced 
using the NWAY \citep{Salvato18} matching algorithm, we 
study the contamination of the cluster catalog by AGN and stars (Section~\ref{sec:contamination}).  In 
general, one can exclude contamination by AGN simply by making tighter cuts on \fcont, which reduces 
the number of random superpositions with physically unassociated optical systems along lines of sight toward the 
2RXS sources.  We find for a sample with \fcont$<0.01$ we have a total of 361 clusters and 33 or 
$\sim$9\% of those have associated NWAY AGN.  Of that subsample, we estimate that 7 show 
evidence of X-ray flux boosting (with respect to the expected X-ray flux given the optical richness 
and redshift).  Accounting for the fact that  \fcont$<0.01$ allows two to three chance superpositions, 
this suggests that for the flux limits and redshift ranges explored in our catalog, 
measurable AGN contamination by AGN in clusters is present in only $\sim$5 out of 361 
or $\sim$1.4\% of the cluster sample. Visual inspection of these sources reveals that three are
AGN likely associated with the rBCG of the cluster, corresponding to a measurable impact by
cluster AGN in $\sim$1\% of our systems.

We study the X-ray to optical offset distribution for our cluster sample, demonstrating median offsets of 
0.21~Mpc, 0.23~$r_{500}$ and 56$''$ for the \fcont$<0.1$ sample. The typical angular offset drops with 
redshift to $z\sim0.3$, beyond which it stays constant as expected if the typical cluster becomes unresolved 
 at that redshift and beyond (see Fig.~\ref{fig:offsetsvsz2}).  Our expectation is that the X-ray to optical offset 
distribution is dominated by 2RXS X-ray positional uncertainties, and we therefore recommend the use 
of the MCMF derived optical cluster positions (discussed in Section~\ref{sec:positions}).

We compare our new cluster catalog to other X-ray  \citep[MCXC;][]{piffaretti11}, optical \citep[RM;][]{rykoff14} and 
SZE selected cluster catalogs (including SPT-SZ \citep[][]{bleem15} and Planck PSZ2 \citep[][]{planck25-27}; 
see Section~\ref{sec:comparison} and the more detailed discussion in Appendix~\ref{app:comparison}).  
We identify redshift outliers in all 
catalogs that are typically due to multiple richness peaks along the line of sight, and we identify a significant 
redshift systematic in the RM catalog at redshifts $z<0.15$.  
We use the MCXC cluster fluxes to derive an aperture correction for our 2RXS fluxes at low redshift. 
The corrected fluxes result in good agreement between our X-ray luminosity based masses and 
those from SPT-SZ \citep{bulbul19} with a ratio of 1.02. This factor does not account for systematic 
differences between X-ray bands of 2RXS (0.1-2.4~keV) and \XMM\ (0.5-2.0~keV). 
Assuming the factor of 1.02 fully results from this difference, we find a factor $f_{0520}=0.96\pm0.02$ to 
convert our 2RXS based, aperture corrected luminosities into 0.5-2.0 keV luminosities that then
yield masses fully consistent with SPT-SZ.
Using these calibrated masses, we find a mass ratio of 1.19 in comparison to the hydrostatic 
equilibrium based masses in the Planck sample.  If as indicated by our analysis, the Planck cluster masses
are on average biased 20\% low, then this would largely resolve the previously reported tension
between Planck SZE selected cluster and CMB anisotropy cosmological constraints.  We refer the reader
to a more complete discussion in Section~ \ref{sec:comparison}  and in
the recent Planck cosmological legacy paper \citep{planck18-cosmolegacy}. 

In comparison to the RM catalog, we find in the overlapping systems excellent agreement in optical centers and richnesses. 
However, there are several advantages to using MCMF to identify optical counterparts of X-ray (or SZE) selected 
sources in comparison to simply cross-matching X-ray (or SZE) and optical cluster catalogs.  
One advantage is that MCMF produces a simple estimator of the probability of random superposition,
which allows one to control the level of (and remove) contamination from 
the resulting cluster catalog.  In addition, the MCMF adoption of an X-ray prior on the sky location and 
the scale of the counterpart search region enables one to deal more easily with 
multiple peaks in redshift, allowing one to better disentangle cases of ambiguous counterparts 
and mismatches with literature redshifts (see SPT-SZ cluster example in Fig.~\ref{fig:SPTdoublepeak}).
 We show that MCMF can be used to de-blend clusters with redshifts of $\Delta 
z=0.15$, even if the cluster centers are separated by only 50\arcsec.  Finally, given the search priors from the X-ray (or SZE), it is 
possible to use MCMF to push the detection limits in the optical data, identifying optical counterparts at 
much higher redshift than the typical redshift limits of the optically selected cluster catalogs.

We explore whether there are missing SPT-SZ clusters in our MARD-Y3 sample or 
vice-versa, confirming the expectation that higher redshift and lower mass SPT-SZ systems are missing in our catalog, 
and that lower mass and low redshift MARD-Y3 systems are missing from SPT-SZ. There are also intriguing
indications for some missing SPT-SZ clusters in regions of mass-redshift space where one would naively
expect counterparts to exist (see Fig.~\ref{fig:SPTmatchesMassRedshift}). 
A detailed investigation that accounts for known selection effects such as scatter in
the X-ray and SZE observable-mass relations is underway and will be reported in a companion paper.
  
Our cluster catalog includes dynamical state estimators derived from the galaxy distribution for each cluster.
We use these estimators to identify interesting merging systems, recovering A514 as one of the most unrelaxed systems 
within our MARD-Y3 catalog.  This system has accompanying \XMM\ observations that provide independent evidence 
of an ongoing merger (see Fig.~\ref{fig:A514}).  We examine the distribution of the combined merger state 
estimator $\gamma$ as a function of redshift, finding no compelling evidence 
that the typical dynamical state in our sample is changing 
with redshift (see Fig.~\ref{fig:dynestvsz4}).  This finding is in agreement with a recent X-ray study of SZE 
selected clusters over a similar redshift range \citep{nurgaliev17}.  Interestingly, we do see evidence for 
more extreme mergers at lower redshifts.

Finally, we perform a first test of consistency between the observed X-ray luminosity function of the MARD-Y3 sample 
and that predicted by 
a standard cosmology with externally calibrated luminosity--mass and richness-mass relations \citep{bulbul19,saro15}.  We find 
reasonable agreement between the predicted and observed luminosity functions out to redshift $z\sim0.9$ 
(see Fig.~\ref{fig:costestLF}), providing a clear indication that the MARD-Y3 cluster catalog selection
can be described by in terms of a cut in X-ray detection significance (used to
produce the 2RXS candidate source catalog) followed by a cut in optical richness (the imposition of an \fcont\ threshold
used to reduce contamination by random superpositions).  A more detailed discussion of the selection function
will appear in a companion paper.

This paper describes the creation of the MCMF confirmed MARD-Y3 cluster catalog from the 2RXS X-ray source catalog.  It
presents the new cluster catalog along with a detailed description of the measurements 
made available for each cluster.  Thereafter, this analysis focuses on tests of the catalog that largely 
demonstrate the performance of the MCMF algorithm and reveal the characteristics of the new cluster catalog.
In forthcoming analyses we will study the X-ray luminosity-based mass-observable 
scaling relation using DES weak lensing and cross-calibration to other previously calibrated samples,  we will carry out a more extensive
cross-comparison with the SPT-SZ catalog, and we will
more precisely investigate the usefulness of the MARD-Y3 cluster catalog as a cosmological probe. 
These studies will further test the performance of MCMF in selecting a clean sample of clusters from what is initially a highly 
contaminated X-ray source catalog, but they will also test the underlying 2RXS catalog and its 
usability for cosmological studies.  We view these studies as useful precursors to the use of MCMF to characterize
future X-ray and SZE selected cluster catalogs from eROSITA \citep{predehl10} and SPT-3G \citep{benson14} using 
the new multiband datasets from Euclid \citep{laureijs11} and LSST \citep{ivezic08}.


\section*{Acknowledgements}

We thank Th. Boller, M. Freyberg and H. Brunner from the MPE high energy group for helpful conversations.  We 
acknowledge the support of the Max Planck Gemeinschaft Faculty Fellowship program and the High Energy Group 
at MPE.  Furthermore, we acknowledge the support of the DFG Cluster of Excellence ``Origin and Structure of the 
Universe'', the Transregio program TR33 ``The Dark Universe'', and the Ludwig-Maximilians-Universit\"at.
 
Funding for the DES Projects has been provided by the U.S. Department of Energy, the U.S. National Science 
Foundation, the Ministry of Science and Education of Spain, the Science and Technology Facilities Council of the 
United Kingdom, the Higher Education Funding Council for England, the National Center for Supercomputing  
Applications at the University of Illinois at Urbana-Champaign, the Kavli Institute of Cosmological Physics at the 
University of Chicago, the Center for Cosmology and Astro-Particle Physics at the Ohio State University, the 
Mitchell Institute for Fundamental Physics and Astronomy at Texas A\&M University, Financiadora de Estudos e 
Projetos, Funda{\c c}{\~a}o Carlos Chagas Filho de Amparo {\`a} Pesquisa do Estado do Rio de Janeiro, Conselho 
Nacional de Desenvolvimento Cient{\'i}fico e Tecnol{\'o}gico and the Minist{\'e}rio da Ci{\^e}ncia, Tecnologia e 
Inova{\c c}{\~a}o, the Deutsche Forschungsgemeinschaft and the Collaborating Institutions in the Dark Energy 
Survey.   The Collaborating Institutions are Argonne National Laboratory, the University of California at Santa Cruz, 
the University of Cambridge, Centro de Investigaciones Energ{\'e}ticas, Medioambientales y Tecnol{\'o}gicas-
Madrid, the University of Chicago, University College London, the DES-Brazil Consortium, the University of 
Edinburgh, the Eidgen{\"o}ssische Technische Hochschule (ETH) Z{\"u}rich, Fermi National Accelerator Laboratory, 
the University of Illinois at Urbana-Champaign, the Institut de Ci{\`e}ncies de l'Espai (IEEC/CSIC), the Institut de 
F{\'i}sica d'Altes Energies, Lawrence Berkeley National Laboratory, the Ludwig-Maximilians Universit{\"a}t M{\"u}
nchen and the associated Excellence Cluster Universe, the University of Michigan, the National Optical Astronomy 
Observatory, the University of Nottingham, The Ohio State University, the University of Pennsylvania, the 
University of Portsmouth, SLAC National Accelerator Laboratory, Stanford University, the University of Sussex, 
Texas A\&M University, and the OzDES Membership Consortium.  The DES data management system is 
supported by the National Science Foundation under Grant Number AST-1138766. The DES participants from 
Spanish institutions are partially supported by MINECO under grants AYA2012-39559, ESP2013-48274, 
FPA2013-47986, and Centro de Excelencia Severo Ochoa SEV-2012-0234.  Research leading to these results has 
received funding from the European Research Council under the European Union's Seventh Framework 
Programme (FP7/2007-2013) including ERC grant agreements  240672, 291329, and 306478.

\section*{Affiliations}
$^{1}$ Faculty of Physics, Ludwig-Maximilians-Universit\"at, Scheinerstr. 1, 81679 Munich, Germany\\
$^{2}$ Max Planck Institute for Extraterrestrial Physics, Giessenbachstrasse, 85748 Garching, Germany\\
$^{3}$ Excellence Cluster Universe, Boltzmannstr.\ 2, 85748 Garching, Germany\\
$^{4}$ Cerro Tololo Inter-American Observatory, National Optical Astronomy Observatory, Casilla 603, La Serena, Chile\\
$^{5}$ Fermi National Accelerator Laboratory, P. O. Box 500, Batavia, IL 60510, USA\\
$^{6}$ Institute of Cosmology and Gravitation, University of Portsmouth, Portsmouth, PO1 3FX, UK\\
$^{7}$ CNRS, UMR 7095, Institut d'Astrophysique de Paris, F-75014, Paris, France\\
$^{8}$ Sorbonne Universit\'es, UPMC Univ Paris 06, UMR 7095, Institut d'Astrophysique de Paris, F-75014, Paris, France\\
$^{9}$ Department of Physics \& Astronomy, University College London, Gower Street, London, WC1E 6BT, UK\\
$^{10}$ Centro de Investigaciones Energ\'eticas, Medioambientales y Tecnol\'ogicas (CIEMAT), Madrid, Spain\\
$^{11}$ Laborat\'orio Interinstitucional de e-Astronomia - LIneA, Rua Gal. Jos\'e Cristino 77, Rio de Janeiro, RJ - 20921-400, Brazil\\
$^{12}$ Department of Astronomy, University of Illinois at Urbana-Champaign, 1002 W. Green Street, Urbana, IL 61801, USA\\
$^{13}$ National Center for Supercomputing Applications, 1205 West Clark St., Urbana, IL 61801, USA\\
$^{14}$ Institut de F\'{\i}sica d'Altes Energies (IFAE), The Barcelona Institute of Science and Technology, Campus UAB, 08193 Bellaterra (Barcelona) Spain\\
$^{15}$ Institut d'Estudis Espacials de Catalunya (IEEC), 08034 Barcelona, Spain\\
$^{16}$ Institute of Space Sciences (ICE, CSIC),  Campus UAB, Carrer de Can Magrans, s/n,  08193 Barcelona, Spain\\
$^{17}$ Kavli Institute for Particle Astrophysics \& Cosmology, P. O. Box 2450, Stanford University, Stanford, CA 94305, USA\\
$^{18}$ Department of Physics and Astronomy, University of Pennsylvania, Philadelphia, PA 19104, USA\\
$^{19}$ Observat\'orio Nacional, Rua Gal. Jos\'e Cristino 77, Rio de Janeiro, RJ - 20921-400, Brazil\\
$^{20}$ Department of Physics, IIT Hyderabad, Kandi, Telangana 502285, India\\
$^{21}$ Department of Astronomy, University of Michigan, Ann Arbor, MI 48109, USA\\
$^{22}$ Department of Physics, University of Michigan, Ann Arbor, MI 48109, USA\\
$^{23}$ Kavli Institute for Cosmological Physics, University of Chicago, Chicago, IL 60637, USA\\
$^{24}$ Instituto de Fisica Teorica UAM/CSIC, Universidad Autonoma de Madrid, 28049 Madrid, Spain\\
$^{25}$ Department of Physics and Astronomy, Pevensey Building, University of Sussex, Brighton, BN1 9QH, UK\\
$^{26}$ Department of Physics, Stanford University, 382 Via Pueblo Mall, Stanford, CA 94305, USA\\
$^{27}$ SLAC National Accelerator Laboratory, Menlo Park, CA 94025, USA\\
$^{28}$ Department of Physics, ETH Zurich, Wolfgang-Pauli-Strasse 16, CH-8093 Zurich, Switzerland\\
$^{29}$ Santa Cruz Institute for Particle Physics, Santa Cruz, CA 95064, USA\\
$^{30}$ Center for Cosmology and Astro-Particle Physics, The Ohio State University, Columbus, OH 43210, USA\\
$^{31}$ Department of Physics, The Ohio State University, Columbus, OH 43210, USA\\
$^{32}$ Universit\"ats-Sternwarte, Fakult\"at f\"ur Physik, Ludwig-Maximilians Universit\"at M\"unchen, Scheinerstr. 1, 81679 M\"unchen, Germany\\
$^{33}$ Harvard-Smithsonian Center for Astrophysics, Cambridge, MA 02138, USA\\
$^{34}$ Australian Astronomical Optics, Macquarie University, North Ryde, NSW 2113, Australia\\
$^{35}$ Departamento de F\'isica Matem\'atica, Instituto de F\'isica, Universidade de S\~ao Paulo, CP 66318, S\~ao Paulo, SP, 05314-970, Brazil\\
$^{36}$ George P. and Cynthia Woods Mitchell Institute for Fundamental Physics and Astronomy, and Department of Physics and Astronomy, Texas A\&M University, College Station, TX 77843,  USA\\
$^{37}$ Instituci\'o Catalana de Recerca i Estudis Avan\c{c}ats, E-08010 Barcelona, Spain\\
$^{38}$ Jet Propulsion Laboratory, California Institute of Technology, 4800 Oak Grove Dr., Pasadena, CA 91109, USA\\
$^{39}$ School of Physics and Astronomy, University of Southampton,  Southampton, SO17 1BJ, UK\\
$^{40}$ Brandeis University, Physics Department, 415 South Street, Waltham MA 02453\\
$^{41}$ Instituto de F\'isica Gleb Wataghin, Universidade Estadual de Campinas, 13083-859, Campinas, SP, Brazil\\
$^{42}$ Computer Science and Mathematics Division, Oak Ridge National Laboratory, Oak Ridge, TN 37831\\
$^{43}$ Argonne National Laboratory, 9700 South Cass Avenue, Lemont, IL 60439, USA\\


\bibliographystyle{mnras}
\bibliography{basic}


\appendix


\section{Cluster catalog comparisons}
\label{app:comparison}
In Section~\ref{app:MCXCcomp} we compare our cluster sample to the MCXC catalog \citep{piffaretti11}, and in the following Sections~\ref{app:RMcomp}, \ref{app:SPTcomp} and \ref{app:Planckcomp} we compare our cluster sample to the 
the redMaPPer \citep{rykoff14}, SPT-SZ \citep{bleem15} and PLANCK PSZ2 \citep{planck25-27} catalogs, respectively. 

\begin{figure}
\includegraphics[keepaspectratio=true,width=\columnwidth]{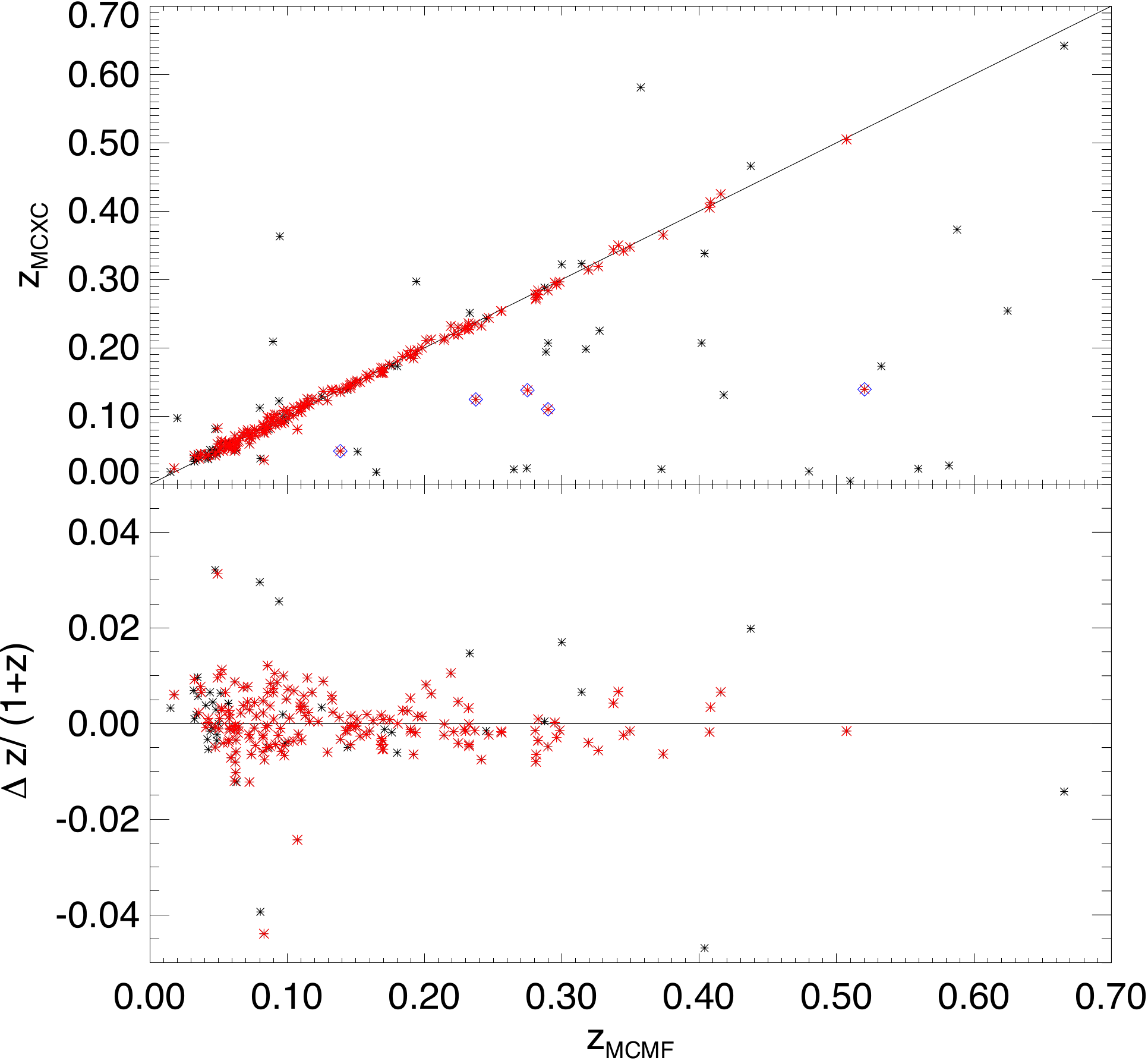}
\vskip-0.10in
\caption{Photo-z comparison between MCXC and MCMF: Red points show matches with \fcont$<0.1$, and
sources with higher \fcont\ are shown in black. Outliers with $>5\%$ redshift offset and 
\fcont$<0.1$ are marked with blue diamonds. We find a scatter of $\sigma_{\Delta z/ (1+z)}=0.0053$.}
\label{fig:comMCMFZ}
\end{figure}

\begin{figure*}
\includegraphics[keepaspectratio=true,width=0.65\columnwidth]{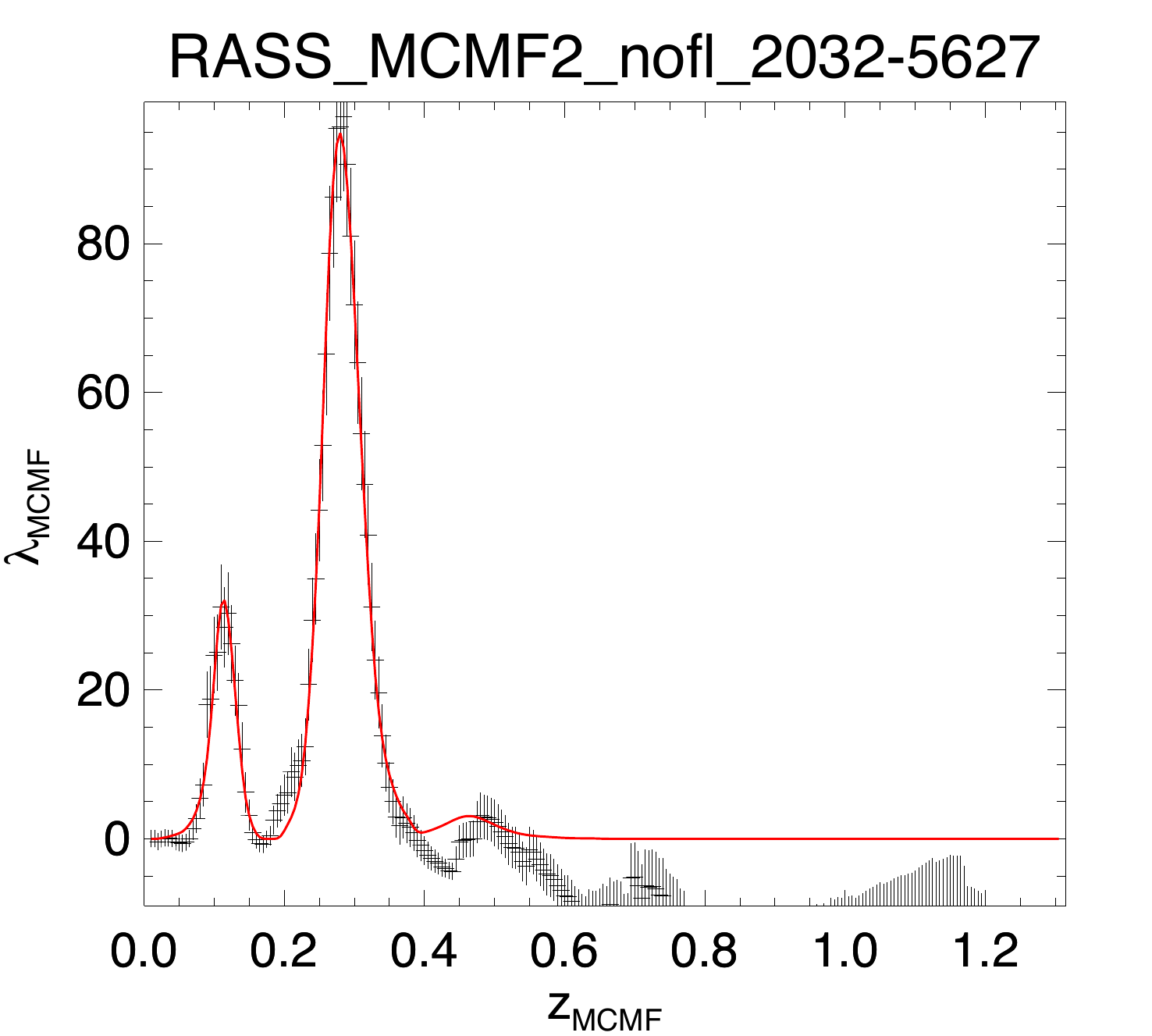}
\includegraphics[keepaspectratio=true,width=0.67\columnwidth]{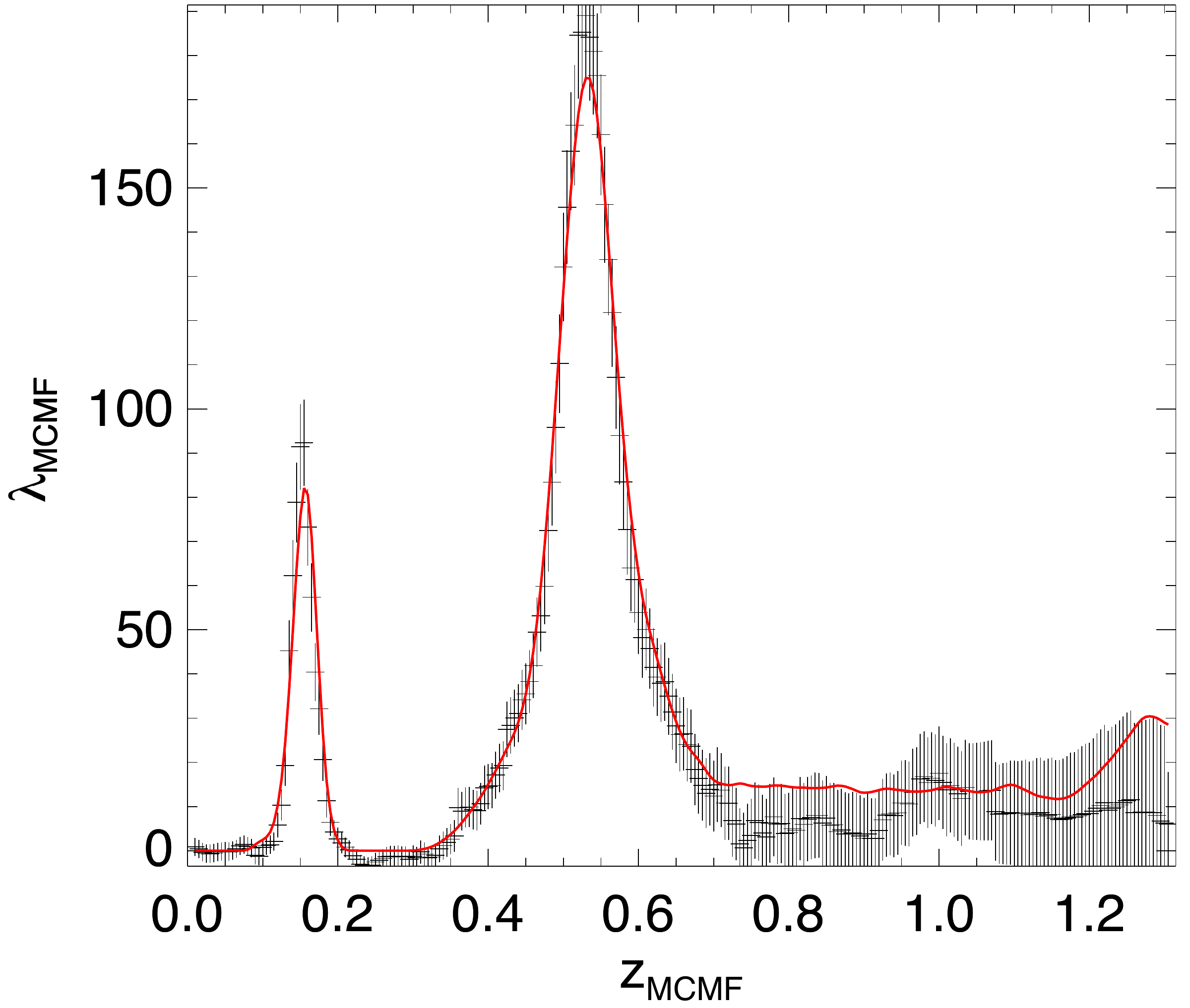}
\includegraphics[keepaspectratio=true,width=0.65\columnwidth]{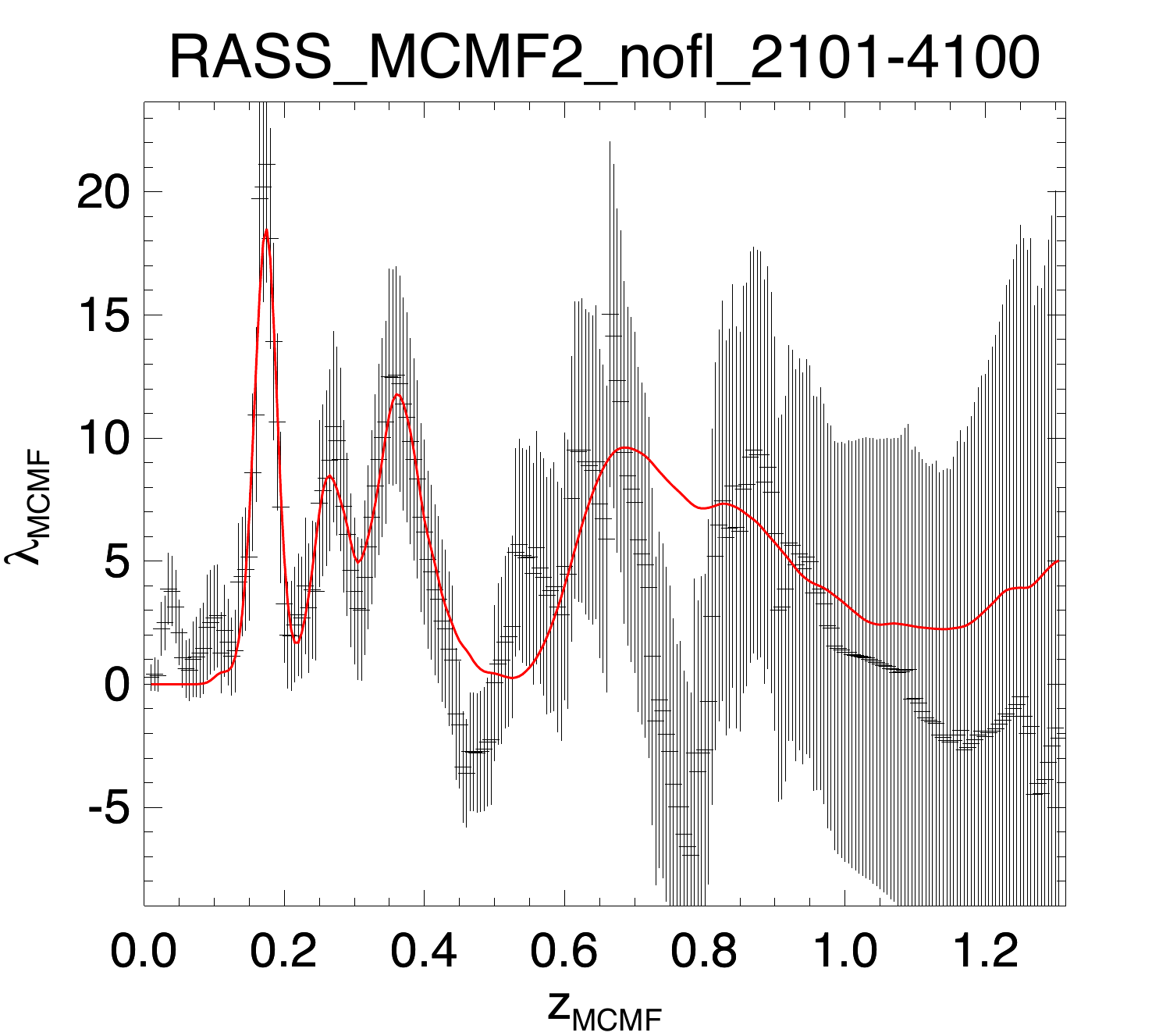}
\includegraphics[keepaspectratio=true,width=0.65\columnwidth]{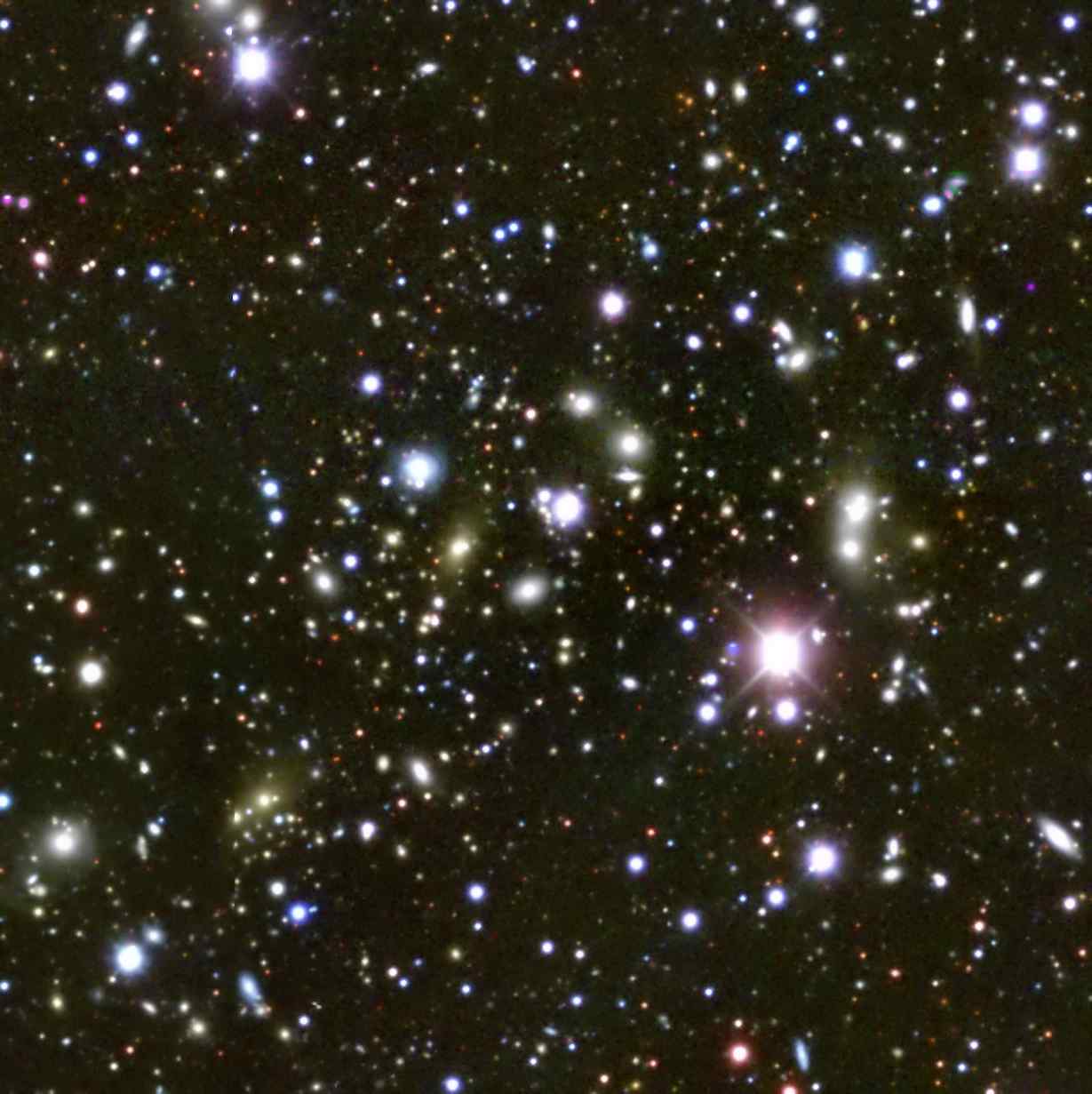}
\vspace{2mm}
\includegraphics[keepaspectratio=true,width=0.685\columnwidth]{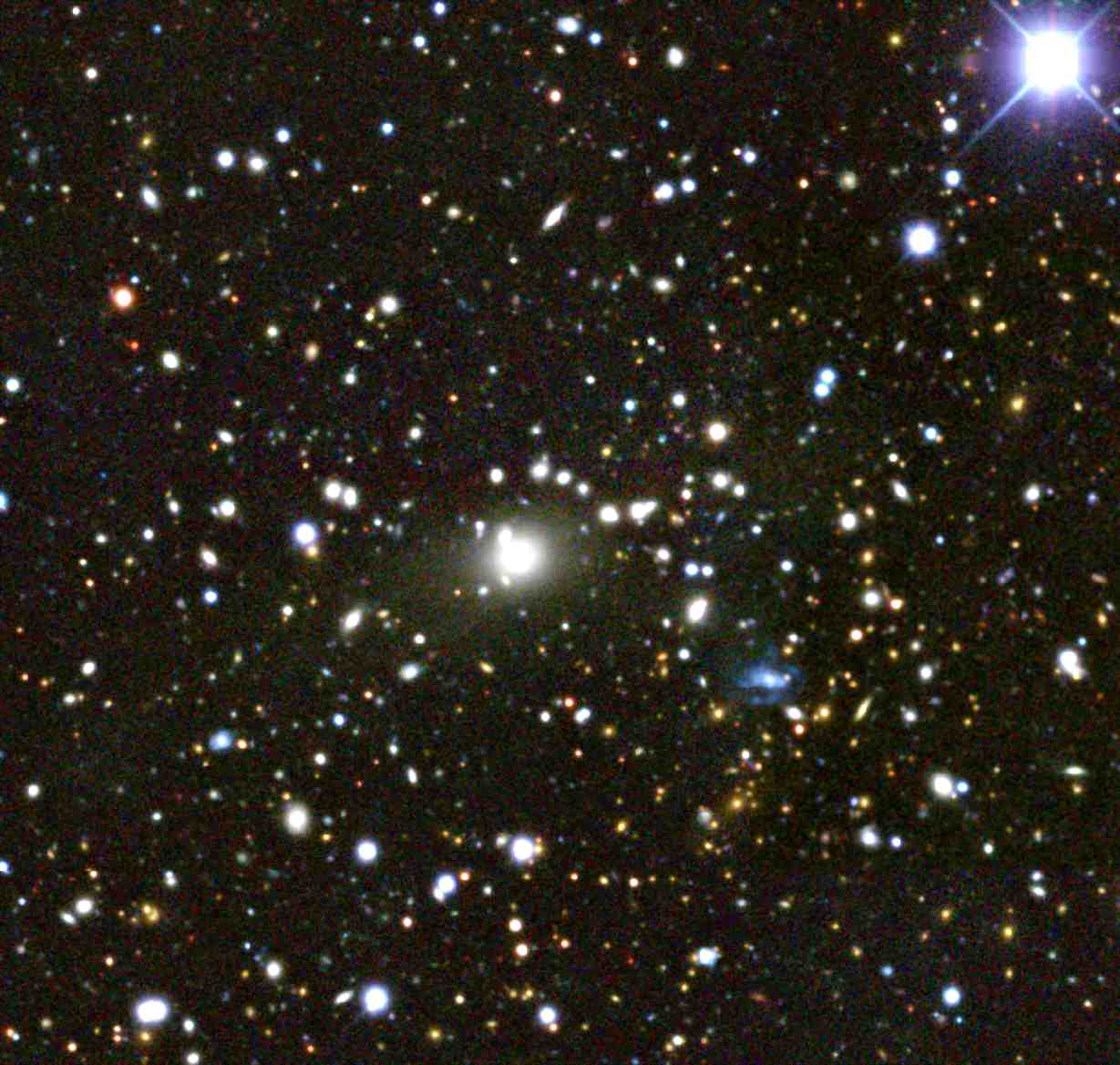}
\vspace{2mm}
\includegraphics[keepaspectratio=true,width=0.655\columnwidth]{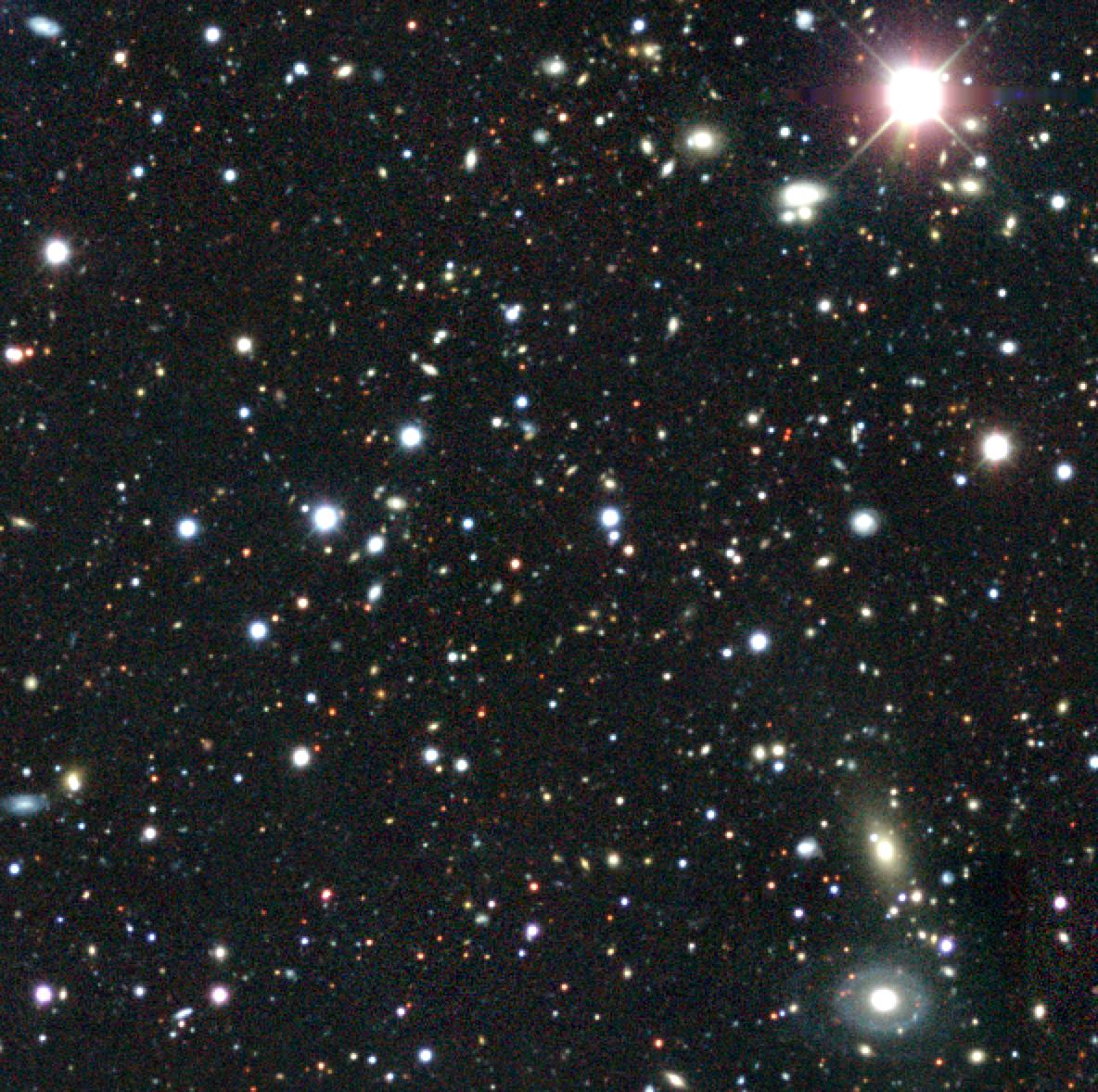}
\vskip-0.15in
\caption{MCXC clusters showing discrepant redshifts or low richness in MCMF: RXC J2032.1-5627 (left), 
RXC J0605.8-3518 (middle) and RXC J2101.4-4100 (right). The richness distributions in redshift are shown in the top 
row, the $grz$ pseudo-color image of the central regions are shown in the bottom row.  The first two show multiple significant $\lambda$ peaks in redshift, and the third shows multiple low significance peaks.}
\label{fig:comMCMFZmultipeak}
\end{figure*}

\begin{figure}
\includegraphics[keepaspectratio=true,width=\columnwidth]{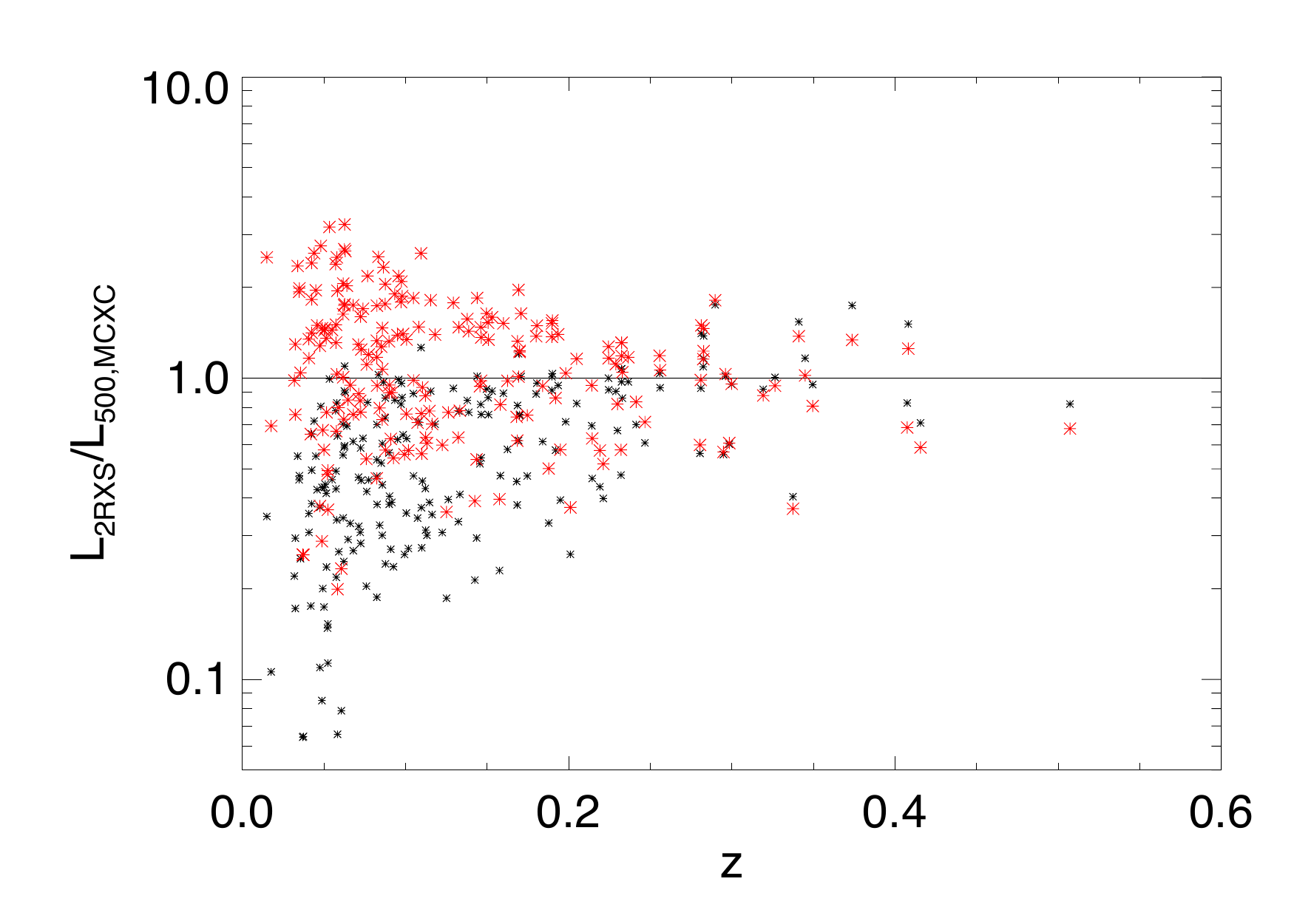}
\includegraphics[keepaspectratio=true,width=\columnwidth]{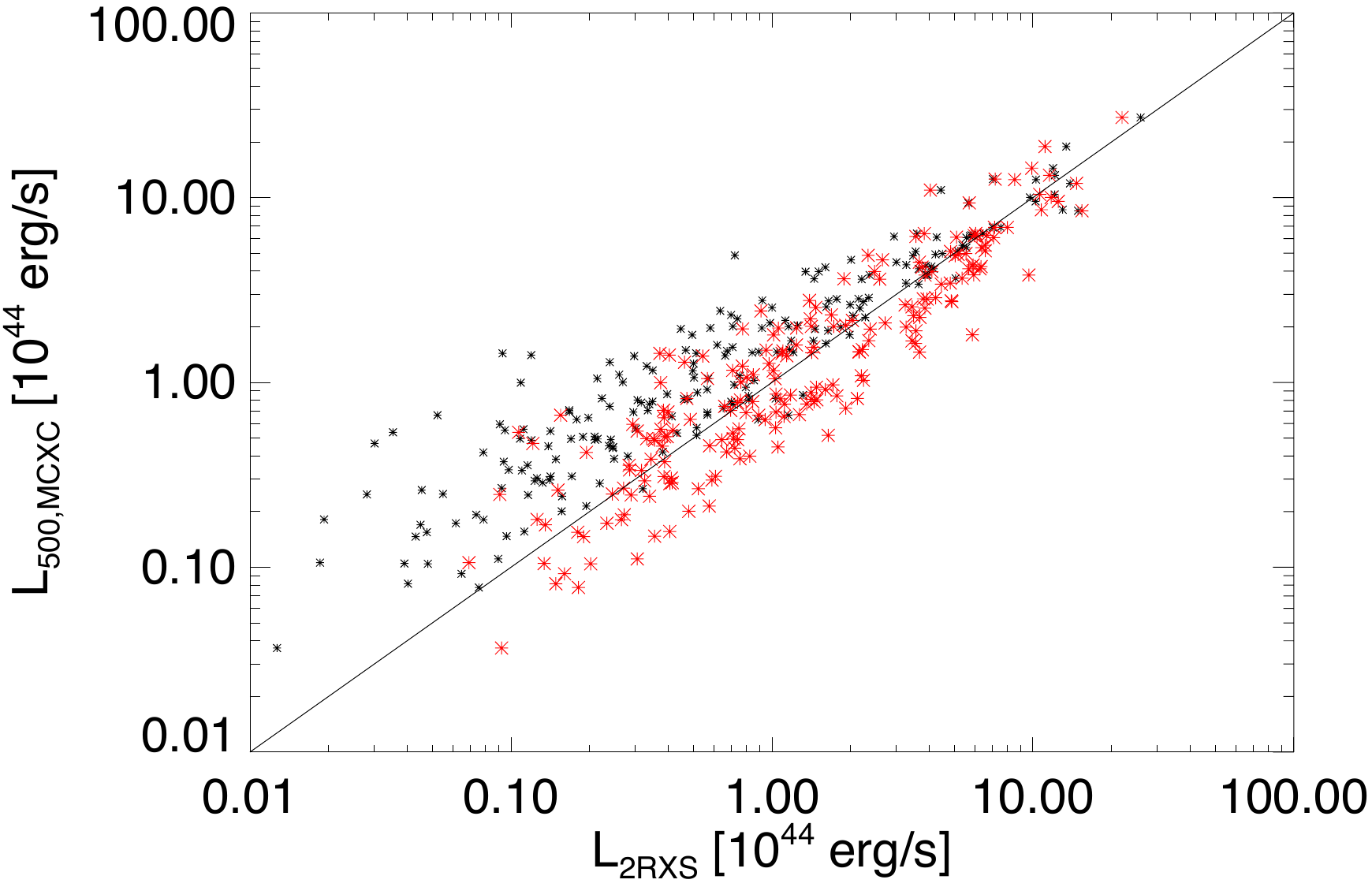}
\vskip-0.10in
\caption{Comparison of MCXC luminosities with corrected (red) and uncorrected (black)  2RXS based luminosities. Corrected luminosities and luminosity based masses are presented in the MARD-Y3 table }
\label{fig:compMCXClum}
\end{figure}

\begin{figure}
\includegraphics[keepaspectratio=true,width=\columnwidth]{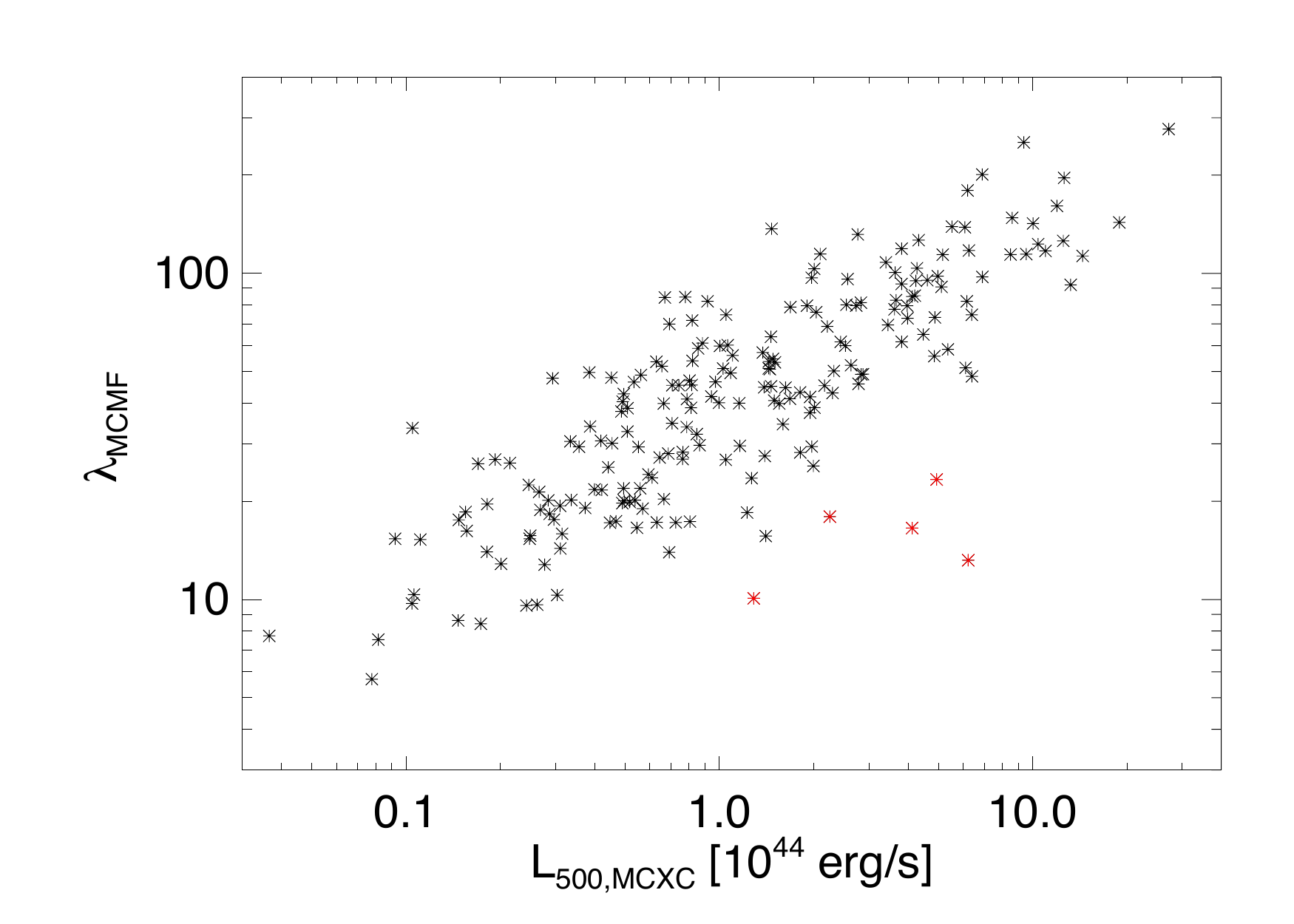}
\vskip-0.10in
\caption{Comparison of MCXC luminosities with richness \LamMCMF. Outliers discussed in the text are marked in red.}
\label{fig:compMCXClamlum}
\end{figure}

\begin{figure*}
\includegraphics[keepaspectratio=true,width=0.8\columnwidth]{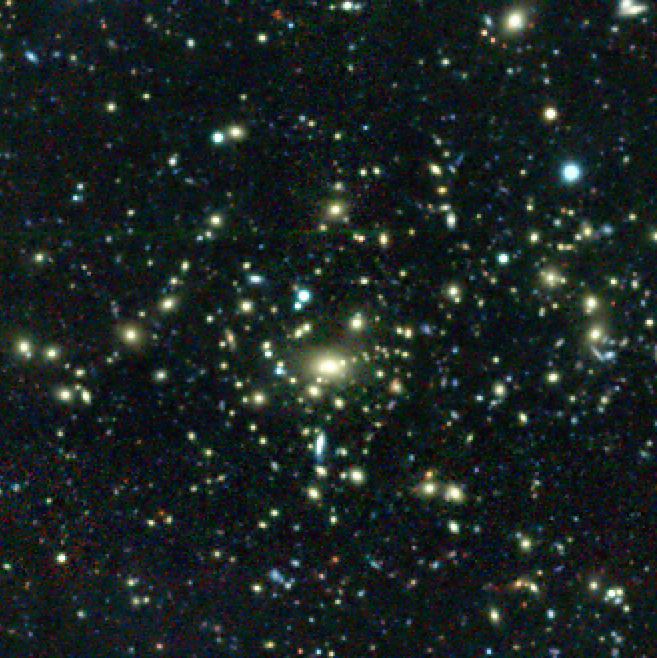}
\includegraphics[keepaspectratio=true,width=1.2\columnwidth]{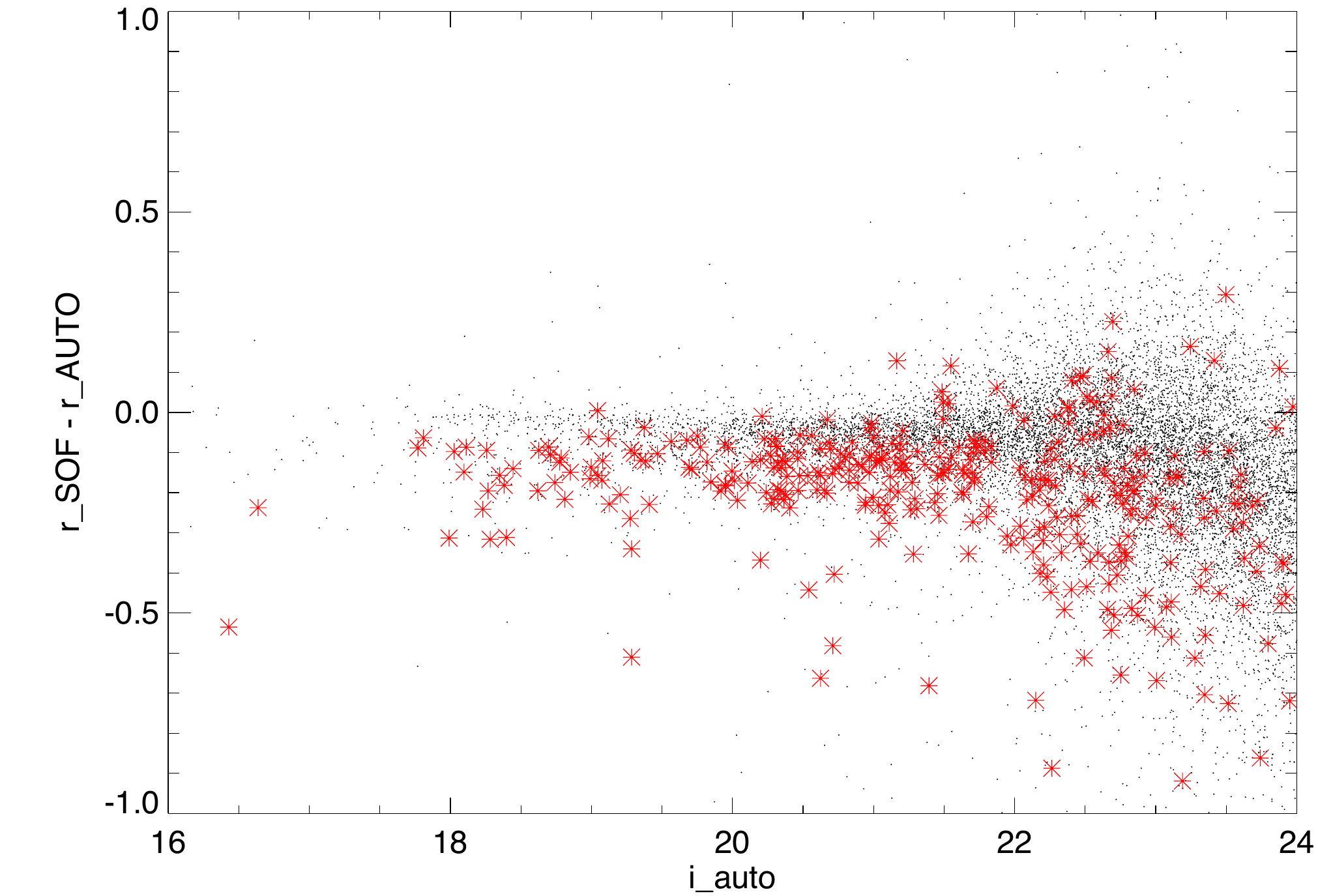}
\caption{MACSJ0257.6-2209: Left: $grz$ pseudo-color image of the central 5x5 arcmin region. Right: Direct 
comparison of $r$-band magnitudes measured based on AUTO and SOF photomety. Galaxies within 120\arcsec\ from 
the cluster center are highlighted in red. Galaxies in the cluster region have systematically different AUTO and 
SOF photometry, a rare calibration error that has been communicated to the DESDM team.}
\label{fig:comMISPLK}
\end{figure*}

\subsection{MCXC clusters}
\label{app:MCXCcomp}
The MCXC \citep{piffaretti11} catalog is a meta-catalog of X-ray detected clusters of galaxies. It combines various 
ROSAT based publicly available catalogs, such as NORAS \citep{2000ApJS..129..435B} the southern pendant 
REFLEX\citep{boehringer04} and MACS \citep{ebeling01}. 

Depending on the characteristic we study, we use different matching criteria to limit the impact of the matching process 
on the result.

We first study the consistency of our photo-z's with the redshifts given in MCXC. For that we use a 
cross matching radius of 150\arcsec. To compare redshifts we use the MCMF redshift that 
corresponds the peak that shows the lowest \fcont. 
Fig~\ref{fig:comMCMFZ} shows the redshift comparison for all sources within 150\arcsec. Highlighted in red 
are sources with  \fcont$<0.1$.
We find a scatter between MCMF and MCXC redshifts of $\sigma_{\Delta z/ (1+z)}=0.0053$ for the  
\fcont$<0.1$ sample.
The MCXC catalog is a mix of mainly spec-z's and some photo-z's. The measured 
scatter can therefore be expected to be slightly higher than that of  a pure spec-z sample.  We do not see 
any significant systematic offsets or biases.

We identify five catastrophic outliers with photo-z offsets of more that 5\%. Two, RXC J0605.8-3518 and 
RXC J2032.1-5627, show a second structure in the line of sight with \fcont$=0.0$ and \fcont$=0.015$. 
The $grz$ psuedo-color images and the \LamMCMF(z) plots of those clusters are shown in 
Fig.~\ref{fig:comMCMFZmultipeak}. The redshifts of those structures are consistent with those listed in 
MCXC. Given the low \fcont\ for the second peak, we can assume that both clusters in each of the two cases significantly 
contribute to the X-ray flux, which means that both redshifts are correctly assigned. 
In three cases, RXC J2103.4-4319 (ABELL 3736), RXC J2135.2+0125 (ABELL 2355) and RXC J2341.1+0018, 
MCMF identifies only one significant peak, suggesting that either the redshift from MCMF or MCXC is wrong. In 
the case of RXC J2135.2+0125 and RXC J2341.1+0018 we find multiple galaxies with spec-z's 
consistent with the MCMF redshift, supporting the MCMF redshift. For RXCJ2103.4-4319, we do not find spec-z's near the cluster 
position. But within 1100\arcsec\ we find another ABELL cluster, ABELL S0919, which does have a spec-z 
that is identical with that of  ABELL 3736. MCMF obtains for that cluster a redshift of $z=0.054$,  consistent with 
the spec-z of z=0.0487. Given the fact the MCMF is capable of ingrecover the redshift of the cluster nearby and that 
the redshift of both ABELL clusters are identical, we conclude that the MCMF redshift is likely correct and that the 
published redshift of ABELL 3736 is likely taken from the same galaxies as for ABELL S0919.

Applying a 150\arcsec\ matching radius with an additional cut on the maximum of the redshift discrepancy of $\Delta 
z<0.03$ allows us to compare the luminosities given in MCXC and our luminosities $L_\mathrm{X}$.
The count rates given in 2RXS, which are used to compute luminosities are derived within a 
fixed aperture of 300\arcsec\ radius. This results in an 
underestimate of the flux (and luminosity) if the cluster extends beyond this radius. 
The measurement of the extent of a source in 2RXS is very noisy, and a flux correction based on that 
doesn't work well. 
Because the main driver of the apparent extent of a cluster is its distance, we use the MCXC clusters to 
estimate a flux aperture correction based 
on the derived redshift. For that we simply measured the median ratio between both luminosities in redshift 
bins and interpolate between bins.
 The plots in Fig.~\ref{fig:compMCXClum} illustrate the difference between luminosities corrected 
 $L_\mathrm{500,2RXS}$ to $L_\mathrm{500,MCXC}$  in comparison to the uncorrected luminosities $L_\mathrm{X}$.

Fig.~\ref{fig:compMCXClamlum} shows the $\lambda-L_{\mathrm{X,MCXC}}$ distribution of redshift matched 
clusters. The most extreme outliers with \fcont$>0.15$ and $M_{\mathrm{500,MCXC}}>2\times10^{14} 
M_{\odot}$ are highlighted in red.
We investigate all five outliers. All these systems show a richness a factor three below the main distribution of 
richnesses given mass.
Optical inspection shows that two systems (RXC J2101.4-4100, RXC J0117.8-5455) seem to be indeed poor 
systems, consistent with the richness given by MCMF.  \Chandra\ data of RXC J0117.8-5455 show a bright point 
source at the cluster candidate position, indicating that the main flux is coming from that point source.
RXCJ2101.4-4100, shown in Fig.~\ref{fig:comMCMFZmultipeak}, shows multiple $6<\lambda<20$ peaks. 
The main peak is at z=0.17 with an rBCG candidate 
about 210\arcsec\ south-west of the X-ray position. The third cluster,
RXC J0157.4-0550, is the outlier with the lowest luminosity. We find two 2RXS sources within 150\arcsec\ of 
the MCXC position, with 111\arcsec\ and 145\arcsec\ distance. The nearer match shows \fcont$=0.33$ the 
second one \fcont$=0.046$. Optical inspection shows that the MCXC position is indeed about 400\arcsec\ off 
from the cluster center, while the 2RXS source closer to the cluster center corresponds to the one with low 
\fcont, which means this cluster is confirmed.

In the fourth system, RXC J0336.3-4037, we find that the central region of the cluster, including the rBCG is not  covered 
by DES data. The confirmation of this cluster is therefore affected by the lack of data, but there are sufficient number of 
cluster members to recover the correct redshift.

The most massive cluster not confirmed with MCMF is MACSJ0257.6-2209 with a mass $M_{500,MCXC}=5.87 \times 
10^{14}$. This cluster is also detected by Planck with a mass of  $M_{500,Planck}=6.05 \times 10^{14}$, strongly 
suggesting that this cluster is real and massive. Inspection of the optical images indeed reveals a rich cluster 
suggesting that the richness of this system is heavily underestimated.
A detailed study of the photometry of this system shows that SOF based magnitudes of the $r$-band are 
systematically off by 0.15 mag at the location of the cluster. 
The systematic offset causes the RS models be a bad fit in $g-r$ and $r-i$. Simpler RS methods using 
only one color at a time might be less effected but would provide a biased redshift.
Fig.~\ref{fig:comMISPLK} shows the photometric properties and the $g$, $r$, $z$ psuedo-color image of 
MACS J0257.6-2209.
Investigations have shown that the reason for the failing of the SOF measurements 
might be caused  by a bad PSF model at the position of the 
cluster.  At the location of the cluster only one imaging layer is available which is largely filled by the cluster 
members. A leakage of cluster members into the sample of stars used for the PSF 
modelling might be the source of the bad PSF model.
 Further inspection of this error within DES  indicates that the number PSF failures with similarly 
 strong impact are generally small,
 affecting less than 0.25\% of the data. A connection to the presence of 
 a cluster could not be shown in other cases with PSF failures.
 We check further each outlier discussed in this section for this effect, finding 
 no evidence that this effect has caused similar 
problems in other cluster fields.

As shown in Fig.~\ref{fig:compMCXClum} the fixed aperture used for the flux measurement causes a systematic under estimation of the flux at low redshifts. Further it is expected that this flux estimate causes additional scatter between the corrected 2RXS luminosities and true  $L_{X,500}$.  A first estimate of the amount of this additional scatter can be derived from the scatter between corrected 2RXS based luminosity and those from MCXC. We split the matched sampe in three, equally populated redshift bins of $z<0.07$, $0.07<z<0.15$ and $z>0.15$. We find a total scatter of 77, 50 and $40\%$ for the different redshift ranges. MCXC as well 2RXS are greatly based on the same ROSAT data, the measured total scatter should therefore roughly show the increased scatter due to the 2RXS flux measurement. 

As a second way to estimate the scatter is to measure the scatter of the luminosity-richness relation for the same set of clusters. We find a total scatter of $66.7$, $70.0$ and $67.2\%$ between for MCXC, while $86.7$, $84.0$ and $79.0 \%$ for 2RXS based luminosities. Ignoring the correlation introduced by the definition of our aperture for $\lambda$, the increased scatter seen in luminosity-richness for the 2RXS case, can be described by an additional scatter in luminosity of $55$, $46.4$ and $41.5\%$. These number are close to those of the direct comparison of luminosities and indicates that the fixed aperture causes additional scatter of $40-77\%$ depending on redshift.
We expect the scatter to further decrease with redshift as clusters become less and less resolved. From Fig.\ref{fig:offsetsvsz2} we expect that clusters become unresolved at a redshift above $z=0.3$.

\begin{figure}
\includegraphics[keepaspectratio=true,width=\columnwidth]{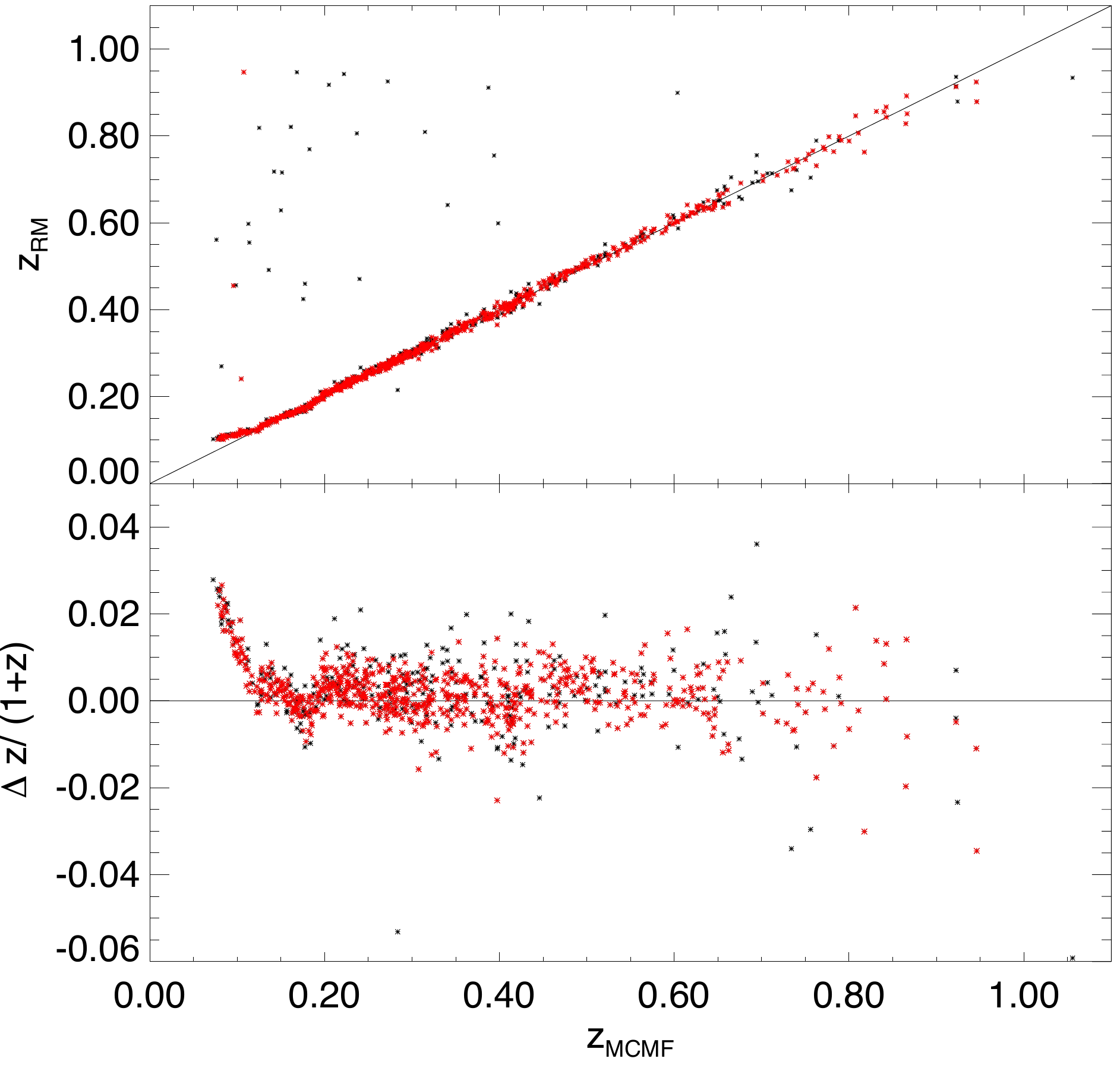}
\includegraphics[keepaspectratio=true,width=\columnwidth]{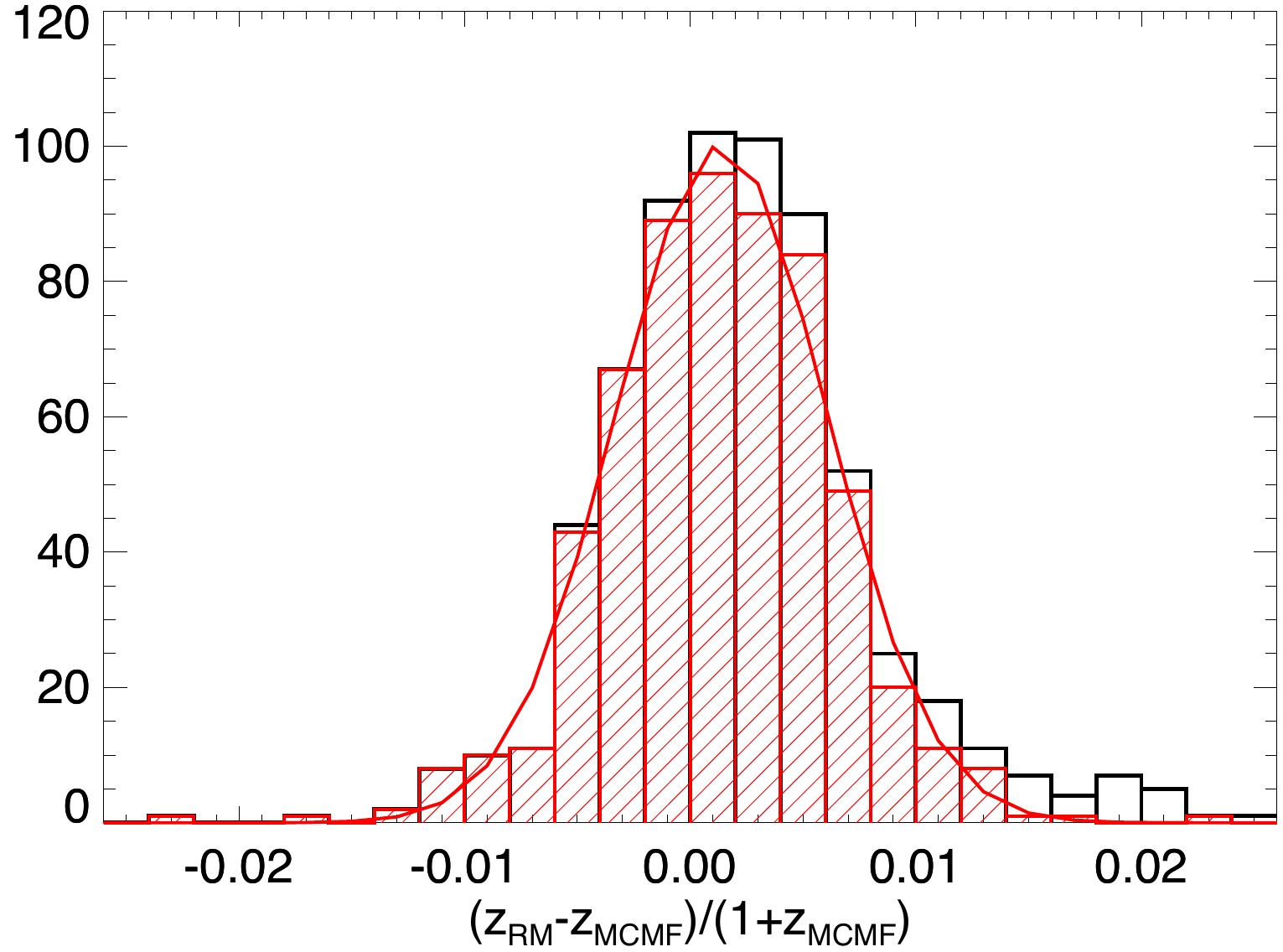}
\vskip-0.1in
\caption{Photo-z comparison between RM and MCMF: Photo-z versus photo-z and the residuals are shown in the 
top panel. Sources with richness $\lambda_\mathrm{RM}>20$ are shown in red.  Note the systematic error in RM redshifts at $z_\mathrm{MCMF}<0.15$.  The bottom panel shows the 
histogram of the residuals between MCMF and RM redshifts. The red histogram uses a $z_\mathrm{MCMF}>0.15$ 
cut. The fitted Gaussian function shows a standard deviation of $\sigma=0.0047$.}
\label{fig:compRedmap}
\end{figure}

\begin{figure}
\includegraphics[keepaspectratio=true,width=\columnwidth]{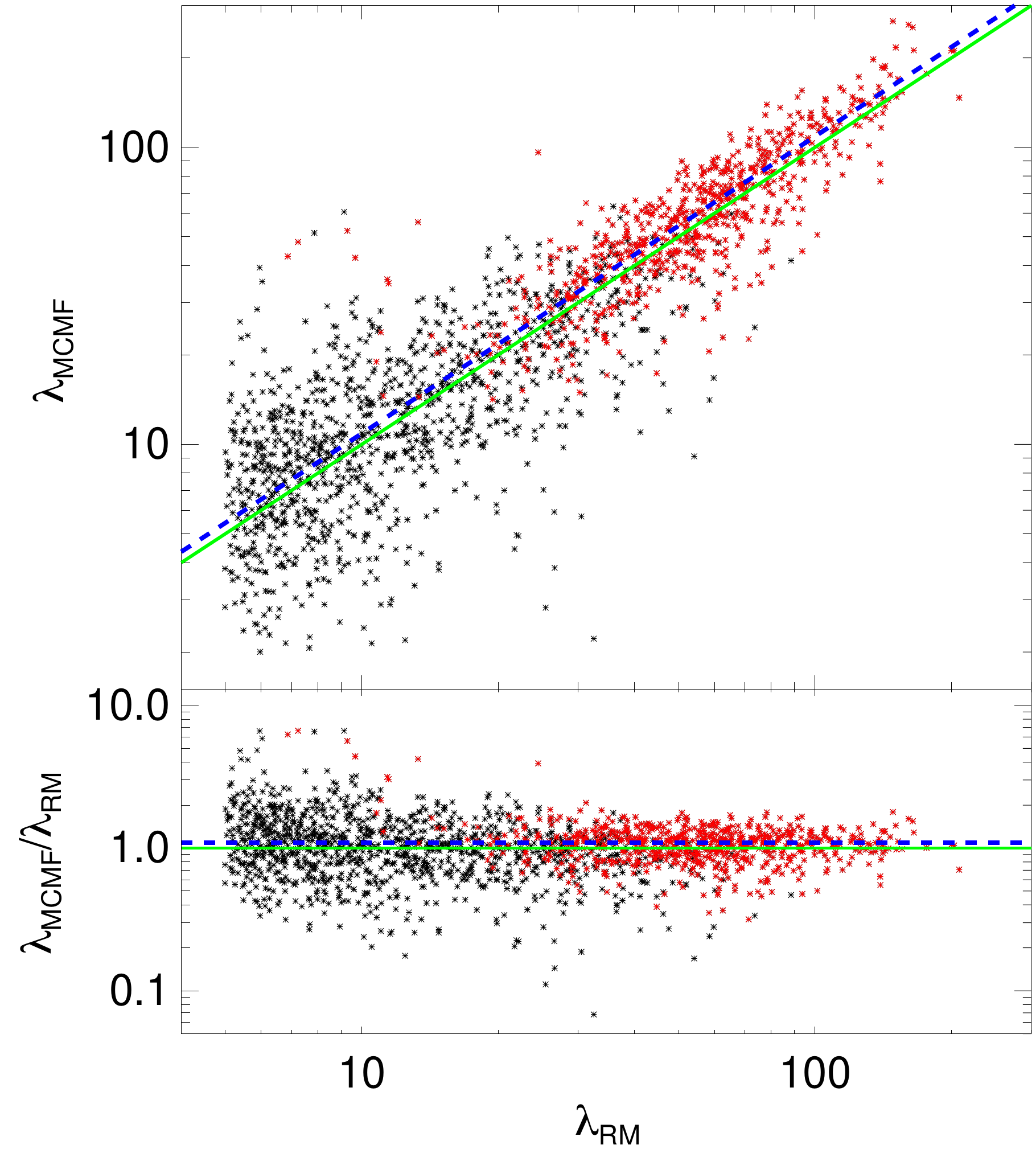}
\vskip-0.1in
\caption{Richness comparison between RM and MCMF. Top panel: The MCMF based richness ($
\lambda_\mathrm{MCMF}$) versus RM richness  ($\lambda_\mathrm{RM}$) for radial and redshift matched 
systems . Bottom panel: Richness ratio $\lambda_\mathrm{MCMF}/\lambda_\mathrm{RM}$  versus redmapper 
richness. Sources with \fcont$<0.1$ are highlighted in red. The one to one relation is shown as green 
continuous line, the
median ratio of 1.088 is shown as blue dashed line.}
\label{fig:compRMlambda}
\end{figure}

\begin{figure}
\includegraphics[keepaspectratio=true,width=\columnwidth]
{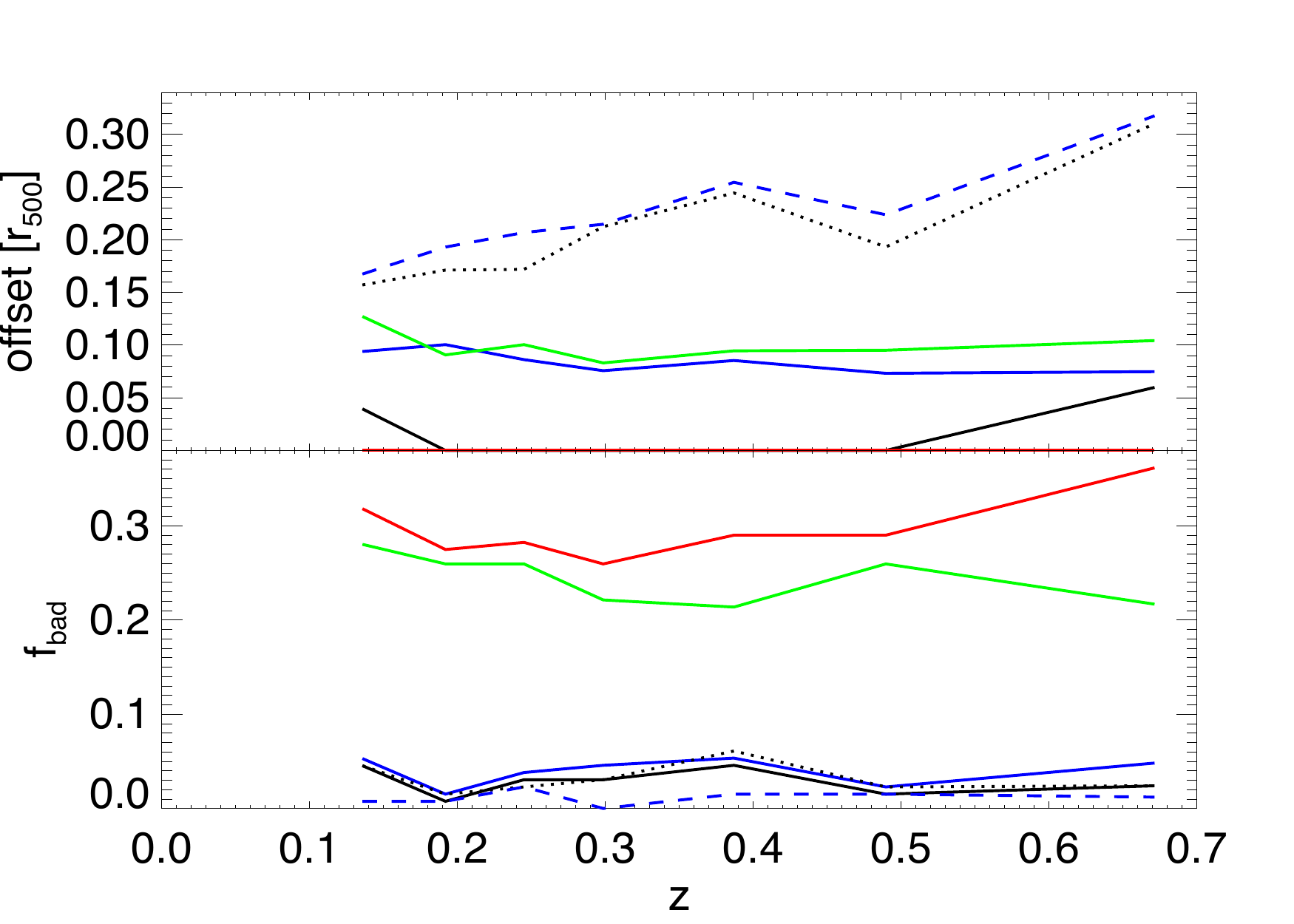}
\vskip-0.10in
\caption{Above is plotted the redshift dependence of the median offset between RM and MCMF centers for different 
center estimators. The offset using the galaxy density peak is shown in blue, the rBCG estimate in red and the 2D 
fit position in green continuous  lines. The default center is shown as a black line. Black dotted line shows the X-ray 
to RM offset, the blue dashed line the default MCMF center to X-ray center offset of the same clusters. Bottom: 
Below we show the fraction of clusters with large ($>r_{500}$) offsets or unsuccessfully measured centers.}
\label{fig:offsetsvsz3}
\end{figure}

 \begin{figure}
 \centering
\includegraphics[keepaspectratio=true,width=0.9\columnwidth]
{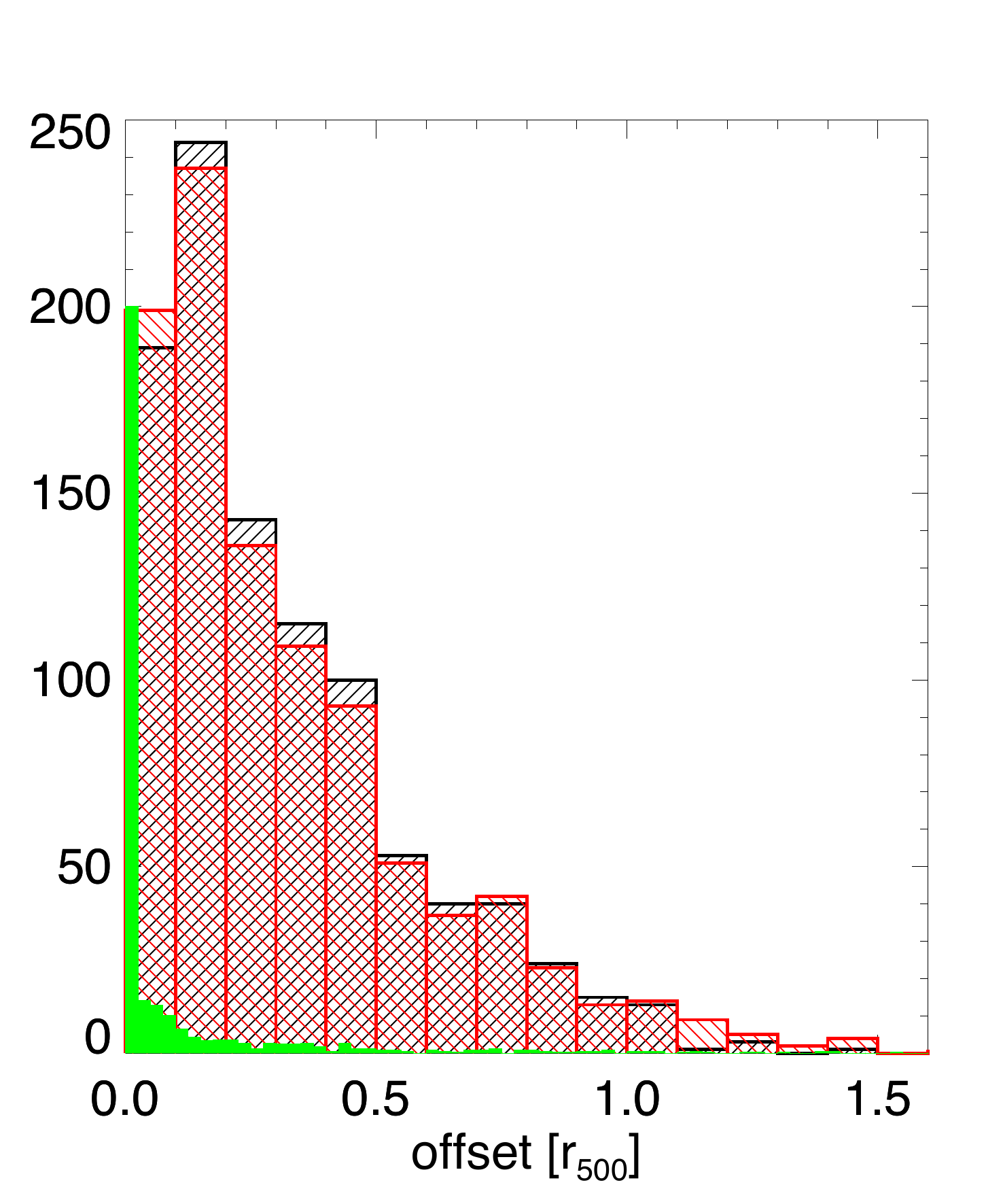}
\vskip-0.10in
\caption{Offset distribution of RM matched systems between default MCMF and 2RXS (black), RM and 2RXS (red) 
and default MCMF and RM (filled, green). The green histogram was scaled down by a factor of 2.3 to fit into the plot.
}
\label{fig:offsetsvsz4}
\end{figure} 

\subsection{RedMaPPer clusters}  \label{app:RMcomp}
The RM algorithm \citep{rykoff14} relies on a RS technique to identify clusters using optical imaging 
data with multi-band photometry. Within DES the RM cluster catalog is the most widely used cluster catalog based 
on DES data alone and will be the basis of a cluster based cosmological analysis \citep{mcclintock19}.
 MCMF and RM share the same input data (DES) and both rely on the RS technique. However, they do not share software and RS models,
 and use different ways to derive redshifts and richnesses. The usage of prior information from 
2RXS and less stringent cuts on the optical data could cause additional differences in performance.
 
Given the high source density of the RM catalog and the positional accuracy of 2RXS, mismatches 
between 2RXS clusters and the true RM counterpart will result. We therefore use different matching constraints in our comparisons, 
depending on which 
quantity is investigated.

We first test the photo-z's for consistency. For that we match the 2RXS catalog with \fcont$
<0.1$ with RM systems within 200\arcsec\ from the X-ray and 75\arcsec\ from the optical position. To further reduce 
the contamination by mismatches we additionally require the richness given by RM (\LamRM) 
to be larger than 20. Fig.~\ref{fig:compRedmap} shows the redshift comparison of both matched samples. The most 
obvious feature that can be seen is that mismatches tend to result in matches with higher redshift than that given 
by MCMF. This can be explained with the larger volume that is searched for a possible counterpart.
At low redshifts, bellow $z=0.15$ we see a systematic offset between RM and MCMF redshifts. This is partially due to 
the lower limit in redshift of the RM catalog, causing z<0.1 sources scattered up into the RM sample. 
From testing against spec-z's we know that MCMF does not show a significant bias at this redshift range. A potential bias in 
RM might persist, but will likely not affect the majority of science drawn from RM clusters because the suggested 
redshift cut for RM is z>0.2.
Applying a cut at $z_{\mathrm{MCMF}}>0.15$ we find a nearly Gaussian offset distribution between MCMF and 
RM of $\sigma_{\Delta z/ (1+z)}=0.0047$.

The richness of RM clusters is the primary mass proxy for DES cluster cosmology. A special 
focus has therefore been placed on a well behaved low scatter richness estimator. The richness provided by MCMF is only an 
additional mass proxy, due to the fact that the X-rays as a primary mass proxy are used to detect the clusters. The 
main purpose of the MCMF richness is therefore to allow reasonable cleaning of the input cluster candidate list. 
Because this is based on comparing richnesses of observed with those of random sources, potential undesired
dependencies on redshifts or X-ray count rate cancel out in the computation of \fcont. The requirements on the richness 
estimate are therefore lower for MCMF than for RM.

For comparing RM and MCMF richness we make use of the good agreement between RM and MCMF redshifts. 
We match all sources with redshift differences $\Delta z/(1+z)<0.01$ and RM position within 200\arcsec\ of the X-
ray and optical position without a cut in richness. The richness estimate is largely unaffected
by the probability of a cluster candidate to be indeed an X-ray detected cluster 
or a chance superposition. We therefore show the matches with \fcont$<0.1$ as 
well as \fcont$<1.0$.
As can be seen in Fig.~\ref{fig:compRMlambda} \LamMCMF\ scales well with \LamRM, even in the cases 
where the X-ray measurements that define the aperture for \LamMCMF\ are likely not related to 
the cluster. The median ratio\LamMCMF/\LamRM\ for \fcont$<0.1$ is found to be 1.088.

After comparing the redshift and richness estimates between MCMF and RM we now investigate the centering 
performance. For that we match the MCMF \fcont$<0.1$ catalog with the RM using a richness cut of 
\LamRM$>15$, a redshift difference of $\Delta z/(1+z)<0.015$ and an initial cross match radius of 
1100\arcsec\ (2~Mpc at $z=0.1$). The redshift and richness cut allows us to use a large matching 
radius to avoid artificially truncating the offset distribution.
Similar to Fig.~\ref{fig:offsetsvsz1}, Fig.~\ref{fig:offsetsvsz3} shows the median offset within $r_{500}$ between RM 
center and various MCMF centers.
A zero offset means that the default MCMF center is identical in more than 50\% of the cases with the RM center. In 68\% of the cases 
our automatic rBCG selection identifies the same source as BCG as found in RM as the most central galaxy.
The red line, representing the
BCG positions agree in more then 50\% of the cases over all redshifts, while the 2D fit and galaxy density center show a median 
offset of 0.1 $r_{500}$.
The dotted line in the upper panel of  Fig.~\ref{fig:offsetsvsz3} shows the X-ray to RM offset for the matched 
systems, the blue dashed line the default MCMF to 2RXS positions for the same sources.
Here the RM center seems to perform slightly better than the default MCMF center, but the difference is small and 
does not exceed 0.05 $r_{500}$.  The bottom panel of the same figure shows again the fraction of sources with 
offsets larger $r_{500}$ using the same color coding as in the top panel. The default MCMF center shows only a 
small fraction of sources with offsets larger than $r_{500}$.
The lower fraction of bad centers of MCMF to 2RXS centers compared to RM to 2RXS centers might be caused by 
the follow-up nature of MCMF compared to an independent center search by RM.

We do not see a significant redshift trend of the centering performance comparison between RM and optical MCMF. 
However we do see the same redshift dependent trend between optical and X-ray centers for both RM and MCMF, 
(see Fig.~\ref{fig:offsetsvsz1}).
 
In addition to the compressed information of Fig.~\ref{fig:offsetsvsz1}, we additionally show the full distribution of RM to 
2RXS, default MCMF to 2RXS and default MCMF to RM for all matches over the the full redshift range in 
Fig.~\ref{fig:offsetsvsz4}. We find that 71\% of the clusters show an offset between MCMF and RM center of less than $0.05 r_{500}$.

As a final test we investigate all clusters with $0.2<z<0.9$ and $M_{500}>3.5\times10^{14}$\Msun~ that do not have a RM counterpart within 200\arcsec. We find all clusters without nearby RM counterpart are within or near a region that is flagged by the DES foreground flag or has regions masked in its vicinity. 
The DES RM catalog follows stricter masking and flagging requirements than MCMF, and this explains the missing clusters in the RM catalog. 
Optical investigation of the clusters does not show that the imaging quality (background and colors) is significantly off from the average.
We therefore do not consider stricter masking and flagging requirements for MCMF.

\begin{figure}
\includegraphics[keepaspectratio=true,width=\columnwidth]{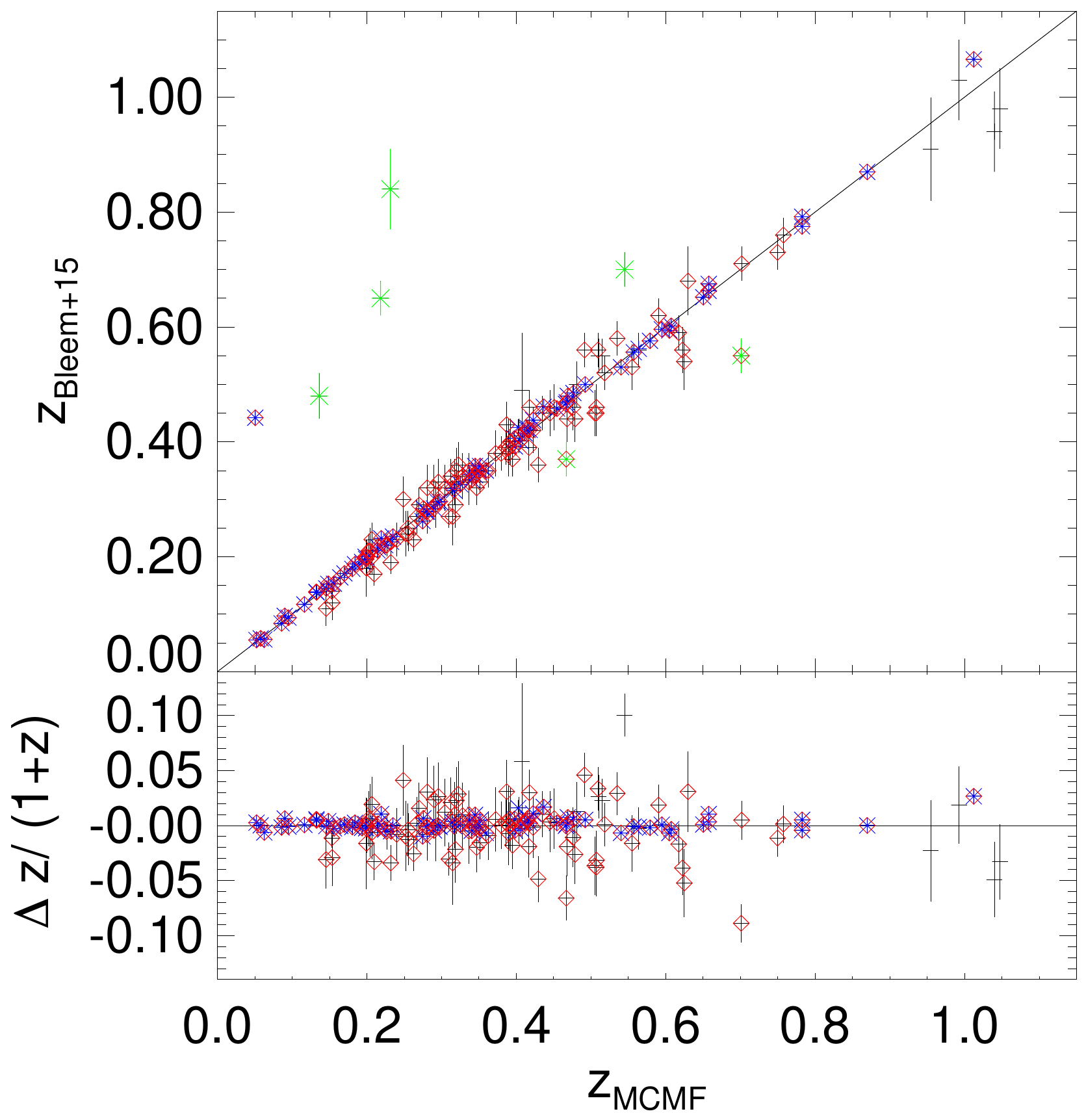}
\vskip-0.1in
\caption{Photo-z comparison between SPT-SZ and MCMF: Cluster with spec-z's are shown in blue, 
clusters with \fcont$<0.1$ are marked with red diamonds. Outliers with photometric redshifts are 
highlighted in green.}
\label{fig:compSPTz}
\end{figure}

\begin{figure}
\includegraphics[keepaspectratio=true,width=\columnwidth]{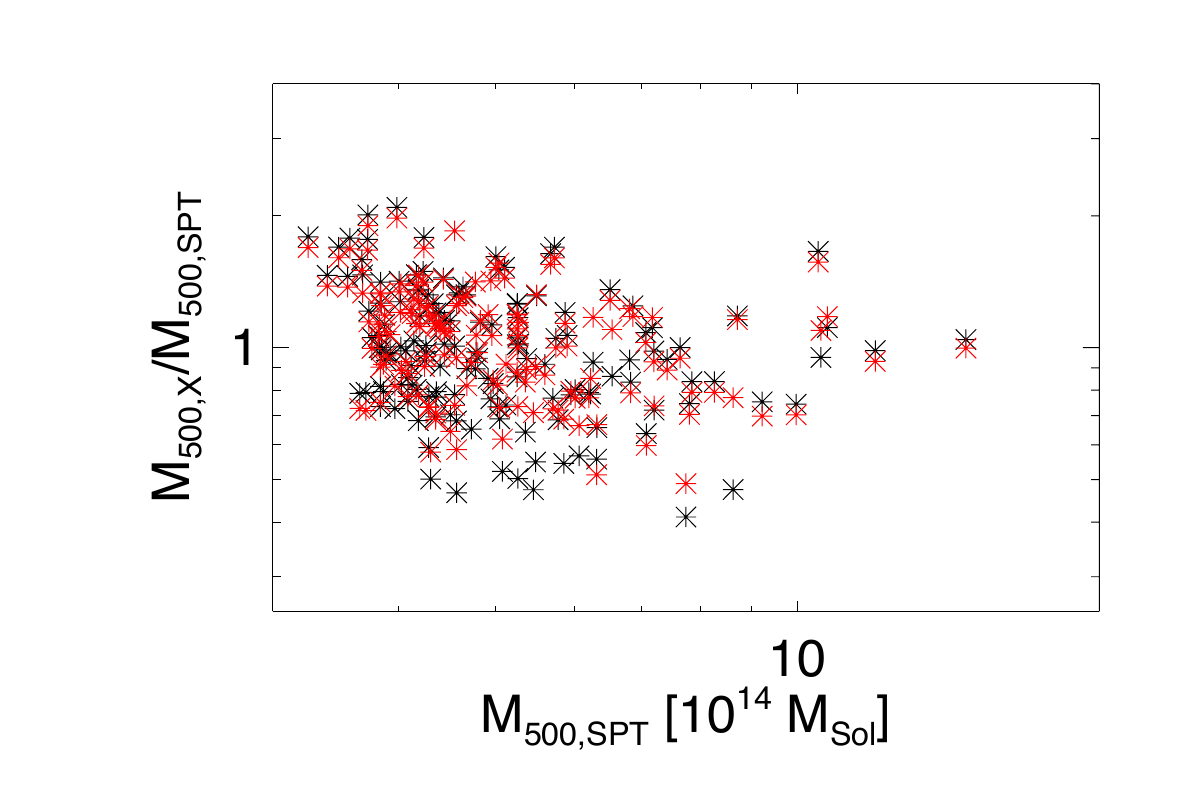}
\vskip-0.10in
\caption{Mass comparison between X-ray based mass and SZE based mass after correcting for a calibration factor in luminosity. Shown are only matches with \fcont$<0.1$, z>0.2 and $|z_\mathrm{MCMF}-z_\mathrm{SPT}|<0.1$. Mass estimates including the redshift dependent aperture correction are shown in red.)}
\label{fig:correctedMass}
\end{figure}

\begin{figure}
\centering
\includegraphics[keepaspectratio=true, width=0.8\columnwidth]{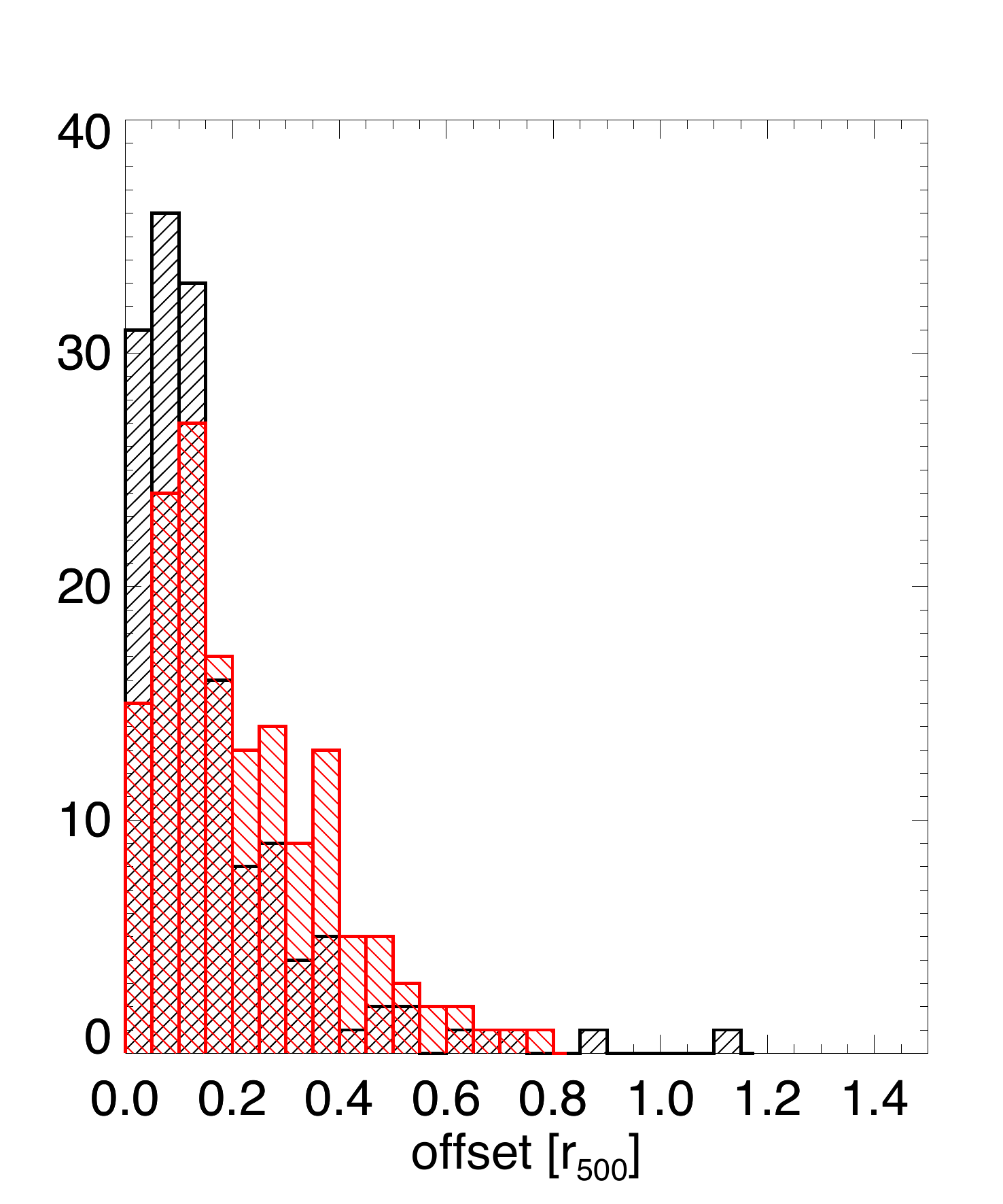}
\vskip-0.10in
\caption{The offset distribution between default MCMF center and SPT-SZ center in black. The offset distribution 
between 2RXS position and SPT-SZ position is shown in red.}
\label{fig:offsetsSPT}
\end{figure}

\subsection{SPT-SZ clusters}\label{app:SPTcomp}

Using a 150\arcsec\ cross-match radius, we find 168 matches with the SPT-SZ cluster catalog \citep{bleem15}. 153 
of those sources have  \fcont$<0.1$, indicating that the majority of these systems are indeed true X-ray 
detected clusters.
For a subset of 74 systems, listed as clusters with a spec-z, we find a scatter in $\Delta z/ (1+z)$ of 0.005. 
Fig.~\ref{fig:compSPTz} shows the photo-z comparison between MCMF and SPT clusters. Blue points show 
clusters with spec-z's, and clusters falling into our \fcont$<0.1$ selection are marked with red 
diamonds.

We find one outlier with a spec-z (SPT-CL J0330-5228) to be off by $\Delta z=0.4$. Similar to previous 
comparisons to other catalogs we find two peaks in redshift with low \fcont\ of 0.003 and 0.010. The 
second peak with $z_\mathrm{MCMF}=0.437$ is consistent with the spec-z of z=0.442.

Further, we investigate all clusters whose photo-z given in \citet{bleem15} differs by more than three sigma from our 
redshifts.  We find seven outliers, only two of them with \fcont$<0.1$.
Out of those seven, only one (SPT-CL J0306-4749) shows a single richness peak in MCMF,  $\Delta z=0.097$ away from the 
redshift given in \citet{bleem15}. After inspection of this cluster we could not find any reason why the MCMF 
redshift is off by this amount. In fact at the redshifts in question, $z=0.467$ in MCMF and $z=0.37$ in \citet{bleem15}, it is barely possible to shift the redshift estimate by this amount and to obtain a reasonable peak profile in MCMF. Similar to the case of MACSJ0257.6-2209 one would expect that the cluster signal would be dramatically reduced if one of the colors would be off by the amount that is needed to shift the redshift estimate by 0.097. But without spectroscopic data as a final proof the discrepancy stays unresolved.  
 
The second outlier with \fcont$<0.1$ is indeed interesting. Thanks to the de-blending 
capabilities of MCMF in redshift space, we find two peaks at z=0.54 and z=0.70, and
both have similar \fcont\ va,ues of 0.0479 and 0.0486. Both peaks are therefore equally good 
counterparts given the X-ray information. Fig.~\ref{fig:SPTdoublepeak} shows the $\lambda$ versus z distribution 
for the 2RXS source together with the a $grz$ pseudo-color image of the central region.
The separation between both clusters  is less than 25\arcsec, the distance between the 2RXS and the 
SPT-SZ position is $50\arcsec$.
We note here that for being at the same \fcont, the cluster at higher redshift needs be at higher richness. 
For an MCMF follow-up of SZE selected sources, \fcont\ would likely behave somewhat differently with redshift.
In extreme cases this could lead to different counterparts for SZE and X-ray based follow-up,
even if the candidate position is the same. MCMF follow-up of X-ray based catalogs will tend to select 
lower redshift counterparts while MCMF follow-up of SPT-SZ like catalogs will prefer 
the more massive counterpart.
We find that about 2-3\% of MARD-Y3 clusters show a second peak in redshift 
with \fcont\ less than 0.1 higher than the main counterpart (i.e., with only a 10\% higher chance of being a random
superposition than the primary peak).

The remaining four outliers with \fcont$>0.1$ all show multiple peaks in redshift and offsets ranging 
from 80\arcsec\ to 170\arcsec\ from the 2RXS position. In each case a richness peak consistent with the photo-z given in  
\citet{bleem15} is found.

Besides redshifts, the SPT-SZ cluster catalog offers SZE based mass proxies, which were also used for the X-ray 
scaling relation presented in \citet{bulbul19} that is used in this work to define the region of interest.
For defining the region of interest, we ignored the impact of the fixed aperture on the fraction of the cluster 
flux measured within that aperture, as well as the fact that the \citet{bulbul19} scaling relations were 
obtained using the 0.5-2.0 keV rest frame energy band, instead of 0.1-2.4 keV used here.
With a matched sample of clusters one could ask how wrong these assumptions are. Unfortunately the SPT-SZ masses 
are also less reliable for z<0.2 due to the fact that cluster size becomes increasingly comparable with the primary CMB fluctuations.
We are therefore not able to constrain the aperture size effect. Therefore, we restrict ourselves to measuring the median mass offset between
those mass estimates. Additionally, we derive a factor $f_{0520}$ which translates the $0.1-2.4$ keV X-ray luminosity calibrated to MCXC ($L_\mathrm{500,2RXS}$) to $L_\mathrm{500,0.5-2.0}$, the luminosity in the 0.5-2.0~keV band that provides consistent masses
 to SPT-SZ and therefore should provide consistent luminosities to \citet{bulbul19}.
We expect the value of this correction to be different from 1 not only because of the different spectral range, 
but also because of systematic differences between mass estimates in MCXC and SPT-SZ. 
The conversion factor therefore converts luminosities widely used in RASS based analysis to 0.5-2.0~keV luminosities 
within $r_{500}$ derived from state of the art scaling relations. A direct matching of MARD-Y3 sources with SPT-SZ 
sources used in \citet{bulbul19} provides just a rough estimate of $f_{0520}=L_\mathrm{500,2RXS}/L_\mathrm{500,0.5-2.0}\approx0.75$ 
due to the small number of matches and the large scatter between single measurements.

Selecting a sample with \fcont$<0.1$, $z>0.2$, $M_{500,\mathrm{SPT}}>4\times10^{14} M_\odot$ and 
$|z_\mathrm{MCMF}-z_\mathrm{SPT}|<0.1$, we find a median mass ratio 
$M_{500,\mathrm{L,x}}/M_{500,\mathrm{SPT}}$ of 1.12 for the uncorrected MARD-Y3 luminosity $L_\mathrm{X}$
and 1.07 for the aperture correct luminosity ($L_\mathrm{500,2RXS}$). 
For $M_{500,\mathrm{SPT}}$ we make use of the scaling relation given in \cite{dehaan16} that is derived using SPT-SZ cluster counts together
with external constraints from BAO studies as well as priors from BBN and direct H0 measurements.
The conversion factor to get from the corrected luminosity to $L_\mathrm{500,0.5-2.0}$ is  $f_{0520}=0.87\pm0.02$. 
The factor going from the uncorrected luminosity $L_\mathrm{X}$ to  $L_\mathrm{500,0.5-2.0}$ is 0.80. 
The mass ratios as a function of mass is shown in Fig.~\ref{fig:correctedMass}, using the listed correction terms.
Throughout this paper we are using the masses obtained using $L_\mathrm{500,2RXS}*f_{0520}$. We further note that 
the biased mass by factor 1.12 for the $L_\mathrm{X}$ would lead to an overestimate by 4\% of the MCMF followup region of interest 
as compared to the true $r_{500}$.
 
This simple mass calibration does not replace a real scaling relation analysis between observables 
($L_\mathrm{X}$, \LamMCMF) and mass. A detailed study of these scaling relations is the subject of a forthcoming paper 
(Paulus et al.,in prep) that will use weak lensing masses derived from DES data.

The masses from SPT-SZ and the well known selection function of the catalog in general allows us to pose the question: 
do we see the expected number of matches and non-matches given both SZE and X-ray selection functions and the corresponding 
mass proxies?
Answering this question is part of a forthcoming paper (Grandis et al.,in prep.) as part of a detailed 
examination of the usefulness of the MARD-Y3 catalog for cosmology.

We therefore ask here only the simpler question: given a match between a 2RXS source and an SPT-SZ cluster, 
does MCMF provide a reasonable confirmation?
In Fig.~\ref{fig:SPTmatchesMassRedshift} we show the distribution of clusters in the mass-redshift plane. 
The top panel shows  X-ray luminosity based masses and MCMF redshifts. The mass $M_\mathrm{500,X}$ uses aperture corrected luminosities. 
The full \fcont$<0.2$ sample is shown in gray, and the SPT-SZ matches are shown in blue. Unconfirmed 2RXS matches 
(with \fcont$>0.2$ above the cut used to produce the MARD-Y3 sample) are shown in red.
From this plot we can already see that all unconfirmed matches are 
at the lower limit or below the mass range probed by 2RXS at a given cluster redshift.
The bottom plot in Fig.~\ref{fig:SPTmatchesMassRedshift} shows the same 
MARD-Y3 clusters in gray, but uses SPT-SZ masses instead of X-ray masses for the SPT-SZ systems. 
The intrinsic mass to observable as well as the absolute scatter of the mass proxy 
is typically smaller for the SPT-SZ mass proxy. 
Additionally we show SPT-SZ clusters that do not have a 2RXS match but lie within the DES footprint (black).

We find that three out of five unconfirmed matches are consistently below the mass limit of 
RASS using both mass proxies. A fourth source is a match to SPT-CL J0459-4947 (z>1.5), 
which is too high in redshift and too low in mass to have been realistically detected by RASS or MCMF using DES. 
In this case MCMF finds a low-z (z=0.25) counterpart, which is still too low in mass to be a reasonable match.
Only one source shows a SPT-SZ mass at a redshift such that it might be a reasonable match. 
Our investigation shows that the offset between the 2RXS source and the  SPT-SZ cluster 
is $\sim$1.2$r_{500}$,  using the X-ray mass. The separation is therefore larger than the aperture
used to confirm the 2RXS source. Such a large offset is indicative of this not being a true match.

As a last comparison (in this case one similar to the comparison to RM in subsection~\ref{app:RMcomp}), 
we also investigate the offset distribution 
between our MARD-Y3 clusters and the SPT-SZ clusters. We use an extended 500\arcsec\  matching radius together with 
$\Delta z<0.1$ redshift cut and match to the \fcont$<0.1$ catalog.  The result is shown in 
Fig.~\ref{fig:offsetsSPT}. The cross match radius corresponds to $\sim r_{500}$ at z=0.1 
and $\sim 4*r_{500}$ at $z=0.9$. 
The median offset between default MCMF center and SPT-SZ center is 0.12 $r_{500}$. The median offset between 
2RXS and SPT-SZ  position is 0.18 $r_{500}$, which is the same as the default MCMF to 2RXS offset distribution for 
the matches sources.

\begin{figure}
\centering
\includegraphics[keepaspectratio=true, width=0.9\columnwidth]{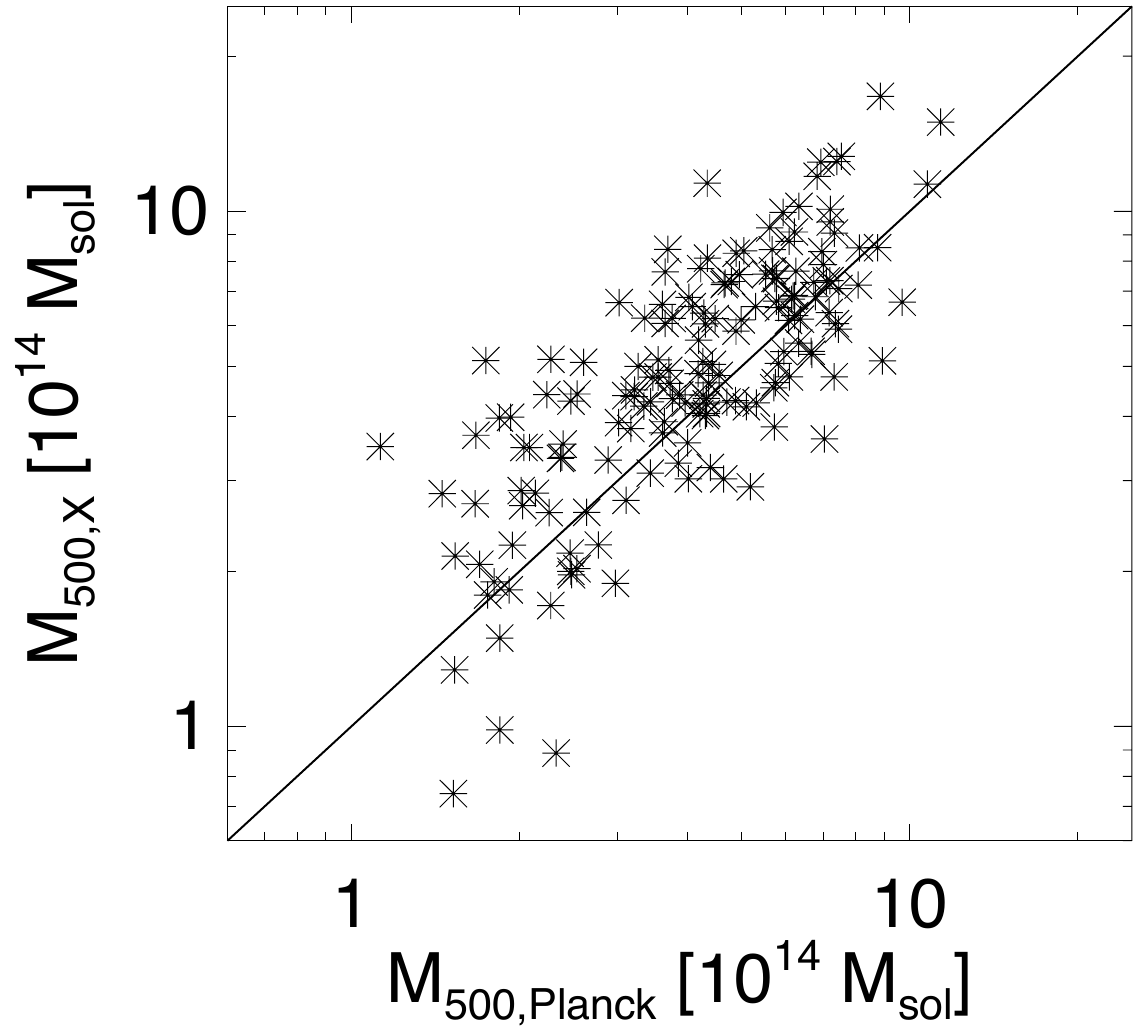}
\vskip-0.10in
\caption{Comparison of mass estimates from Planck to our X-ray based mass estimates. The X-ray masses do include the calibrations from sec.\ref{app:SPTcomp} and ~\ref{app:MCXCcomp}. The black line indicates the one to one relation. The median mass ratio  $M_\mathrm{500,Planck}/M_\mathrm{500,X}$ is 0.84.}
\label{fig:PlanckMasscomp}
\end{figure}

\subsection{Planck clusters}  
\label{app:Planckcomp}

The Planck PSZ2 cluster catalog \citep{planck25-27} is an all-sky catalog of SZE sources using the full 29 months of Planck observations.
We find 266 clusters within the DES footprint, 211 of which have redshifts listed in the PSZ2 catalog and 227 have external validation.
Given the large positional uncertainties for both RASS and PSZ2 clusters we use a 360\arcsec\ cross matching radius, 
within which we find 201 matches. From those, only five sources show \fcont$>0.2$.
Optical investigation shows that those two are indeed not matches to 2RXS, with one lying in a masked region. The two remaining clusters are MACSJ0257.6-2209, which was discussed in sec.\ref{app:MCXCcomp}, and PSZ2 G258.33-38.54. The latter has \fcont$=0.23$ and '
the X-ray based mass is close to the limit of MARD-Y3 clusters at the cluster redshift. 
We therefore assume that this is just a border-line case, and the \fcont\ estimate appears reasonable.

The redshifts given in PSZ2 are a collection from various different sources and are not split into photo-z or spec-z based estimates. 
Given the extensive tests on photo-z performance in previous sections we do not expect to learn much from a PSZ2 to MCMF comparison. 
We find a reasonable scatter of $\Delta z/(1+z_\mathrm{MCMF})=0.006$ based on 173 matches, 
indicating that the majority of the PSZ2 sources within DES do have reasonable redshifts.
We find seven sources with more than a 5\% offset in redshift. Investigation shows two to be positional mismatches, 
four to have photo-z and the last one is RXC J0605.8-3518, discussed in sec.\ref{app:MCXCcomp}.
We further find 21 sources with \fcont$<0.2$ that do not have redshifts listed in PSZ2.

Finally we compare the mass estimates obtained using the X-ray luminosities, including corrections to SPT-SZ clusters 
and those provided in the PSZ2 catalog. We restrict the comparison to matches with redshift offsets smaller 
than $\Delta z=0.05$, \fcont$<0.1$ and AGN rejection, which results in 162 clusters. 
As shown in Fig.~\ref{fig:PlanckMasscomp}, the masses do follow a reasonable scaling, but with a mass offset. 
We find a median mass ratio $M_\mathrm{500,X}/M_\mathrm{500,Planck}$ of 1.19.  We note that the masses given in 
PSZ2 are without applying a correction for the hydrostatic mass bias and that it is therefore expected that these masses
would be systematically low compared to the  SPT-SZ calibrated X-ray masses.

\begin{landscape}
\begin{table}
 	\centering
 	\caption{First twenty entries of the MARD-Y3, limited to the most important columns of the catalog. The full catalog ( $f_\mathrm{cont}\leq0.2$) including additional columns will be available online via VizieR server at CDS (\textit{http://vizier.u-strasbg.fr}). The first three columns show the cluster name and the position of the  X-ray source. The following columns are MCMF derived quantities related to the redshift peak that provides the lowest $f_\mathrm{cont,r}$ (LFCR). The columns show the optical centre, $f_\mathrm{cont,r}$,  redshift (z), richness ($\lambda$).
 	The luminosity corrected for fixed aperture (L\_LFCR\_COR), the mass assumed for the aperture definition to obtain $\lambda$ ( $\mathrm{M}_{500}$\_LFCR) and the mass accounting for aperture effects ($\mathrm{M}_{500}$\_LFCR\_COR). Masses are given in $10^{14} \mathrm{M}_\odot$. MULTI is equal one if a X-ray source is likely a multiple detection of one optical cluster and is not the most likely counter part. The last entry, $\mathrm{D}_\mathrm{sig}$\_LFCR, provides the distance to the observed lambda to X-ray based mass relation in terms of sigma for sources that do have a NWAY match. Sources without NWAY match are set to -99.9. The last column, DFC\_LFCR, provides the difference between the lowest  $f_\mathrm{cont,r}$ and the second lowest  $f_\mathrm{cont,r}$ and indicates the importance of a second structure at different redshift.	To reject AGN contamination and multiple detection as described in this paper we recommend to require MULTI = 0 and $\mathrm{D}_\mathrm{sig}$\_LFCR$ < 2.0$. Similar columns exist in the full catalog for up to three peaks found in the $\lambda$ vs. z distribution.}
 	\label{tab:clust_table}
 	\setlength\tabcolsep{3.0pt}
 	\begin{tabular}{l c c c c c c r c c c c r r } 
 		\hline
 		NAME & RA & DEC &  RA\_LFCR & DEC\_LFCR & \fcont\_LFCR & z\_LFCR & $\lambda$\_LFCR & L\_LFCR\_COR & $\mathrm{M}_{500}$\_LFCR & $\mathrm{M}_{500}$\_LFCR\_COR & MULTI & $\mathrm{D}_\mathrm{sig}$\_LFCR & DFC\_LFCR \\ 	 		
 		\hline 		
 MARD	J013916.2-073326	&	24.81285	&	-7.55064	&	24.81285	&	-7.55064	&	0.06	&	0.272	&	33.6	&	1.772	&	4.32	&	4.22	&	0	&	1.44	&	0.49	\\
MARD	J013916.2-073933	&	24.81545	&	-7.66385	&	24.81545	&	-7.66385	&	0.01	&	0.271	&	57.3	&	1.471	&	3.92	&	3.81	&	0	&	-99.90	&	98.99	\\
MARD	J013240.5-080406	&	23.22015	&	-7.98288	&	23.22015	&	-7.98288	&	0.06	&	0.150	&	25.0	&	3.140	&	4.81	&	6.13	&	0	&	3.29	&	0.58	\\
MARD	J013636.0-080614	&	24.14816	&	-8.10180	&	24.14816	&	-8.10180	&	0.01	&	0.145	&	43.4	&	0.988	&	2.60	&	3.35	&	0	&	-99.90	&	0.32	\\
MARD	J014632.6-081332	&	26.67190	&	-8.22717	&	26.67190	&	-8.22717	&	0.10	&	0.434	&	32.4	&	1.207	&	3.79	&	2.98	&	0	&	-99.90	&	0.72	\\
MARD	J013149.8-081912	&	23.00393	&	-8.35457	&	23.00393	&	-8.35457	&	0.02	&	0.148	&	24.8	&	0.283	&	1.36	&	1.69	&	0	&	-99.90	&	98.98	\\
MARD	J014719.6-082234	&	26.85084	&	-8.36657	&	26.85084	&	-8.36657	&	0.00	&	0.789	&	111.3	&	9.532	&	9.21	&	6.98	&	0	&	-99.90	&	0.59	\\
MARD	J015442.7-082606	&	28.68288	&	-8.42619	&	28.68288	&	-8.42619	&	0.00	&	0.207	&	49.4	&	1.123	&	3.11	&	3.40	&	0	&	-99.90	&	0.64	\\
MARD	J025149.7-081342	&	42.94518	&	-8.17774	&	42.94518	&	-8.17774	&	0.05	&	0.317	&	44.7	&	2.266	&	5.18	&	4.51	&	0	&	1.20	&	0.68	\\
MARD	J013724.6-082744	&	24.35405	&	-8.45635	&	24.35405	&	-8.45635	&	0.00	&	0.564	&	130.4	&	3.393	&	6.07	&	4.76	&	0	&	-99.90	&	1.00	\\
MARD	J010846.5-151430	&	17.19710	&	-15.41502	&	17.19710	&	-15.41502	&	0.19	&	0.039	&	8.6	&	0.041	&	0.34	&	0.66	&	1	&	-2.79	&	0.54	\\
MARD	J011304.7-151600	&	18.28428	&	-15.29383	&	18.28428	&	-15.29383	&	0.07	&	0.110	&	14.4	&	0.188	&	1.04	&	1.41	&	0	&	-0.59	&	98.93	\\
MARD	J012928.6-152254	&	22.36509	&	-15.37388	&	22.36509	&	-15.37388	&	0.20	&	0.387	&	21.6	&	0.824	&	3.28	&	2.55	&	0	&	-99.90	&	0.50	\\
MARD	J010847.9-152438	&	17.19764	&	-15.41452	&	17.19764	&	-15.41452	&	0.00	&	0.054	&	39.7	&	0.648	&	1.62	&	2.83	&	0	&	-99.90	&	0.01	\\
MARD	J010904.9-152752	&	17.19743	&	-15.41506	&	17.19743	&	-15.41506	&	0.01	&	0.052	&	28.9	&	0.167	&	0.79	&	1.38	&	1	&	-99.90	&	0.01	\\
MARD	J014925.0-143854	&	27.36511	&	-14.65254	&	27.36511	&	-14.65254	&	0.03	&	0.132	&	20.3	&	0.185	&	1.06	&	1.37	&	0	&	-99.90	&	0.60	\\
MARD	J014132.6-152757	&	25.41928	&	-15.53233	&	25.41928	&	-15.53233	&	0.07	&	0.234	&	35.8	&	8.122	&	9.29	&	9.60	&	0	&	4.12	&	98.93	\\
MARD	J015750.2-151751	&	29.45277	&	-15.30753	&	29.45277	&	-15.30753	&	0.08	&	0.176	&	20.3	&	0.641	&	2.20	&	2.62	&	0	&	0.82	&	98.92	\\
MARD	J022320.3-144734	&	35.83870	&	-14.79612	&	35.83870	&	-14.79612	&	0.00	&	0.865	&	90.1	&	7.282	&	7.67	&	5.53	&	0	&	-99.90	&	99.00	\\
MARD	J025733.5-150827	&	44.37622	&	-15.15300	&	44.37622	&	-15.15300	&	0.04	&	0.366	&	52.8	&	2.049	&	5.35	&	4.14	&	0	&	-99.90	&	98.96	\\
\hline
 	\end{tabular}
 \end{table} 

\end{landscape}

\label{lastpage}
\end{document}